\documentclass[superscriptaddress,amsmath,amssymb,prd,twocolumn,nofootinbib]{revtex4-1}
\pdfoutput=1
\usepackage{natbib}
\usepackage{graphics}
\usepackage{multirow}
\usepackage{amsbsy}
\usepackage{footmisc}
\usepackage{hyperref}
\usepackage{amsmath}
\usepackage{aas_macros}

\usepackage{array}
\usepackage[mathscr]{eucal}
\usepackage{morefloats}
\usepackage{amssymb}
\usepackage{txfonts}
\usepackage{placeins}
\usepackage{natbib}

\DeclareMathAlphabet{\mathbfsf}{\encodingdefault}{\sfdefault}{bx}{n}

\usepackage{verbatim}
\usepackage{graphicx}
\input{epsf}
\usepackage{subfigure}
\usepackage{color}
\usepackage{threeparttable}
\usepackage{comment}
\usepackage{epsfig}
\usepackage{xspace}
\usepackage{enumitem}
\usepackage{hyperref}
\usepackage{ulem}
\usepackage{courier}
\usepackage{appendix}
\usepackage{comment}
\usepackage[usenames,dvipsnames,svgnames]{xcolor}
\usepackage{latexsym}
\usepackage{url}

\newcommand{\lsim}{\,{}^<_{\sim}\,}

\hypersetup{
  colorlinks = true,
  urlcolor = blue,
  linkcolor=blue,
  citecolor=blue
}

\usepackage[applemac]{inputenc}

\newcommand{\mksym}[1]{\ifmmode {\rm #1}\else #1\fi}

\def\epstwo@scaling{1.1}

\def\showtwo#1#2{
  \centering
  \leavevmode
  \epsfxsize=\epstwo@scaling\linewidth
  \epsfbox{#1.eps} \hfil
  \epsfxsize=\epstwo@scaling\linewidth
  \epsfbox{#2.eps}
}

\def\showthree#1#2#3{
  \centering
  \leavevmode
  \epsfxsize=\epstwo@scaling\linewidth
  \epsfbox{#1.pdf} \hfil
  \epsfxsize=\epstwo@scaling\linewidth
  \epsfbox{#2.pdf} \hfil
  \epsfxsize=\epstwo@scaling\linewidth
  \epsfbox{#3.pdf}  
}

\def\showfour#1#2#3#4{
  \centering
  \leavevmode
  \epsfxsize=\epstwo@scaling\linewidth
  \epsfbox{#1.eps} \hfil
  \epsfxsize=\epstwo@scaling\linewidth
  \epsfbox{#2.eps} \hfil
  \vskip 0.2in
  \epsfxsize=\epstwo@scaling\linewidth
  \epsfbox{#3.eps} \hfil
  \epsfxsize=\epstwo@scaling\linewidth
  \epsfbox{#4.eps} \hfil  
}

\def\showfourvertical#1#2#3#4{
  \centering
  \leavevmode
  \epsfxsize=\epstwo@scaling\linewidth
  \epsfbox{#1.eps} \hfil
  \vskip -0.1in
  \epsfxsize=\epstwo@scaling\linewidth
  \epsfbox{#2.eps} \hfil
  \vskip -0.1in
  \epsfxsize=\epstwo@scaling\linewidth
  \epsfbox{#3.eps} \hfil
  \vskip -0.1in
  \epsfxsize=\epstwo@scaling\linewidth
  \epsfbox{#4.eps} \hfil  
}

\newcommand{\yah}[1]{{\textcolor{blue}{#1}}}
\newcommand{\yahc}[1]{{\textcolor{red}{#1}}}
\newcommand{\av}[1]{\langle {#1} \rangle}
\newcommand{\beq}{\begin{equation}}
\newcommand{\eeq}{\end{equation}}
\newcommand{\barr}{\begin{eqnarray}}
\newcommand{\earr}{\end{eqnarray}}

\begin{document}
\title{Cosmic Recombination in the Presence of Primordial Magnetic Fields}

\author{Karsten Jedamzik}
\email[]{karsten.jedamzik@umontpellier.fr}
\affiliation{Laboratoire Univers et Particules de Montpellier (LUPM), Universit\'e de Montpellier (UMR-5299) \& CNRS, Place Eug\`ene Bataillon, F-34095 Montpellier Cedex 05, France}

\author{Tom Abel}
\email[]{tabel@stanford.edu}
\affiliation{Kavli Institut for Particle Astrophysics and Cosmology,
Stanford University, 452 Lomita Mall, Stanford, CA 94305, USA}
\affiliation{
Department of Physics, Stanford University, 382 Via Pueblo Mall, Stanford, CA 94305, USA}
\affiliation{
SLAC National Accelerator Laboratory, 2575 Sand Hill Road, Menlo Park, CA  94025, USA}

\author{Yacine Ali-Ha\"imoud}
\email{yah2@nyu.edu}
\affiliation{Center for Cosmology and Particle Physics, Department of Physics, New York University, New York, NY, USA}

\date{\today}
\begin{abstract}

Primordial magnetic fields (PMFs) may explain observations of magnetic fields
on extragalactic scales. They are most cleanly constrained by measurements of cosmic microwave background radiation (CMB) anisotropies. Their effects on cosmic recombination may even be at the heart of the resolution of the Hubble tension. We present the most detailed analysis of the effects of PMFs on cosmic recombination to date. 
To this end we extend the public magneto-hydrodynamic code {\sl ENZO} with a new cosmic recombination routine, Monte-Carlo simulations of Lyman-$\alpha$ photon transport, and a Compton drag term in the baryon momentum equation. The resulting code allows us, for the first time, to realistically predict the impact of PMFs on the cosmic ionization history and the clumping of baryons during cosmic recombination. Our results identify the importance of mixing of Lyman-$\alpha$ photons between overdense- and underdense- regions for small PMF strength. This mixing speeds up recombination beyond the speed-up due to clumping. We also investigate the effects of pecuilar flows on the recombination rate and find it to be small for small PMF strengths.
For non-helical PMFs with a Batchelor spectrum we find a surprising dependency of results on ultra-violet magnetic modes. 
We further show that the increase in the ionization fraction at low redshift by hydrodynamic baryon heating due to PMF dissipation is completely compensated by the faster recombination from baryon clumping.
The present study shall serve as a theoretical foundation for a future precise comparison of recombination with PMFs to CMB data.
\end{abstract}
\maketitle

\section{Introduction}
\label{sec:introduction}

The local Universe seems to be permeated by magnetic fields in virtually
all astrophysical environments observed \cite{Subramanian:2015lua}. Fields with large coherence scales
of $\sim \mu$G magnitude are present in the Milky Way and other galaxies.
Though details are not clear, these fields may be due to the action of a large-scale dynamo amplifying an initially small seed field \cite{Brandenburg:2004jv}. Fields of similar
strengths are also found in higher redshift galaxies \cite{1998A&A...329..809A,Wolfe:2008nk}, where
the fewer rotations those galaxies have undergone should make the dynamo less
efficient. Magnetic fields of $\mu$G strengths are observed in clusters of 
galaxies and could be explained by outflows of the fields in galaxies within the cluster \cite{Aramburo-Garcia:2022gzn}. More difficult to explain are the fields of $\sim$ 50 nG strengths recently observed in filaments of the cosmic structure \cite{2019Sci...364..981G,Vernstrom:2021hru,Carretti:2022tbj}, in factor $\sim$ ten overdense regions, and it has been argued that a primordial origin may be the best explanation \cite{Carretti:2022fqk}. It came to a surprise in 2010 
that observations of TeV and GeV $\gamma$-rays from blazars seem be most
straightforwardly interpreted by an almost volume-filling cosmic magnetic field in the extragalactic medium, albeit of potentially weak strength $B_0 >  10^{-15}$G \cite{Neronov:1900zz,Tavecchio:2010mk,Tavecchio:2010ja,Taylor:2011bn,Vovk:2011aa,Dolag:2010ni}
(see also \cite{Podlesnyi:2022ydu,MAGIC:2022piy,Xia:2022uua,Dzhatdoev:2023opo,HESS:2023zwb,Huang:2023bod,Vovk:2023qfk} for more recent observations).
It may well
be that all of these fields have a purely astrophysical origin, nevertheless,
the community is far from a detailed understanding. 

Alternatively, it has been proposed that magnetic fields may be a remnant
of the early Universe. A plethora of magnetogenesis scenarios exist
(for reviews cf \cite{Durrer:2013pga,Subramanian:2015lua,Vachaspati:2020blt}). In case such a primordial magnetic field (hereafter, PMF)
had ever been in equipartition with radiation, such as plausible during 
cosmic phase transitions (e.g. the electroweak transition), the smallest fraction
$\sim 10^{-10}$ of this magnetic energy density surviving to the present, may
be sufficient to explain all present day fields without dynamo action. In such
scenarios the bulk of the magnetic energy density is indeed dissipated \cite{Banerjee:2004df}
with important details of such dissipation currently 
under debate \cite{Brandenburg:2014mwa,Reppin:2017uud,Hosking:2020wom,Hosking:2022umv,Zhou:2022xhk}, particularly in the non-helical case.
Another possibility is that a PMF is generated right during inflation, in case
conformal invariance of electro-magnetism is broken in the early Universe.
In contrast to phase-transition generated magnetic fields, which have a very
blue "Batchelor" spectrum \cite{Durrer:2003ja,Saveliev:2012ea}, inflationary produced fields have to be 
approximately scale-invariant and dissipation plays a much less important role.

It seems fair to say that the origin of magnetic fields in the present-day Universe is unknown, and further observations and theoretical work is needed.
A seemingly clean argument in favor of PMFs would be their detection at very high redshift, before the epoch of structure formation, which is thought to provide
the seed fields required by astrophysical generation mechanism\footnote{There
are also a number of recent studies attempting to constrain approximately scale-invariant PMFs from the enhanced structure formation on small scales
\cite{Sanati:2020oay,Adi:2023qdf}}. An excellent
probe could be the accurate observations of the spectrum of, and the anisotropies in the cosmic microwave background radiation (hereafter, CMB), which has already led to
precision determination of cosmological parameters in a $\Lambda$CDM model,
as well as its approximate confirmation. Indeed many authors have studied the
effects of PMFs on the CMB. They considered anisotropic expansion \cite{Barrow:1997mj}, spectral $y$ and $\mu$ distortions \cite{Jedamzik:1999bm,Zizzo:2005az,Kunze:2013uja}, anisotropies due to Alfven and slow magnetic waves \cite{Subramanian:1998fn,Subramanian:2002nh,Mack:2001gc,Lewis:2004kg,Kahniashvili:2005xe,Chen:2004nf,Lewis:2004ef,Tashiro:2005hc,Yamazaki:2006bq,Giovannini:2006gz,Kahniashvili:2006hy,Giovannini:2007qn,
 Yamazaki:2010nf,Paoletti:2010rx,
 Shaw:2010ea,Kunze:2010ys,
 Paoletti:2012bb,Ballardini:2014jta,Ade:2015cva,
 Sutton:2017jgr,Zucca:2016iur,Minoda:2020bod}, a changing ionization history due to plasma heating \cite{Sethi:2004pe,Kunze:2013uja,Kunze:2014eka,Ganc:2014wia,Chluba:2015lpa,
 Ade:2015cva}, extra polarization due to PMFs \cite{Durrer:1999bk,Seshadri:2000ky,Mack:2001gc,Subramanian:2003sh,
 Mollerach:2003nq,
 Lewis:2004kg,
 Scoccola:2004ke,
 Kosowsky:2004zh,Kahniashvili:2005xe,Pogosian:2012jd,Kahniashvili:2014dfa,
 Zucca:2016iur,Pogosian:2018vfr}, as well as
non-Gaussianity in bi- and tri-spectrum ~\cite{Brown:2005kr,Seshadri:2009sy,Caprini:2009vk,Cai:2010uw,
 Trivedi:2010gi,Brown:2010jd,Shiraishi:2010yk,Shiraishi:2011dh,Trivedi:2011vt,
 Shiraishi:2013wua,Trivedi:2013wqa,Ade:2015cva}. Most works have not seen evidence for PMFs and have imposed upper limits in the comoving $\sim$ nG range\footnote{The exception here is the limit on PMFs of $B_0 < 0.05nG$
 \cite{Trivedi:2013wqa} on inflationary fields from the non-Gaussianity in the tri-spectrum when the inflationary curvature mode is taken into account.}.
It is noted that these upper limits are much larger than the field strength
$B_0 \sim 0.005\,$nG \cite{Banerjee:2003xk}. 
required for clusters of galaxies to have an exclusively primordial origin.

More recently one particular effect of PMFs on CMB anisotropies has been identified. Independent on
PMFs being of inflationary or phase transition origin, the small-scale
part of the PMFs, dissipating around recombination, induces small-scale, non-linear density fluctuations in the baryons even for fairly weak $B_0\sim 0.01 - 0.05$nG final present day field \cite{Jedamzik:2011cu,Jedamzik:2013gua,Jedamzik:2018itu}. This is possible as on small
scales the photons are free-streaming, and magnetic stresses may induce density
fluctuations only opposed by the comparatively low baryon pressure. Recombination
in such a "clumpy" Universe is sped up, due to high-density regions recombining earlier \cite{Peebles1976}, thereby moving the peaks of the CMB to higher multipoles.
This effect was confirmed in numerical MHD simulations and has been used to impose fairly stringent limits on PMFs \cite{Jedamzik:2018itu}~\footnote{The study used an effective three zone model and employing older CMB data.}.

It has been subsequently shown that an earlier recombination due to PMFs and the
associated baryon clumping,
reduces the sound horizon and seems promising to relieve the Hubble tension \cite{Jedamzik:2020krr,Jedamzik:2023csc}.
The Hubble tension is a $\sim 5\sigma$ tension between the present day
Hubble constant $H_0$ inferred directly from observations of Type 1A supernovae in the local Universe, and $H_0$ from CMB observations employing $\Lambda$CDM and a standard recombination history (see \cite{Abdalla:2022yfr} for a review). 
Using the most recent Planck data,
and three local $H_0$ determinations, a $3-4\sigma$ detection of baryon clumping before recombination was claimed, with an associated increase in $H_0$ to higher values \cite{Jedamzik:2020krr}. When combing with other data sets, such as BAO, Type 1A supernovae and DES, the clumping detection significantly reduced. An analysis without the three local $H_0$ observations, using combinations of Planck data with high multipole
CMB data ACT and SPT ~\cite{Thiele:2021okz,Rashkovetskyi:2021rwg,Galli:2021mxk}, did not show a clear detection of clumping.
Thus CMB data only does not favor clumping but also does not rule it out
to the degree that clumping could enhance the prediction for $H_0$. It is
noted here that the existence of PMFs may be tested by direct observation 
of the remnant recombination radiation~\cite{Lucca:2023cdl} and the formation of dark matter
mini-halos~\cite{Ralegankar:2023pyx}. 

All of the above analysis was performed using ad hoc three-zone
baryon density probability distribution functions (pdfs) describing the clumping by one parameter $b$, the clumping factor, which is the variance of baryon density perturbations. 
Here the ionization history was computed by appropriately averaging the electron fraction of three independent regions with different baryon densities.
Though such models may be good for a first description of clumping,
they are insufficient for an accurate comparison between theory and observations, given the accuracy of current and future CMB data. In particular, one can show that the ionization history does not depend only on the baryon pdf, but
on the evolution of baryon density of each gas element. In realistic scenarios
the baryon pdf also evolves, whereas in three-zone models it is simply assumed constant. Moreover, three-zone models may assume a very unrealistic baryon pdf.
Finally, a direct connection between the
change in ionization history due to PMFs and the present-day leftover PMF field
is hard to accurately establish in three-zone models.

In this paper we wish to address these issues, in order to obtain as precise as
possible predictions. Such a study can only be performed by resorting to a host of full three-dimensional MHD simulations, which include all effects known to date. In the course of our study we analyze other so far unnoted effects,
such as the effect of peculiar motions on the Lyman-$\alpha$ photon escape rate,
and the transport of Lyman-$\alpha$ photons between different regions. We find, that in particular the latter effect is of importance. The goal of the paper is to improve the study of PMFs during recombination significantly
as a preparation for an accurate comparison of PMF theory with CMB data in a future publication.

The outline of the paper is as follows. In Section \ref{sec:overview} we discuss details of the numerical MHD simulations, the employed MHD solver, the physical effects included, and the newly written recombination routine. We then present results of a realistic baryon pdf. Section \ref{sec:transport} reviews standard recombination theory and presents results
of a Monte-Carlo analysis of the propagation of Lyman-alpha photons during recombination. In Section \ref{sec:Recombination} we analyze the effects of Lyman-$\alpha$ photon mixing on the ionization history, presenting a new explicit expression for the recombination rate in the full-mixing limit. Section \ref{sec:realistic} presents results of MHD simulations of PMFs before, during, and after recombination.
Particular emphasis is laid on non-helical PMFs with a Batchelor spectrum
and the change of the global free electron fraction $X_e \equiv \av{n_e}/\av{n_{\rm H}}$ (i.e.~the ratio of the average free-electron abdundance to the average abundance of hydrogen nuclei) compared to a non-magnetized Universe.
In Section \ref{sec:heating} we discuss the combined effect of baryon clumping and hydrodynamic heating by PMF dissipation on $X_e$.
Conclusions are drawn in Section \ref{sec:conclusion}. In Appendix \ref{sec:MC} we present the details of our Lyman-$\alpha$ Monte-Carlo
simulation. Appendix \ref{app:helium} briefly describes helium recombination. Appendix \ref{app:convergence} presents a numerical convergence study.

\section{Numerical Simulations of compressible MHD in the expanding Universe} \label{sec:overview}

\subsection{Physical problem to be solved}

In this paper we want solve the following problem. We assume a Gaussian-distributed primordial magnetic field ${ \bf B}_{\rm prim} = {\bf B}(t \rightarrow 0)$. We then want to follow the time evolution of the magnetic field ${\bf B}$, coupled to the evolution of baryon density and peculiar velocity fields, $\rho_b$ and $\bf{v}$, respectively. The MHD equations in an expanding Universe are given in Refs.~\cite{Brandenburg:1996fc,Banerjee:2004df}. Here we complement them by including photon drag on baryons, assuming photons to be free-streaming (i.e.~have a mean-free path much larger than the scales of interest) thus uniform on the small scales of interest. We moreover neglect gravitational potentials, since we are interested in scales much smaller than the baryon Jeans scale. The equations to be solved are therefore
\barr
a~ \partial_t (a^3 \rho_b) + {\bf \nabla}\cdot(a^3 \rho_b {\bf v}) &=& 0,\\
\partial_t(a {\bf v}) +  {(\bf v \cdot \nabla) v}  + \alpha~ a~ {\bf v} &=& -  \frac{{\bf \nabla} (c_s^2 \rho_b)}{\rho_b} + \frac{\bf B \times (\nabla \times B)}{4 \pi \rho_b},\label{eq:momentum}\\
\partial_t (a^2 {\bf B}) &=& a {\bf \nabla \times (v \times B)},
\earr
where all gradients are comoving, and $a$ is the scale factor, whose time evolution is given by $\dot{a}/a = H$, where $H$ is the Hubble rate. Note that the equation for the magnetic field does not account for the Biermann-battery effect, which is relevant at much later times than of interest here \cite{Naoz_2013, Lee_2024}. In the Euler equation \eqref{eq:momentum}, the (spatially-varying) baryon sound speed is 
\beq
c_s^2 = \frac{n_{\rm tot}}{\rho_b} T_b \equiv \frac{n_H (1 + x_e) + n_{He}}{\rho_b} T_b,
\eeq
where $T_b$ is the baryon temperature, $x_e \equiv n_e /n_H$ is the (spatially-varying) free-electron fraction, with $n_e$ the electron density and $n_H = n_p + n_{H^0}$ the sum of proton and neutral hydrogen density, and $n_{He} \approx 0.08 ~n_H$ is the total helium density, in both neutral and ionized forms. Even though photons are free-streaming, it is well known that in this limit
the abundant photons still play an important role, as they induce a drag force on the baryons \cite{Peebles1976} proportional to the drag rate $\alpha$, given by
\beq
\alpha = \frac43 \frac{n_e \sigma_{\rm Th}  \rho_\gamma}{\rho_b}, \label{eq:alpha-def}
\eeq
where $\sigma_{\rm Th}$ is the Thomson cross section, $\rho_{\gamma}$ is the (uniform) photon energy density and $m_p$ the proton mass. Before recombination $\alpha \gg H$ so photon drag forces the MHD evolution in the viscous regime, i.e.~with Reynolds number of order unity. After recombination, the MHD evolution becomes turbulent.

We see that the MHD equations must be complemented by an evolution equation for the local ionization fraction $x_e$, relevant to both photon drag and baryon pressure. The evolution of $x_e$ takes the following form:
\beq
\partial_t x_e + a^{-1}{\bf v \cdot \nabla} x_e = \frac{\dot{n}_e|_{\rm rec}}{n_{\rm H}},
\eeq
where $\dot{n}_e|_{\rm rec}$ is the (local) rate of change of free-electron density due to cosmological recombination, which we will discuss in more detail in the following sections. The most relevant quantity for CMB anisotropies is the \emph{global} free-electron fraction $X_e$, defined as
\beq
X_e \equiv \frac{\av{n_e}}{\av{n_H}} = \frac{\av{n_H x_e}}{\av{n_H}} = \frac{\av{\rho_b x_e}}{\av{\rho_b}},
\eeq
where $\av{...}$ denotes a spatial average. Accurately computing the effect of PMFs on $X_e$ is one of the main goals of this work.

Though the baryon temperature $T_b$ is to excellent approximation given by the CMB temperature $T_{\rm CMB}$ due to Thomson scattering for redshifts $z\gtrsim 1000$, it may deviate at lower redshifts. In particular, $T_b$ may exceed 
$T_{\rm CMB}$ due to the energy released into the plasma by magnetic field dissipation \cite{Sethi:2004pe}. The evolution of the local baryon temperature
is governed by the following equation
\begin{equation}
\frac{d T_b}{dt} = \frac{2}{3}\frac{d\rho_b/dt}{\rho_b} + \frac{8\sigma_{Th} x_e n_H\rho_{\gamma}}
{3m_e c n_{\rm tot}}\bigl(T_{\rm CMB}-T_b)+\frac{2\Gamma}{3 n_{\rm tot}}\, , \label{eq:Tb-evolution}
\end{equation}
where the the terms on the RHS are due to adiabatic cooling/heating, CMB cooling/heating, and heating due to magnetic field dissipation, with $\Gamma$ the volumetric heating rate. We will discuss $\Gamma$ and the effects of this dissipation in Section \ref{sec:heating}.

\subsection{Numerical implementation}

For the numerical simulations of PMFs before, during and after recombination
we used the publicly available code {\sl ENZO} \cite{ENZO:2013hhu}. We chose the 
MUSCL solvers with Dedner divergence cleaning described in \cite{Wang:2007qk}.
Through rescaling all physical 
quantities by appropriate powers of the scale factor $a$, the MHD equations given above can be reformulated as the MHD equations in a Minkowski, static
metric, with an additional drag term proportional to the Hubble rate. The Minkowski metric is chosen to coincide with the CMB rest frame. 
Specifically, we defined the rescaled variables
\beq
d\tilde{t} \equiv a^{-3/2}dt,  \ \ \ \ \tilde{\rho}_b \equiv a^3 \rho_b, \ \ \ \ \tilde{{\bf v}} \equiv a^{1/2} {\bf v},  \ \ \ \ \tilde{\bf B} \equiv a^2 {\bf B}.
\eeq
In terms of these variables, the MHD equations become
\barr
\frac{\partial\tilde{\rho}_b}{\partial \tilde{t}} + {\bf\nabla}\cdot \bigl(\tilde{\rho}_b \tilde{\rm \bf v}\bigr)
& = & 0, \\
\frac{\partial {\bf \tilde{v}}}{\partial \tilde{t}} 
+ \bigl( \tilde{\rm\bf v} \cdot {\bf\nabla}\bigr)\tilde{\rm\bf v} 
+ \left(\tilde{\alpha} + \frac{\tilde{H}}{2}\right){\tilde{\bf v}} & = & -\frac{{{\bf\nabla} (\tilde{c}_s^2 \tilde{\rho}_b})}{\tilde{\rho}_b}
- \frac{\tilde{\bf B}\times \bigl({\bf\nabla}\times \tilde{\bf B}\bigr)}{4\pi\tilde{\rho}_b},~~~~ \label{eq:Euler}\\
\frac{\partial \tilde{\bf B}}{\partial \tilde{t}} & = & {\bf\nabla}\times 
\bigl(\tilde{\rm\bf v}\times {\tilde{\bf B}}\bigr),
\earr
where we have defined the rescaled sound speed, photon drag rate and expansion rate as 
\beq
\tilde{c}_s \equiv a^{1/2} c_s, \ \ \ \ \ \tilde{\alpha}\equiv a^{3/2} \alpha, \ \ \ \ \ \tilde{H} \equiv a^{3/2}H.
\eeq
For details the reader is referred to Appendix B
of \cite{Banerjee:2004df}. The rescaling induces only one extra Hubble redshift term in the Euler equation, such that standard MHD codes in Minkowski space may be used when this term is included. It is
noted that this rescaling may be applied during matter domination as well as
radiation domination. 
As $T_b = T_{\rm CMB}\propto 1/a$ for $z\gtrsim 1000$, up to the dependence on $x_e$ we have $c_s\propto 1/a^{1/2}$ which implies that $\tilde{c}_s$ is approximately constant. Nevertheless, $\tilde{c}_s$ drops by an approximate factor $1/\sqrt{2}$ during recombination due to the factor two smaller total 
particle number density when $x_e\to 0$.

In the rescaled variables, the free-electron fraction $x_e$ evolves according to 
\beq
\frac{\partial x_e}{\partial \tilde{t}} + \tilde{{\bf v}} \cdot \nabla x_e = a^{3/2} \frac{\dot{n}_e|_{\rm rec}}{n_H}, \ \ \ \ \ \ \ \tilde{n}_H \equiv a^3 n_H.
\eeq
To evolve $x_e$, we coupled the MHD simulations to a "chemical" solver which computes the abundances
of singly ionised hydrogen and helium at each time step.
The existing chemistry solver in {\sl ENZO} \cite{Anninos:1996kf} was optimized
for non-equilibrium gas chemistry in the coronal limit \cite{Abel:1996kh} and is not suited for this study. We instead developed a new routine using a 6th order Runge-Kutta solver. We will discuss the recombination rate $\dot{n}_e|_{\rm rec}$ in great detail in Sections \ref{sec:transport} and \ref{sec:Recombination}, but in a nutshell, we include the most
relevant processes during cosmological recombination and develop a new analytic result valid in the limit of scales much smaller than the mixing length of  Lyman-$\alpha$ photons. We checked that our routine reaches $0.5-1.5\%$ accuracy in comparison with Recfast \cite{Seager:1999bc} when computing recombination in a homogeneous Universe. 
Our routine does not, however, reach the sub-percent accuracy of state-of-the art cosmological recombination codes \textsc{Hyrec} \cite{Ali-Haimoud:2010hou,Lee:2020obi} and \textsc{Cosmorec} \cite{Chluba:2010ca}, which include many other physical effects. As we will see, PMFs may easily change the free-electron fraction at the $\sim 20\%$ level, and such an extreme accuracy is therefore not required for our purposes.

Let us emphasize that our implementation of recombination within the MHD simulations is an important novel aspect of this work, which goes well beyond the ``three-zone model" implementations of past works, as we will discuss again later.


\subsection{Results}

\begin{figure*}
\includegraphics[width = \columnwidth]{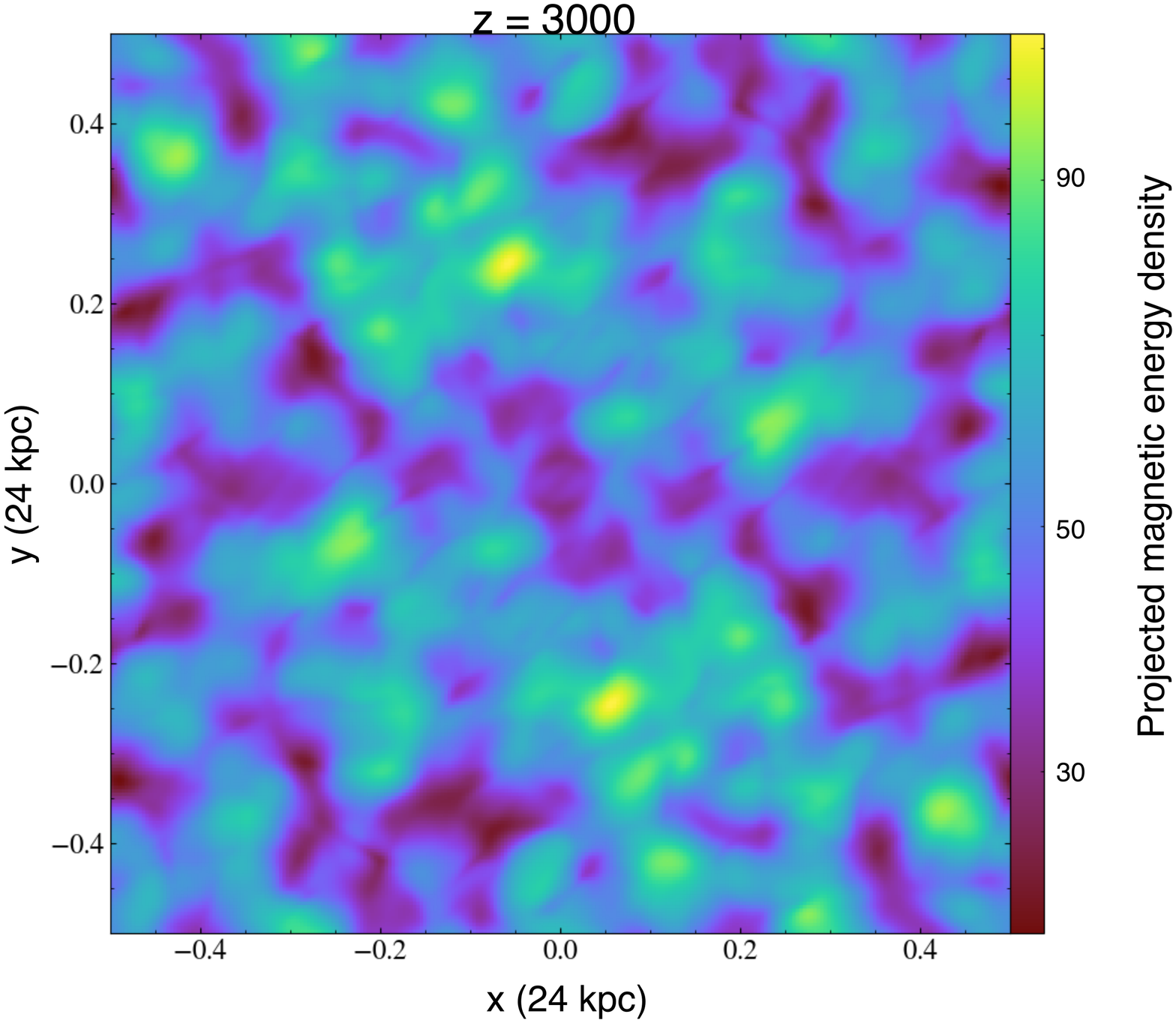}
\includegraphics[width = \columnwidth]{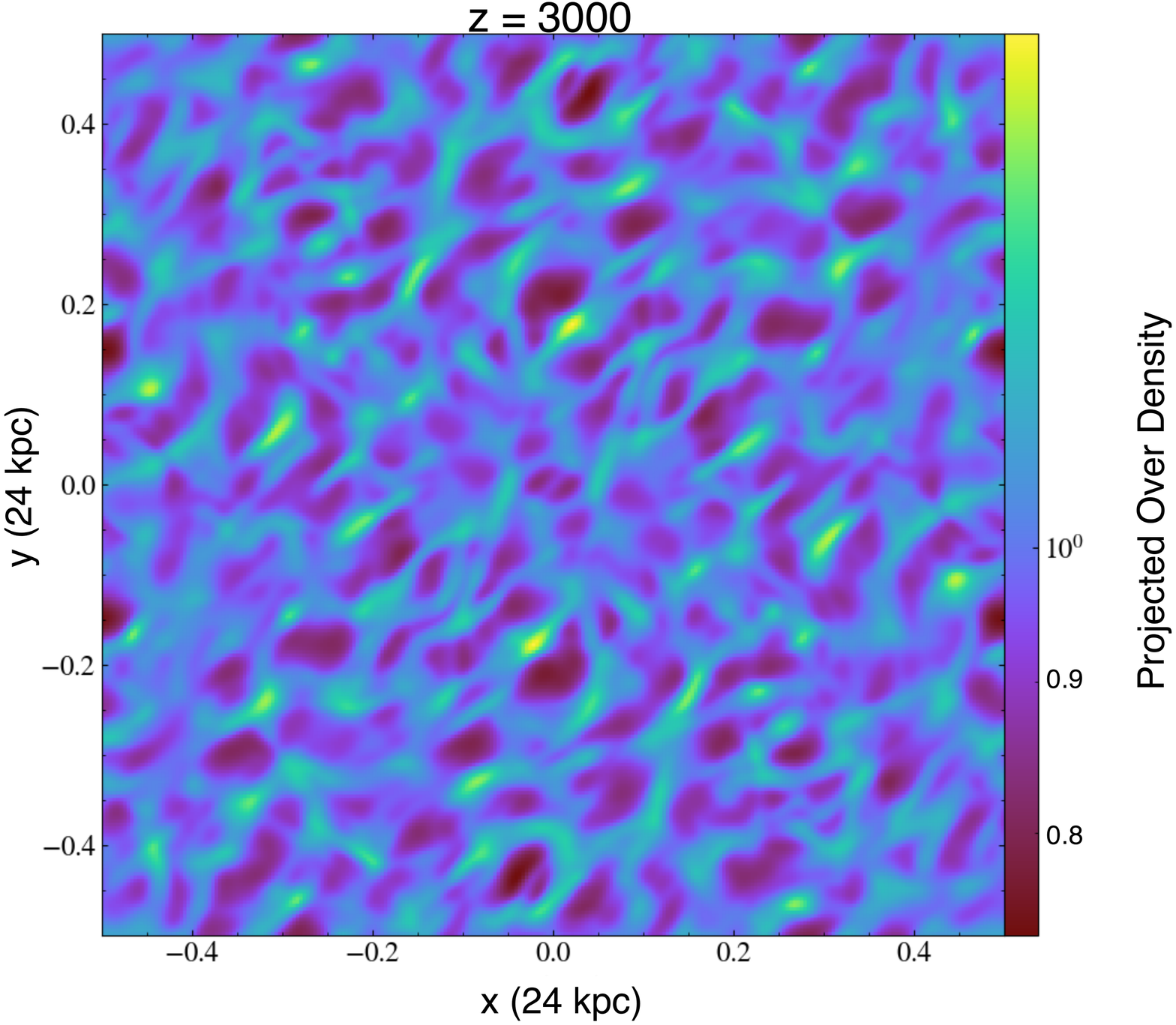}
\includegraphics[width = \columnwidth]{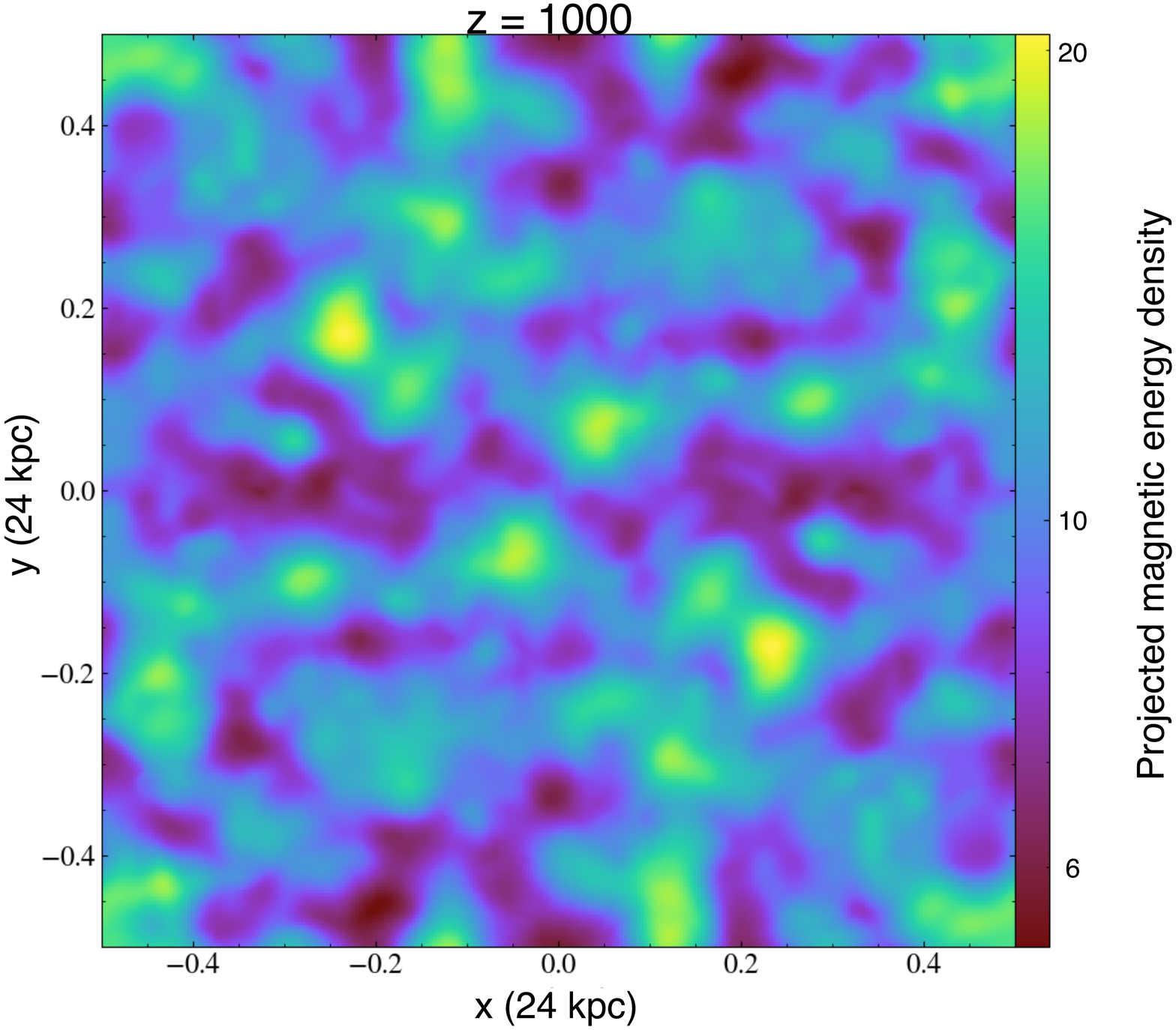}
\includegraphics[width = \columnwidth]{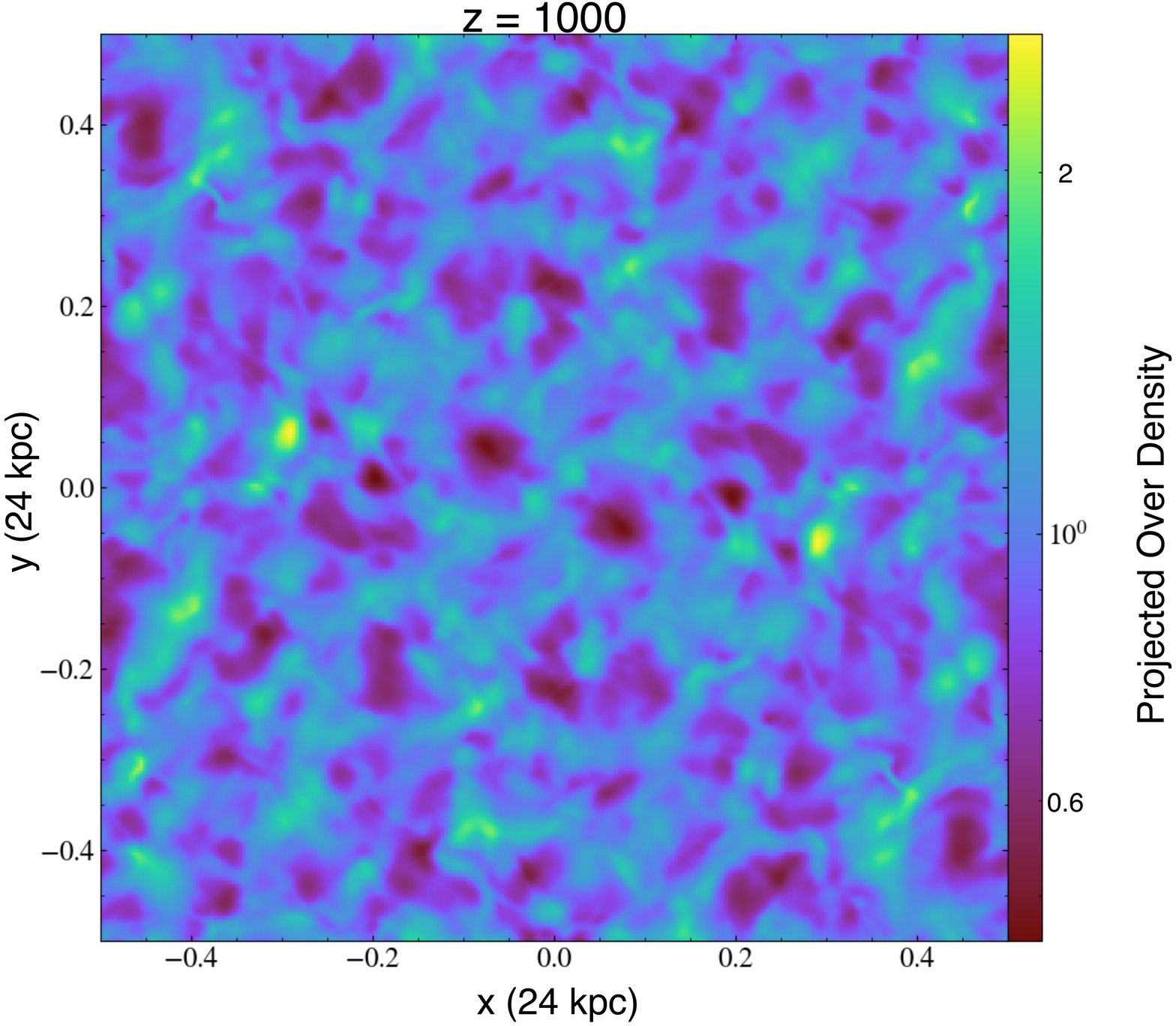}
\includegraphics[width = \columnwidth]{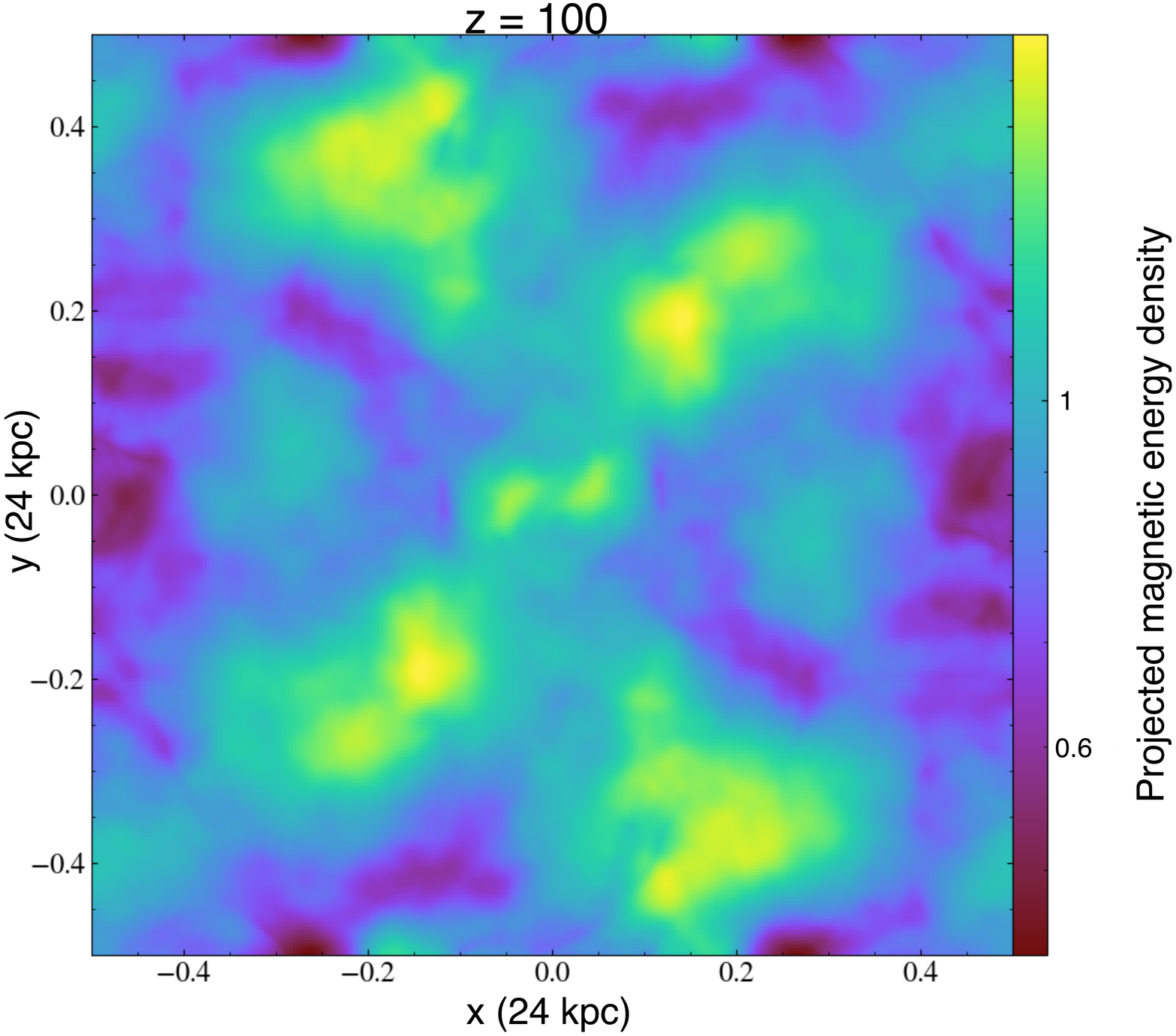}
\includegraphics[width = \columnwidth]{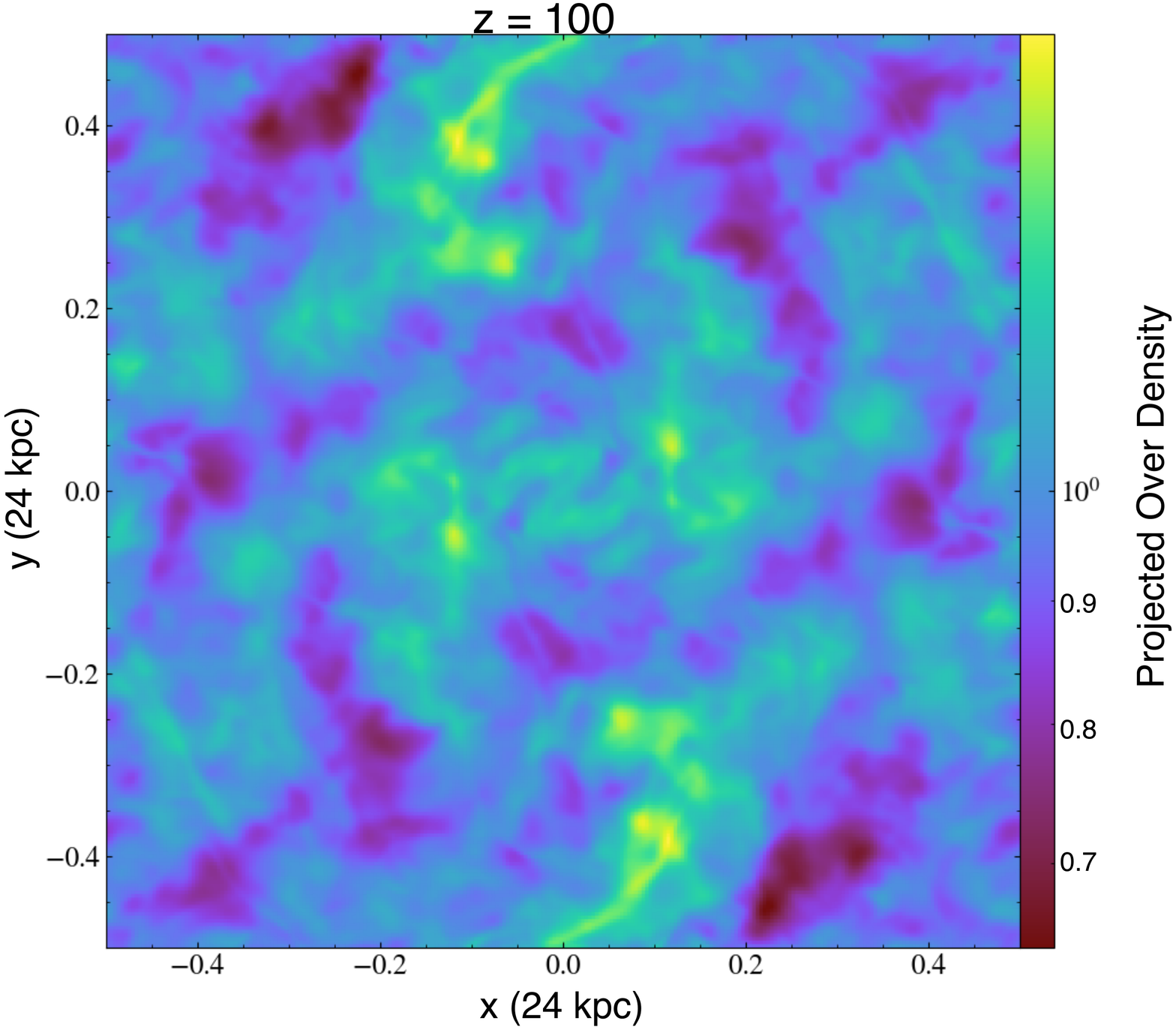}
\caption{Two dimensional visualisations of the evolution of a PMF and the generated baryon density fluctuations in a $256^3$ numerical MHD simulation of a non-helical PMF with Batchelor spectrum. Initial conditions at redshift $z = 4500$ were chosen as 
$V_{A,{\rm rms}} = 12 c_s$ (i.e. comoving 0.526 nG)
and vanishing density baryon fluctuations.
The left panels show the projected $V_{A,{\rm rms}}^2$ whereas the right 
panels show the projected baryon overdensity $\Delta \equiv \rho_b/\av{\rho_b}$ at three different redshifts $z=3000, 1000$ and $100$.}
\label{fig:2D}
\end{figure*}

All MHD simulations in this paper use default cosmological values of 
$h = 0.67$, $\Omega_bh^2 = 0.0224$, $\Omega_ch^2 = 0.12$, and $Y_p=0.24$
for Hubble parameter, baryon density, CDM density, and helium mass fraction,
respectively. In Fig.~\ref{fig:2D} 
results of a numerical simulation of a PMF in a comoving $(24{\rm kpc})^3$ box with resolution $256^3$ are shown. In particular, the left panels show projected magnetic energy density at redshifts $z=3000$, $z=1000$, and $z=100$, respectively, whereas the right panels show projected baryon overdensity $\Delta \equiv \rho_b/\av{\rho_b}$ for the same redhifts. As initial conditions we chose a particular random configuration of a non-helical PMF with Batchelor spectrum
\begin{equation}
B(L) \sim L^{-5/2}
\label{eq:Batchelor}
\end{equation}
where $L$ is length scale~\footnote{Eq.~\ref{eq:Batchelor} should be taken illustrative, meaning that the rms magnetic field strength when smoothed over  $L$ scales as $L^{-5/2}$. In praxis, to assure a divergence-free field $\bf{\nabla 
 B}=0$, the Fourier modes
of the vector potential ${\bf A}$ with ${\bf B= \nabla}\times {\bf A}$ are excited with random amplitudes drawn from a Gaussian with width 
$\sigma_k\sim (k/k_0)^{5/2 -3/2 -1}$, where $k$ is wavevector and $k_0$ a
reference scale. Here the exponent $-3/2$ stems from phase space, i.e. 
$\langle B^2 \rangle = k^3 \langle\tilde{B}_k^2\rangle$ where $\tilde{B}_k$
is the Fourier amplitude, and the exponent $-1$ from the relationship between
$\bf \tilde{B}_k = {\bf k} \times\bf \tilde{A}$. Finally, one can show that a completely non-helical field is attained when setting all phases to zero in the Fourier decomposition.}.
All modes with wavelength in the range ${\rm 24 kpc} \geq\lambda\geq {\rm 3 kpc}$ were excited. The $rms$ magnetic field strength in the simulation box has been normalized to $\tilde{B}_{rms} = 0.525\,$nG comoving. This corresponds to
an Alfven velocity twelve times as large as the speed of sound of singly 
ionized hydrogen and helium~\footnote{The speed of sound of singly ionized hydrogen and helium and the Alfven velocity at redshift $z = 1090$ are $c_s \approx 6.33 {\rm km/s}$ and ${\rm v}_A = 4.34 {\rm km/s}[B_0/(0.03 {\rm nG}])$,
respectively.}.
The simulation started at redshift $z=4500$ with vanishing
peculiar flows and uniform baryon density. Note that such initial conditions
are very realistic. At early times when the photon mean free path 
$l_{\gamma} < L$, the effective speed of sound is large, $c_s = c/\sqrt{3}$,
such that the fluid is incompressible and no substantial density fluctuations
can be generated by relatively weak PMFs. When $l_{\gamma}\sim L$ photon dissipation
is so strong that all pre-existing peculiar velocities will be erased.
Only when $l_{\gamma} \gg L$ at lower redshift will magnetic stresses slowly
be able to accelerate the fluid and build up density perturbations. For details
of the evolution of PMFs in the early Universe the reader is referred to 
\cite{Banerjee:2004df}.

It is seen that by redshift $z=3000$ the PMF has generated slightly non-linear
density fluctuations on $\sim\,$kpc scales. These overdensities seem to follow a filamentary structure. At redshift $z = 1000$, close to the maximum clumping
produced in baryons, structures appear more fuzzy. This is due to the
fluid leaving the viscous regime and transitioning into turbulence, due to the drop of electron density and the associated significant reduction of drag force.
By redshift $z = 100$, when the clumping has significantly diminished, structures are even more fuzzy. At all three redshifts magnetic structures are
more extended.

\begin{figure}[htbp]
\centering
\includegraphics[width=0.4\textwidth]{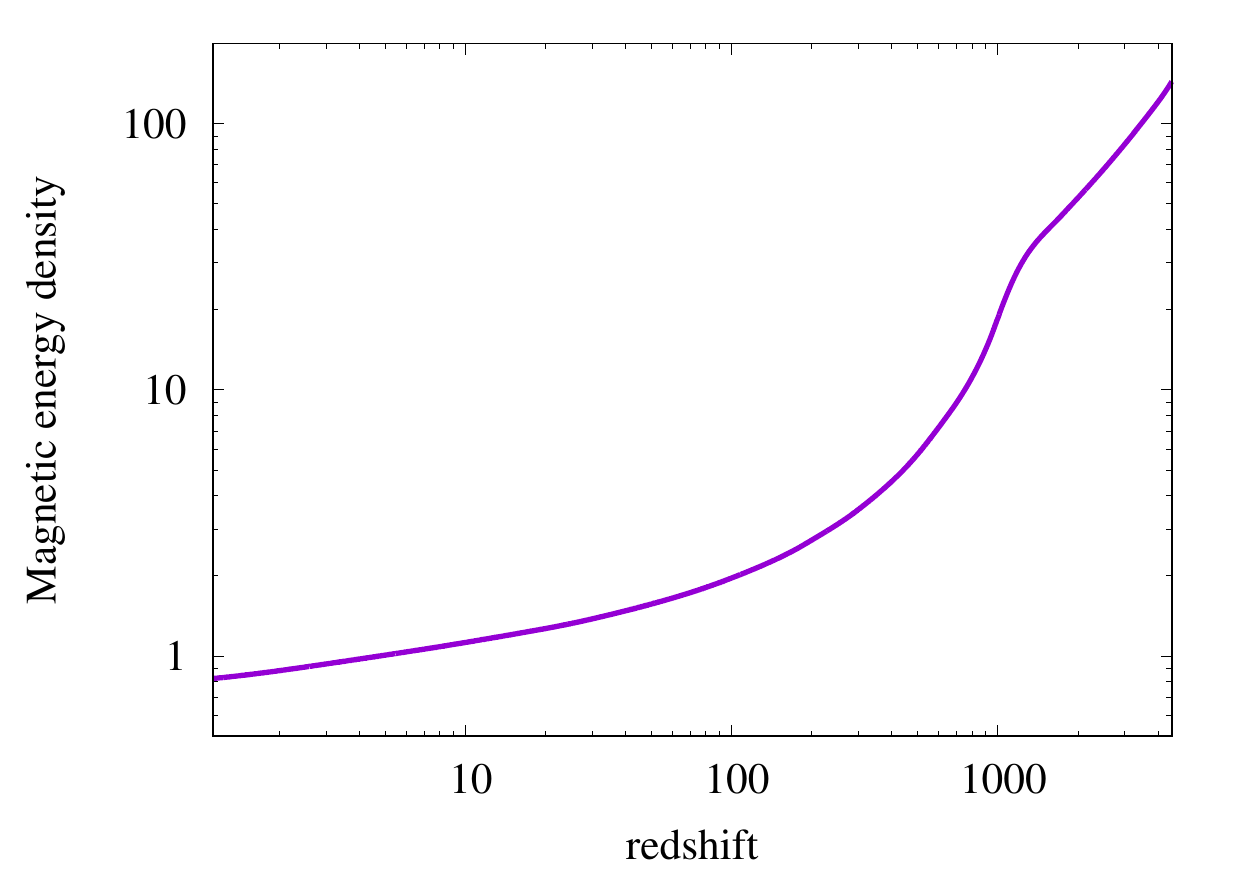}
\includegraphics[width=0.4\textwidth]{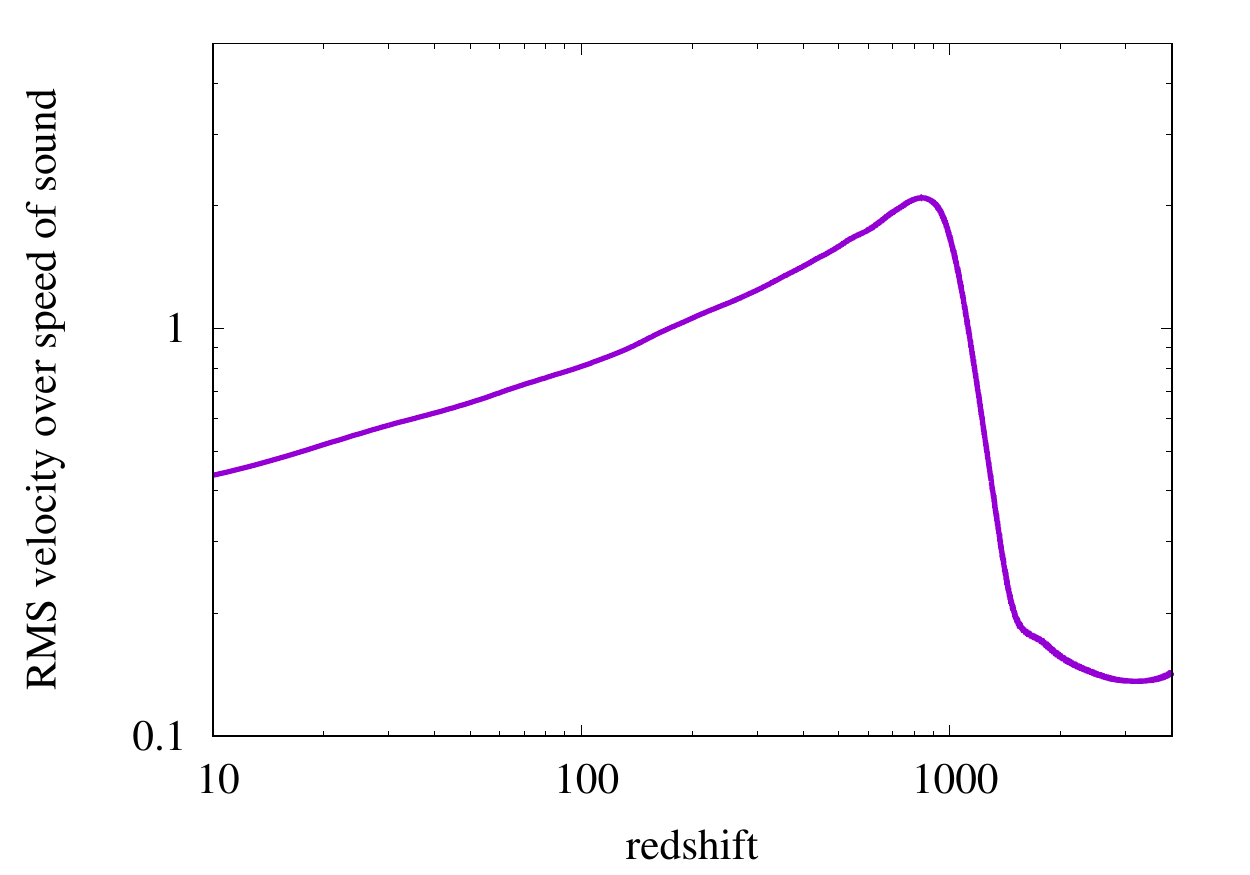}
\includegraphics[width=0.4\textwidth]{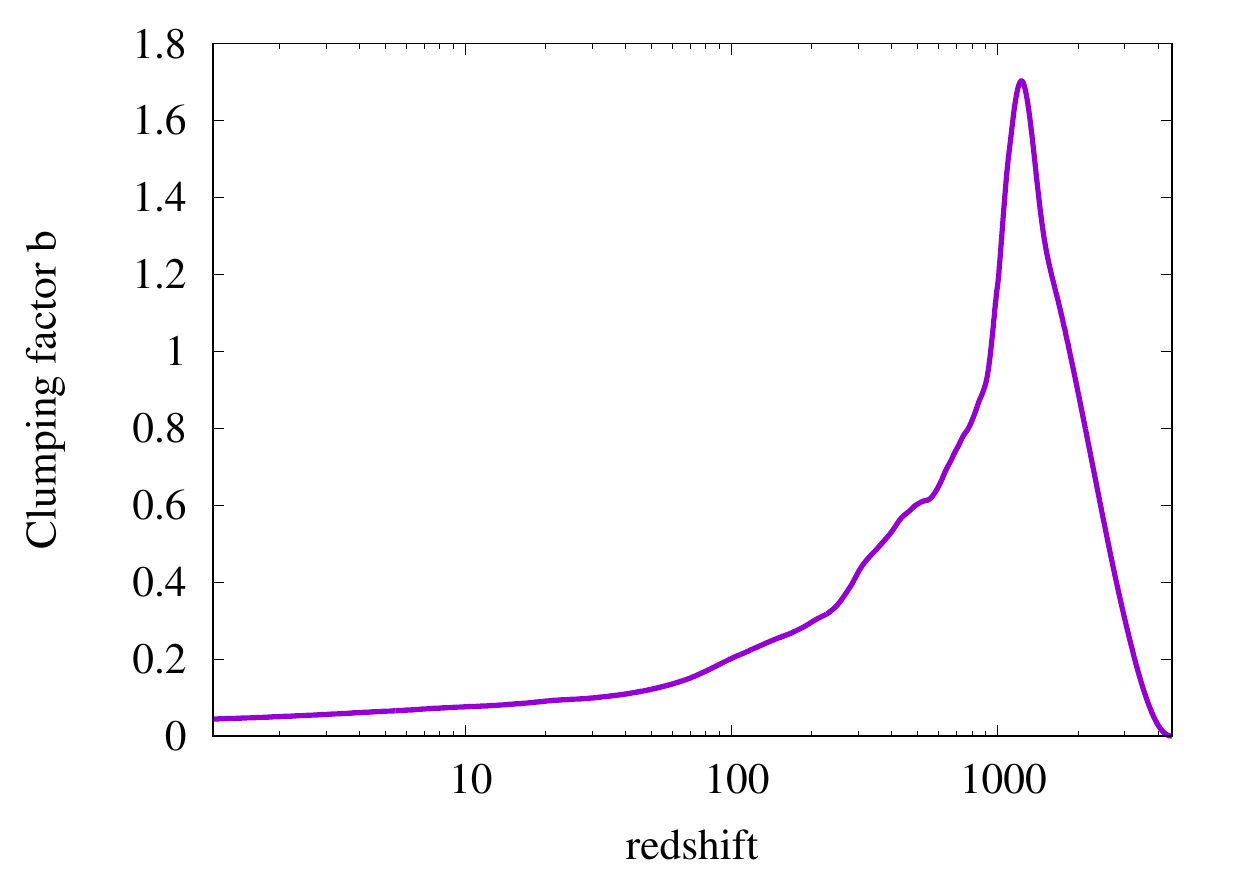}
\caption{Redshift evolution of the comoving magnetic energy density $\epsilon_B = \tilde{B}_{rms}^2/8\pi$, normalized to unity for $\tilde{B}_{rms} = 4.38\times 10^{-2}$nG (\emph{top}), of the root-mean-square velocity over the speed of sound (\emph{middle}), and of the baryon clumping factor $b$ given in Eq.~\eqref{eq:clumpingfactor} (\emph{bottom}). Baryon density fluctuations are generated before recombination by the Lorentz force of the stochastic magnetic field. All quantities are evaluated for the numerical simulation shown in Fig.~\ref{fig:2D}.}
\label{fig:Bdecay}
\end{figure}

In Fig.~\ref{fig:Bdecay} the evolution of comoving magnetic energy density,
root-mean-square velocity, and clumping factor
\begin{equation}
b \equiv \frac{\langle \rho_b^2\rangle - \langle \rho_b\rangle^2}{ \langle \rho_b\rangle^2}
\label{eq:clumpingfactor}
\end{equation}
are shown for this simulation. During the freely decaying evolution of the PMF the magnetic energy density is reduced by more than two orders of magnitude, leading to a final comoving rms field of $B_0 \approx 4.38\times 10^{-2}$nG. Here the most rapid reduction occurs right around recombination, as dissipation which could not occur before due to the large photon drag occurs when the fluid becomes turbulent. One may also observe the transition from viscous MHD to turbulent MHD by the increase of the root-mean-square velocity from subsonic values to slightly supersonic values during recombination. The PMF evolution time scale during recombination is thus not governed by the Hubble time, as during most other periods of the early Universe, but the shorter time scale of change in $X_e$. It is noted that even after recombination there is still some
further decay. Simple analytic estimates \cite{Banerjee:2004df} predict only logarithmic decay after recombination, but since the redshift range between recombination and the present is substantial, even logarithmic decay is still notable. The clumping factor of the baryons increases from zero to a maximum of $b \approx 1.7$ at redshift $z\approx 1250$ to then decrease to much lower values $b\approx 0.1$ at
redshift $z\approx 10$. Due to the reduction of the speed of sound and the beginning of turbulent evolution right at recombination, clumping quickly increases due to magnetic stresses  and then decreases due to baryonic pressure forces at recombination. The box size employed is such as to have the maximum effect at recombination, in order to induce the maximum change in ionisation history $\Delta X_e$ observable by the CMB anisotropies. Much smaller modes (simulation boxes) would induce the peak in $b$ at higher redshifts, whereas much large modes would induce a peak in $b$ at lower redshifts. We discuss this point further in Section \ref{sec:optimize}.

\begin{figure*}[htbp]
\centering
\includegraphics[width=0.48\textwidth]{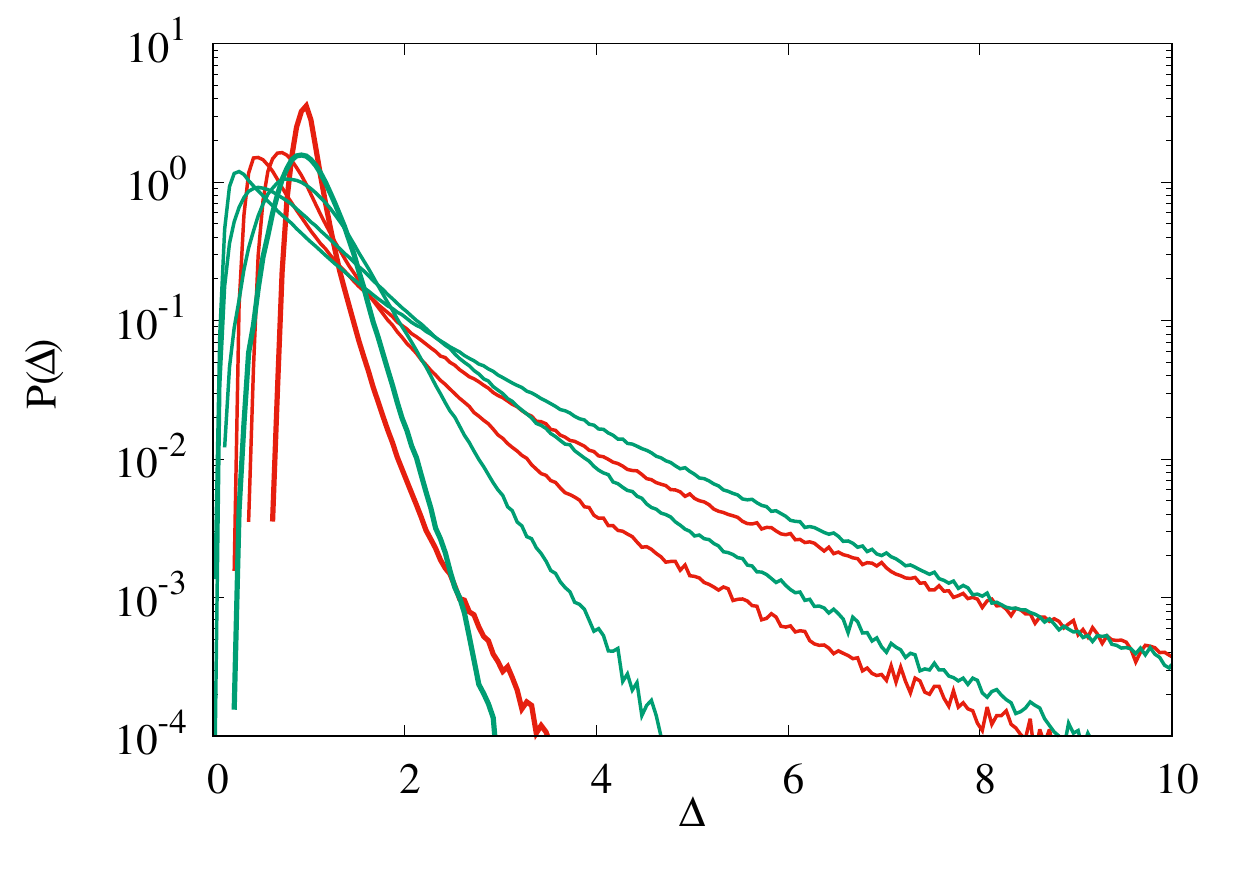}
\includegraphics[width=0.48\textwidth]{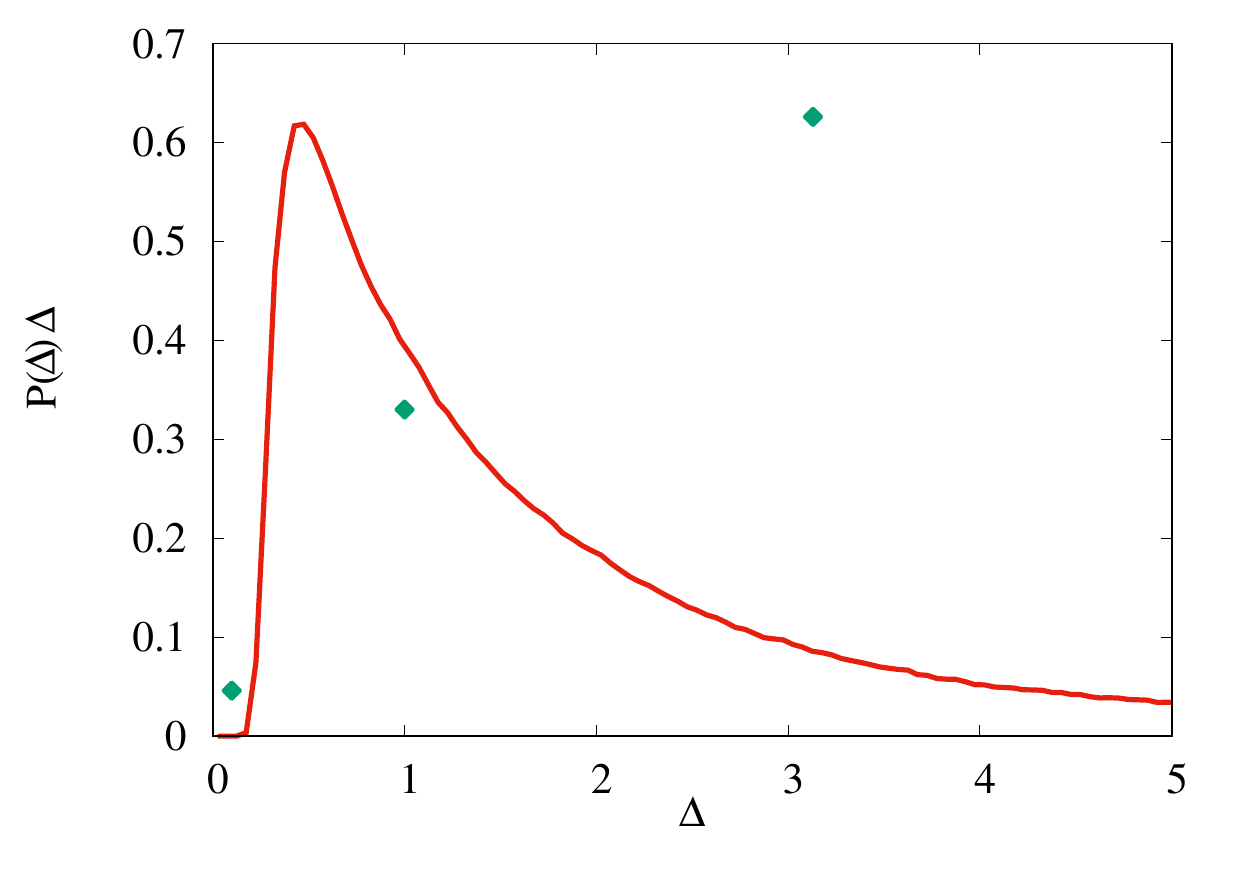}
\caption{\emph{Left:} Probability $P(\Delta )$ to find a volume element at density between $\Delta$ and $\Delta + d\Delta$, where $\Delta \equiv \rho_b/\langle \rho_b\rangle$. The redshift evolution for the numerical simulation shown in Fig.~\ref{fig:2D} at $z = 4000,
3000, 2000, 1000, 500, 100$ and $10$ is shown. $P(\Delta )$ is shown by red (green) lines before (after) the maximum of the clumping factor occurs at $z_{\rm max}\approx 1250$. The lines for $P(\Delta )$ at redshifts $z=4000$ and $z=10$ are slightly thicker.
For $z > z_{\rm max}$ the
maximum moves to lower densities and very high density regions get more and more probable, whereas for lower redshifts $z < z_{\rm max}$ the maximum moves to higher densities and very high density regions get less and less probable. \emph{Right:} The probability $P(\Delta )\Delta$ to
find a baryon at density between $\Delta$ and $\Delta + {\rm d}\Delta$ at redshift $z = 1500$, compared with the analogous quantity 
$f_V^i\Delta_i$ for the M1 three-zone model at the same clumping factor $b = 1.28$, shown with green dots. This illustrates that three zone models
do not capture the baryon probability function correctly.}
\label{fig:pdf}
\end{figure*}

The left panel of Fig.~\ref{fig:pdf} shows the pdf of baryon overdensity $P(\Delta )$ for a large number of redshifts. Red lines are for the redshift range $z= 4400 - 1500$ with the pdf becoming continuously harder, while green lines are for the redshift range 
$z = 1000 - 10$ with the pdf becoming softer again. At the peak of clumping a small fraction of volume $\sim 10^{-3}$ exists at extreme overdensities of $\Delta > 10$, while most of the volume is at $\Delta \approx 0.2$. The effects
of PMFs on the baryons even for a final magnetic field strength as low as $\sim 0.05\,$nG are thus quite drastic. We note that those trends are not so well
observed in Fig.~\ref{fig:2D}, as there projected density is shown. The existence of very high-density zones poses a problem for the computation of the clumping factor. As we show in Appendix B, in order to correctly resolve these high-density regions one needs very high resolution or adaptive mesh. The computed clumping factor $b$ increases with resolution. However, also shown in Appendix B is that the quantity which is really of interest, the ionization fraction $X_e$, converges even at lower resolutions. In three zone models a classification of
the density fluctuations by $b$ was employed. In realistic MHD simulations as performed here, the clumping factor should not really be used as a quantitative
measure to obtain the connection between $X_e$ and PMF strength $B_0$.

For comparison, ``three-zone models" used in previous analyses assume an unrealistic PDF $P(\Delta) = \sum_{i = 1}^3 f_V^i~ \delta_{\rm D}(\Delta - \Delta_i)$, where the three overdensities $\Delta_i$ and volume-filling fractions $f_V^i$ are constant in time, and satisfy $\sum_i f_V^i = 1 = \sum_i f_V^i \Delta_i$, and $\sum_i f_V^i \Delta_i^2 = 1+ b$. Here $b$ is the
clumping factor. This problem is underconstrained, so different models (M1, M2, etc...) make some extra ad-hoc assumptions to fix all six parameters $\{ f_i, \Delta_i\}$. In the right panel of Fig.~\ref{fig:pdf} we compare the numerically-obtained pdf of baryon density at $z = 1500$ to the M1 three-zone model with the same clumping factor $b = 1.28$. It is seen that
the M1 model gives a particularly poor representation of the baryon pdf. The same
holds for the M2 model. Both predict a much larger fraction of baryons at $\Delta > 1$ and a lower fraction of underdense regions than the simulations. 
We thus suspect that three-zone models overestimate the effect on 
$\Delta X_e$, or vice versa, for a similar $\Delta X_e$ much larger clumping factors will occur in the realistic case, with a significant fraction of the contribution to the clumping factor from very high-density regions.

\section{Lyman-$\alpha$ photon transport during recombination} \label{sec:transport}
%
In what follows we give a brief review of cosmological recombination and motivate the need for a Monte-Carlo computation of Lyman-$\alpha$ photon transport in an inhomogeneous Universe. We focus on Hydrogen recombination, which is most relevant to the MHD evolution (during the second Helium recombination, $x_e$ is always in the range $(1, 1.08)$, regardless of the details of helium recombination), as well as to CMB anisotropies. For completeness, we briefly describe our simplified model for Helium recombination in Appendix \ref{app:helium}.


\subsection{Review of standard Hydrogen recombination}

\subsubsection{The standard effective 3-level atom model}

We assume that helium has fully recombined by the time hydrogen recombination starts, which is a very accurate approximation (see e.g.~Ref.~\cite{Kholupenko:2007qs}). As a consequence the abundances of free electrons and protons are equal by charge neutrality, $n_e = n_p$. We denote by $n_{1s}$ the abundance of neutral hydrogen in its ground state, which to excellent accuracy is equal to the total abundance of neutral hydrogen, $n_{1s} \approx n_{H^0} = n_{H}  - n_e$, where $n_H$ is the total abundance of hydrogen, ionized and neutral. To simplify the calculation, we assume that angular momentum substates within a given energy shell are in statistical equilibrium, so that, in particular, the abundance of hydrogen in the $2p$ and $2s$ states are related through $n_{2p} = 3 n_{2s}$. 

Direct recombinations to the ground state are highly inefficient due to the very short mean-free-path of Lyman-continuum photons. As a consequence, recombination proceeds through the excited states, with a rate given by \cite{Peebles:1968ja}
\begin{equation}
\dot{n}_e|_{\rm rec} = -\alpha_e n_e^2 +\beta_e n_{2s},
\label{eq:rec0}
\end{equation}
where $\alpha_e$ is the case-B radiative-recombination coefficient and $\beta_e$ is the corresponding photoionization rate per atom in the $2s$ state\footnote{We choose to write all the rates in terms of the abundance of hydrogen in the $2s$ state, instead of the total abundance of excited hydrogen $n_2 = 4 n_{2s}$. Therefore, $\beta_e$ is 4 times the photoionization rate per atom in the $n = 2$ state.}. Note that instead of the exact effective recombination coefficient $\mathcal{A}_B(T_\gamma, T_b)$ computed in Ref.~\cite{Ali-Haimoud:2010hou}, where $T_\gamma$ and $T_b$ are the photon and baryon temperatures, respectively, we use the RecFast approximation \cite{Seager:1999bc, Seager:1999km, Wong:2007ym} $\alpha_e(T_b) = 1.14~ \mathcal{A}_B(T_\gamma = 0, T_b)$, where $\mathcal{A}_B(0, T_b)$ is obtained from Ref.~\cite{Pequignot_1991}, and obtain $\beta_e$ by detailed balance. 

This equation must be complemented by an evolution equation for the abundance of excited hydrogen $n_{2} = n_{2s} + n_{2p} = 4 n_{2s}$. It evolves due to three processes: radiative recombinations (and the corresponding photoionization), two-photon transitions from the $2s$ state, and Lyman-$\alpha$ transitions from the $2p$ state:
\beq
\dot{n}_{2} = -\dot{n}_e|_{\rm rec} + \dot{n}_{2s}|_{2 \gamma} + \dot{n}_{2p}|_{\rm Ly \alpha}.
\eeq
Up to small corrections, the two-photon term is
\beq
\dot{n}_{2s}|_{2 \gamma} = - \Lambda_{2 \gamma}\left(n_{2s} - n_{1s} e^{- E_\alpha/T_\gamma}\right), 
\eeq
where $\Lambda_{2\gamma} \approx 8.22$ s$^{-1}$ is the $2s\rightarrow 1s$ two-photon spontaneous decay rate and $E_\alpha = 10.2$ eV is the energy between the ground and first excited state. 

Naively, the net decay rate in the Lyman-alpha transition would take a similar form, with the much larger $2p\rightarrow 1s$ spontaneous decay rate $A_{2p, 1s} \approx 6.3 \times 10^8$ s$^{-1}$. However, because the Lyman-$\alpha$ transition is optically thick, only a small fraction of emitted photons redshift out of the line and escape reabsorption. In the Sobolev approximation, this fraction is $P_{\rm esc} \approx 8 \pi H/(3 \lambda_\alpha^3 n_{1s} A_{2p, 1s}) \ll 1$, where $\lambda_\alpha$ is the Lyman-$\alpha$ wavelength and $H$ is the expansion rate. The net decay rate in the Lyman-$\alpha$ transition is thus
\barr
\dot{n}_{2p}|_{\rm Ly \alpha} &=& - P_{\rm esc} A_{2p, 1s} \left(n_{2p} - 3 n_{1s}e^{- E_\alpha/T_\gamma} \right) \nonumber\\
&=& R_\alpha \left(n_{2s} - n_{1s}e^{- E_\alpha/T_\gamma} \right), \label{eq:Ly-alpha-std}
\earr
where 
\beq
    R_\alpha = 3 P_{\rm esc} A_{2p, 1s} = \frac{8 \pi H}{\lambda_\alpha^3 n_{1s}}. \label{eq:R-alpha}
\eeq 

In practice, we may solve for $n_{2s}$ in the quasi-steady state approximation, taking advantage of the very short timescale for transitions into and out of the excited states relative to the expansion rate:
\begin{equation}
    \dot{n}_{2s} \approx 0 = \alpha_e n_e^2 - \beta_e n_{2s} - \left(\Lambda_{2\gamma} + R_\alpha\right)\left(n_{2s} - n_{1s} e^{- E_\alpha/T_\gamma}\right), \label{eq:n2s-ss}
\end{equation}
implying
\beq
n_{2s} = \frac{\alpha_e n_e^2 + (\Lambda_{2 \gamma} + R_\alpha) n_{1s} e^{- E_\alpha/T_\gamma}}{\beta_e + \Lambda_{2 \gamma} + R_\alpha}. \label{eq:n2s-sol}
\eeq
Inserting this result into Eq.~\eqref{eq:rec0}, one obtains the following net rate of recombinations:
\begin{equation}
    \dot{n}_e|_{\rm rec} = - C\left(\alpha_e n_e^2 - \beta_e n_{1s} e^{- E_\alpha/T_\gamma}\right),\label{rec}
\end{equation}
where the dimensionless parameter $C$ is the famous Peebles suppression factor
\begin{equation}
C = \frac{\Lambda_{2\gamma} + R_\alpha}{\beta_e + \Lambda_{2\gamma} + R_\alpha},
\label{Cfactor}
\end{equation}
which quantifies the efficiency of net transitions to the ground state (accounting for the suppression of Lyman-$\alpha$ transitions due to their large optical depth) relative to the total net rate of transitions out of the excited state. The Peebles $C$ factor is much less than unity for $z \gtrsim 1000$ (of order $C \sim 10^{-2}$ near the peak of the visibility function at $z \approx 1100$), and becomes close to unity for $z \lesssim 800$.

In the absence of large inhomogeneities, the background ionization history is obtained by solving Eq.~\eqref{rec} with the background abundances $n_e = \langle n_e \rangle$ and $n_{1s} = \langle n_{1s} \rangle$.


\subsubsection{Perturbed recombination in the no-mixing limit}

The generalization of the recombination rate to a perturbed Universe is well-known in the limit that the characteristic distance $D_\alpha$ traveled by Lyman-$\alpha$ photons before re-absorption (as opposed to merely scattering) is much shorter than the length scale $\lambda$ of perturbations of interest \cite{Senatore:2008vi,Lee:2021bmn}. In that case, recombination is entirely local\footnote{This limit also requires the mean-free path of Lyman-continuum photons to be short relative to the scales of interest. This mean-free path is of order a comoving parsec \cite{Venumadhav:2014aka}, much smaller than the scales we consider here, so we need not consider the finite propagation of Lyman-continuum photons.}. The local net recombination rate is then still given by Eq.~\eqref{rec}, where $n_e$ and $n_{1s}$ are now the local ionized and neutral hydrogen densities, and with a modified net rate of Lyman-$\alpha$ loss obtained from substituting $H \rightarrow H + \frac13 {\bf\nabla \cdot v}$ in Eq.~\eqref{eq:R-alpha}, where ${\bf\nabla \cdot v}$ is the local divergence of baryon peculiar velocities. The latter substitution accounts for the enhanced (or decreased) redshifting of Lyman-$\alpha$ photons out of resonance for diverging (or converging) local bulk flows. Note that this substitution is exact (in the limit that $D_\alpha \ll \lambda$) for large diverging local bulk flows with ${\bf\nabla \cdot v} > 0$, as long as the local optical depth for true Lyman-$\alpha$ absorption, inversely proportional to $(H + {\bf\nabla \cdot v}/3)$, remains large, which in practice means as long ${\bf\nabla \cdot v}/3 \lesssim 10^4 H$. However, for converging local bulk flows with ${\bf\nabla \cdot v} < 0$, the expression only holds as long as ${\bf\nabla \cdot v}/3 > - H$, and special care should be given for more negative local velocity divergence.

In the case of non-linear baryon perturbations sourced by primordial magnetic fields of the characteristic strengths we consider in this paper, {we find root-mean-square values for the local divergence of peculiar flows of $({\bf\nabla \cdot v})_{\it rms}\sim 20 H$}, 
indicating that bulk flows could in principle have a \emph{very} significant effect on the net rate of Lyman-$\alpha$ decays. However, the problem we consider is \emph{not} within the $D_\alpha \ll \lambda$ limit: as we will see, $D_\alpha$ is of the same order or larger than $\lambda\approx 1$ kpc, 
the typical magnetically induced velocity- and density- fluctuation length. As a consequence, recombination is no longer local, as different patches ``communicate" through Lyman-$\alpha$ photons, and the net rate of recombination must be re-examined.

In the next subsection, we evaluate the typical distance $D_\alpha$ traveled by Lyman-$\alpha$ photons using Monte Carlo simulations.

\subsection{Lyman-$\alpha$ mixing length from Monte Carlo simulations}

In our Monte-Carlo simulations we inject 
Lyman-$\alpha$ photons into the plasma, and follow their evolution in
physical and frequency space during the $\sim 10^5-10^7$ scattering events
on neutral hydrogen, before reionzation by the CMBR black body of the 
excited $n = 2$ state and thus absorption of the Lyman-$\alpha$ photon. Photons which diffuse in frequency too far onto the 
red wing of the Lyman-$\alpha$ resonance line are likely lost by redshifting.
Photons which "diffuse" too far in physical space are subject to different
local physical conditions in ${\bf\nabla v}$ and density.
The details of our Monte Carlo method are presented in Appendix~\ref{sec:MC}.

Fig.~\ref{fig:MCtravel} shows the fractions $f_\alpha(D)$
of Lyman-$\alpha$ photons having traveled
between creation and destruction further than distance $D$. 
In the top panel, we show this function for different redshifts, highlighting the fact
that at lower redshifts Lyman-$\alpha$ photons travel further before being destroyed. This is due to
the decrease in the photoionization rate by CMB photons $\beta_e$, whereas the Lyman-$\alpha$ de-excitation rate $A_{2p, 1s}$ remains constant. It is seen that
appreciable fractions of photons travel further than the typical comoving
fluctuation length $\lambda\approx 1$ kpc. 
Fig.~\ref{fig:MCtravel} also shows by the
dotted line the equivalent distribution at $z=1100$, in case the propagation would be a diffusive random walk. It is seen that simple diffusion does not
apply. As argued in Ref.~\cite{Venumadhav:2014aka} the distribution is more
like a blast wave. Most of the distance is covered during a few $\sim 10^3$ scattering events, when the Lyman-$\alpha$ photon is temporarily on the 
extreme red or blue wing of the line, as here the mean free path drastically increases 
(cf. Fig.~\ref{fig:lalpha}). During the remainder of the $\sim 10^5-10^7$ scatterings, when the Lyman-$\alpha$ photons are in the core of the line, hardly any distance is covered.

As can be observed from the top panel of Fig.~\ref{fig:MCtravel} Lyman-$\alpha$ photon
mixing on $\sim$ kpc scales 
is substantial but not complete, in particular at high redshifts. The bottom panel shows the evolution of $f_{\alpha}(D)$ with
redshift, for three different scales, $D = 0.3, 1$, and 3 kpc, from top to bottom. This fraction can be interpreted as a ``mixing fraction" for each length scale.

\begin{figure}[htbp]
\centering
\includegraphics[width=0.48\textwidth]{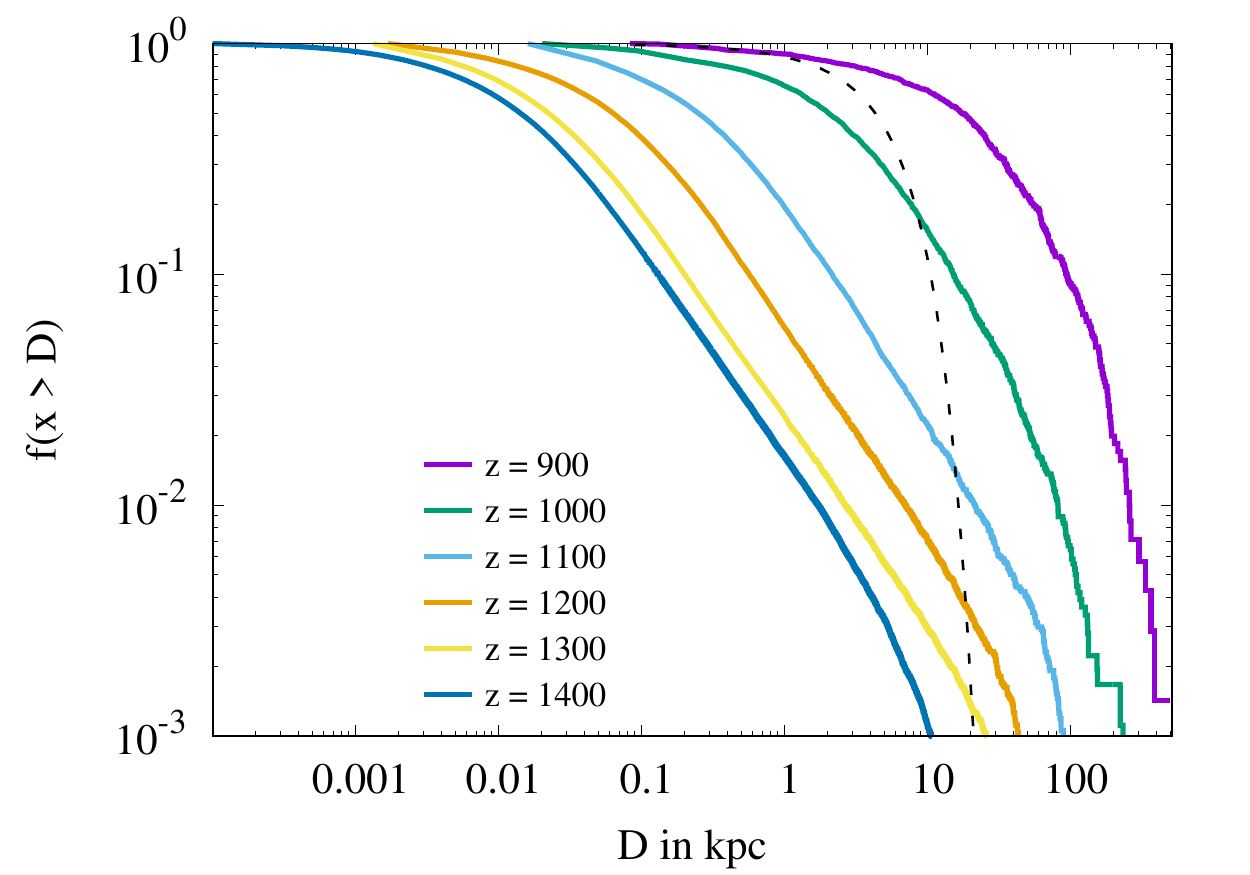}
\includegraphics[width=0.48\textwidth]{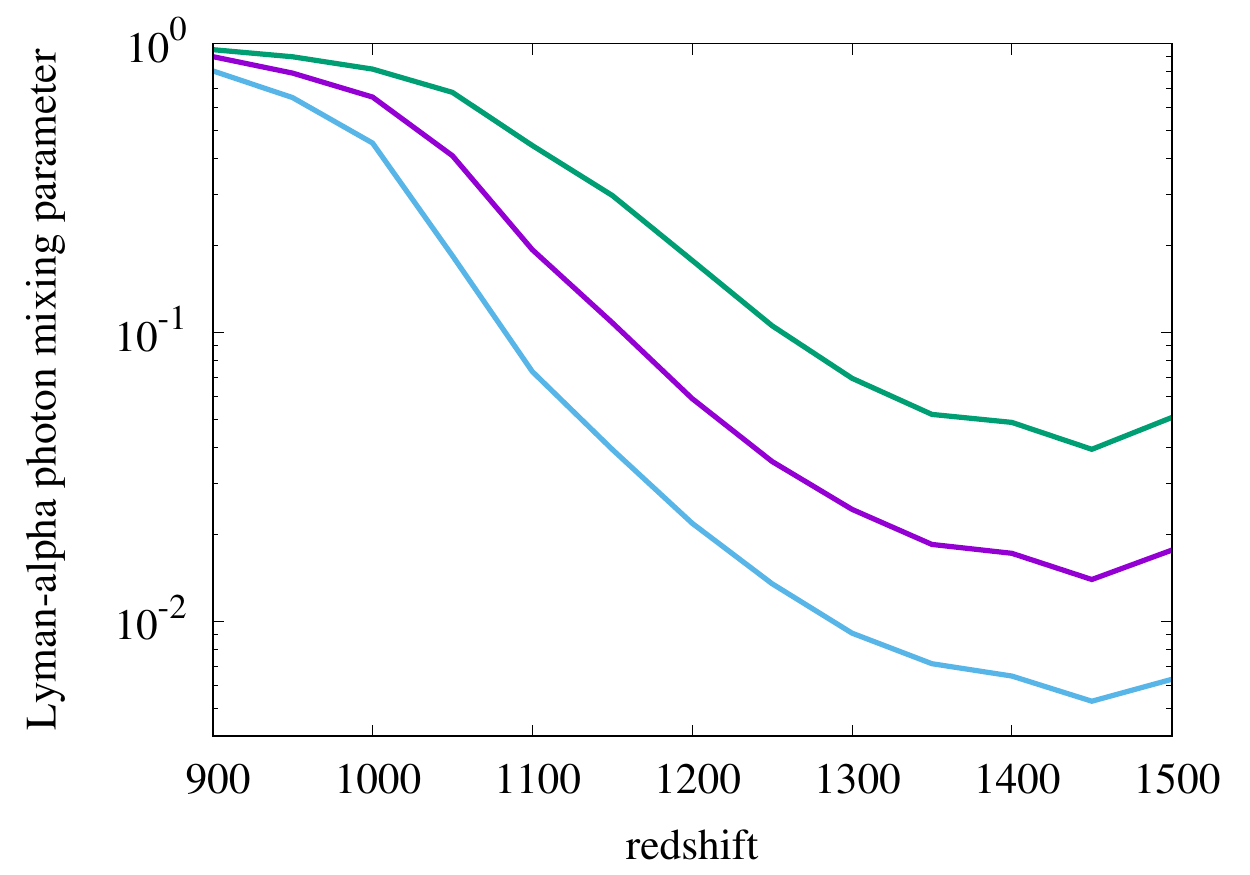}
\caption{Fraction of Lyman-$\alpha$ photons which have traveled further than a given comoving distance $D$ before being destroyed (i.e.~absorbed followed by a photoionization or 2-photon decay). \emph{Top}: This fraction is shown as a function of comoving distance $D$, for several discrete redshifts $900 \leq z \leq 1400$. The light dotted line shows the result for simple
diffusion at redshift $z = 1100$, i.e. a Gaussian distribution with the same
variance as that inferred from the realistic distribution. \emph{Bottom}: This fraction, interpreted as a scale-dependent Lyman-$\alpha$ mixing fraction, is shown as a function of redshift, for comoving scales $D = 0.3, 1$ and 3 kpc, respectively, from top to bottom.}
\label{fig:MCtravel}
\end{figure}

\begin{figure}[htbp]
\centering
\includegraphics[width=0.48\textwidth]{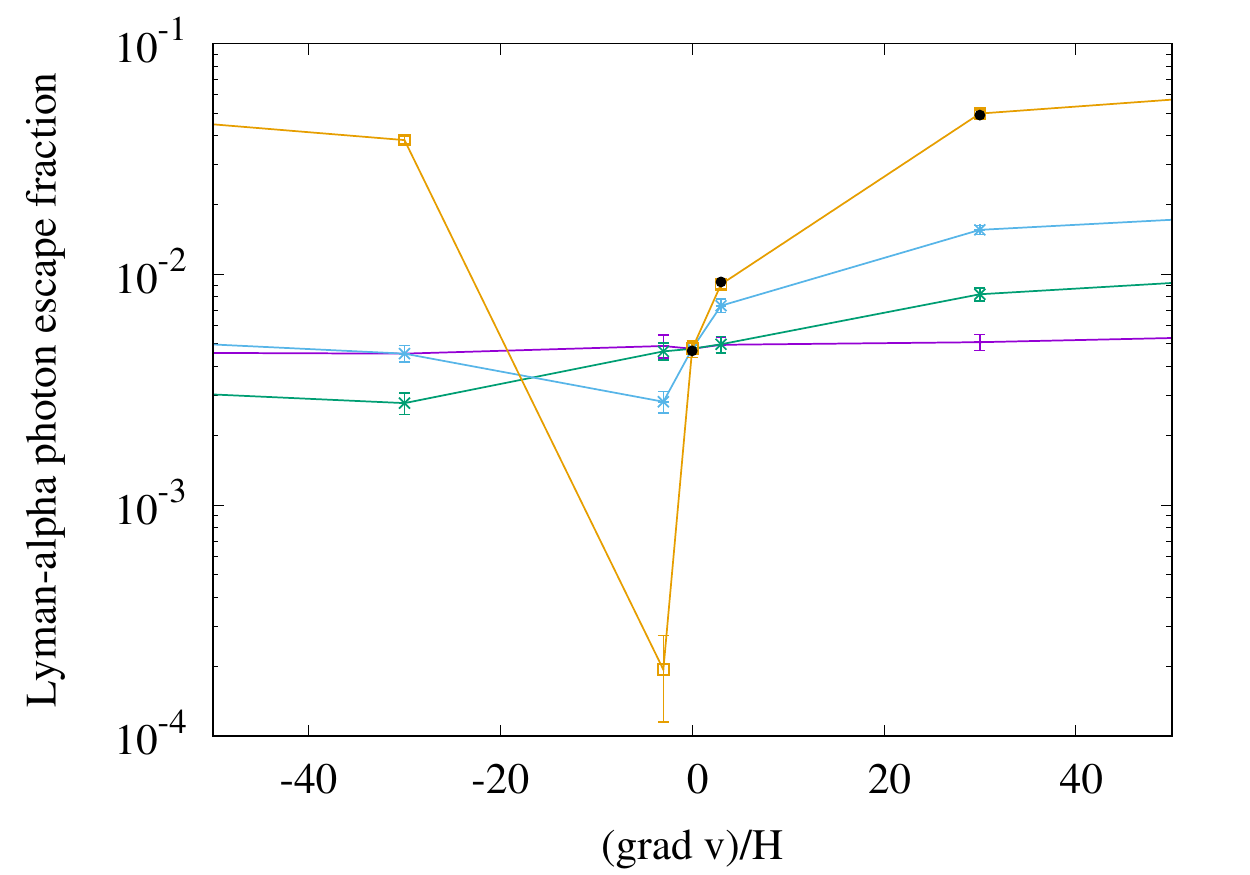}
\caption{The fraction of Lyman-$\alpha$ photons lost due to redshifting as a function of the gradient of the bulk velocity over the Hubble constant ${\bf\nabla \cdot v}/H$ . {Results of the Monte-Carlo simulation are shown by dots with lines as a visual aid. The 3D simulation assumes a bulk flow of form ${\rm v}_b = {\rm v}_0 {\rm sin}(2\pi\, x/\lambda)$ in $x$-direction and photons are injected at $x=0$. Purple, green, blue, and yellow dots are for comoving $\lambda = 0.1\,$kpc $\lambda = 1.\,$kpc, 
$\lambda = 10\,$kpc, and $\lambda = 1\,$Mpc respectively. The two black dots show the
analytical result i.e. $H\to H +{\bf\nabla v}/3$, for coherent bulk flows
on very large scales, illustrating agreement between the Monte-Carlo and the
analytical result. }
The computation was performed at redshift $z = 1100$, the redshift of the approximate peak of the CMB visibility function.}
\label{fig:MCescape}
\end{figure}

Fig.~\ref{fig:MCescape} investigates
the fraction of Lyman-$\alpha$ photons which
is lost due to redshifting, as a function of bulk flows. Without bulk flows, in a Universe with cosmological parameters as inferred from the best-fit to the Planck data, the fraction of photons lost due to redshifting and lost due to two photon decay are
$f_z ({\rm v}=0)\approx 5\times 10^{-3}$ and $f_{2\gamma}\approx 1.5\times 10^{-2}$
at redshift $z \approx 1100$, respectively. 
{In the presence of peculiar flows results are very dependent of the
coherence scale $\lambda$ of the flow. The yellow line shows result for 
$\lambda \gg D_{\alpha}$ and is in full agreement with the analytical result.
However, for smaller $\lambda$ the effect is much diminished as the Lyman-alpha
photons alternatively travel through diverging and converging flows. For
$\lambda = 0.1\,$kpc the result of $f_z ({\rm v}=0)$ is essentially recovered,
whereas for $\lambda = 1\,$kpc there is a small deviation from 
$f_z ({\rm v}=0)$, with however a reduction for 
${\bf \nabla \cdot v} < 0$ and an enhancement for ${\bf \nabla \cdot v} > 0$.
In
Section V we perform MHD simulations for a variety of magnetic field strength.
From these simulations we can infer at typical $\lambda \approx 0.1\,$kpc and 
$1\,$kpc for PMFs with a Batchelor spectrum leading to a "final" field strength
of $20\,$pG and $136\,$pG, respectively. We suspect therefore that for low
final field strengths the effects of peculiar flows on the Lyman-$\alpha$ escape
rate should be negligible, whereas for much stronger PMFs, we expect an 
additional reduction of $X_e$ due to peculiar flows. However, this latter
field strength seems already ruled out if only the effect of clumping is taken into account \cite{Jedamzik:2018itu}.   
}

It is our goal to obtain a very precise prediction of the average ionzation fraction $X_e$. To verify our suspicion that additional losses of Lyman-$\alpha$ photons due to bulk flows are negligible, we have indulged in a more elaborate Monte Carlo simulation of Lyman-$\alpha$ photon loss. In particular, we have used snapshots
of the numerical simulations in order to have a realistic and detailed knowledge of the density- and velocity- distribution at a given redshift. We have then "emitted"
Lyman-$\alpha$ photons from a great number of randomly chosen locations and followed their propagation in physical space and evolution of energy space.
This allowed us to compute the {\it global} Lyman-$\alpha$ photon loss rate numerically ~\footnote{In particular we evaluated the integral $(1/V)
\int C \alpha_e n_e^2$ by Monte Carlo integration and compared it to the
fully Lyman-$\alpha$ mixed (cf. Section \ref{sec:Recombination}) result
$(1/V)\,
C^{\rm mix}\int\alpha_e n_e^2$ which was applied in the simulations.} 
and compare it to the one in a Universe without pecuilar bulk flows and fully
mixed Lyman-$\alpha$ photons. Differences between these two global loss rates are due to two effects: bulk flows
and the effect of partial Lyman-$\alpha$ mixing important at higher redshifts where $D_{\alpha}$ becomes relatively small. To isolate the two effects we
ran our Monte-Carlo simulations with and without pecuilar flows. For a simulation with Batchelor spectrum leading to a final field of 20 pG we did find
an enhancement of the loss reate by only a few percent, with the loss rate at the most important redshifts around recombination $z\sim 1100$ virtually identical. The enhancement of a few per cent was in almost equal parts due to
bulk flows and partial Lyman-$\alpha$ photon mixing. By computing the changes in ionization history due to the small enhancement of Lyman-$\alpha$ photon loss in a homogeneous Universe (for simplicity) we could conclude that
the relative change should be below the one per cent level for those field strengths not already ruled out.
We conclude therefore that negelcting bulk flows and assuming full Lyman-$\alpha$ mixing introduces errors in $X_e$ not larger than one per cent.


\section{Hydrogen recombination with Lyman-$\alpha$ photon mixing} \label{sec:Recombination}

In the previous section we have found that in an inhomogeneous Universe, with inhomogeneities on comoving kpc scales, Lyman-$\alpha$ photons of different regions may actually partially mix. We now investigate the effect on the average ionization fraction during recombination. Our work is mostly concerned with hydrogen recombination, and we briefly describe our approximate treatment of Helium recombination in Appendix \ref{app:helium}.

\subsection{Net Lyman-$\alpha$ transition rate in the full-mixing limit}

\subsubsection{General setup}

In order to treat Lyman-$\alpha$ photon mixing,
we must return to the first-principles derivation of the net decay rate in the Lyman-$\alpha$ transition. We will follow Ref.~\cite{Ali-Haimoud_2010b}.

At a fundamental level, the net decay rate depends on the photon occupation number $f_\nu(\hat{n})$ in the vicinity of the Lyman-$\alpha$ resonance, averaged over photon propagation directions $\hat{n}$ and integrated over the line profile $\phi(\nu)$:
\begin{equation}
    \dot{n}_{2p}|_{\rm Ly \alpha} = - A_{2p, 1s} \left[ n_{2p} - 3 n_{1s} \iint \frac{d^2 \hat{n}}{4 \pi}~ d \nu~ \phi(\nu) f_\nu(\hat{n})\right], \label{eq:dotn1s}
\end{equation}
where the line profile integrates to unity, $\int d \nu ~\phi(\nu) = 1$. To obtain $ \dot{n}_{1s}|_{\rm Ly \alpha}$, we must therefore solve the radiative transfer equation for $f_\nu(t, \vec{x}, \hat{n})$. 

Near the Lyman-$\alpha$ resonance, photons may undergo resonant scattering events $\gamma_{\rm Ly \alpha} + \textrm{H}(1s) \rightarrow \gamma_{\rm Ly \alpha}' + \textrm{H}(1s)$, true absorption events $\gamma_{\rm Ly \alpha} + \gamma_{\rm bb} + \textrm{H}(1s) \rightarrow  \textrm{H}^*$, where H$^*$ is an excited (or ionized) atom in a $s$ of $d$ angular momentum state and $\gamma_{\rm bb}$ is a blackbody photon with energy corresponding to the H($2p) \rightarrow$ H$^*$ transition, or the reverse ``true emission" process, see e.g.~Refs.~\cite{Hirata_08, Ali-Haimoud_2010b}. The radiative transfer equation then takes the form
\begin{equation}
    \partial_t f_\nu + \hat{n} \cdot {\bf \nabla} f_\nu - H \nu ~\partial_\nu f_\nu = \dot{f}_\nu|_{\rm scat} + \dot{f}_{\nu}|_{\rm em, ab}. \label{eq:rad-trans}
\end{equation}
The timescale it takes a photon to redshift across the optically thick part of the line is much shorter than a Hubble time, and one may therefore solve this equation in the quasi-steady-state approximation, i.e.~neglecting the partial time derivative $\partial_t f_\nu$.

Before describing the scattering and emission/absorption terms, we first define $p_{\rm sc} \equiv A_{2p, 1s}/\Gamma_{2p}$, where $\Gamma_{2p}$ is the total rate of all transitions out of the $2p$ state. The quantity $p_{\rm sc}$ is the probability that an atom in the $2p$ state decays to the ground state rather than be excited or ionized by a blackbody photon. We denote its complementary by $p_{\rm ab} = 1 - p_{\rm sc}$. These two probabilities only depend on photon temperature and are therefore nearly homogeneous.

The scattering term $\dot{f}_{\nu}|_{\rm scat}$ does not change photon number, and only changes photon direction and frequency. The latter effect is due to the thermal motions of scattering atoms. It is not essential to our derivation, and we shall therefore ignore atomic velocities as a first pass, and come back to this point at the end of the derivation. In that limit, resonant scattering simply changes photon propagation directions, while preserving the angle-integrated photon occupation number: 
\begin{equation}
    \int d^2 \hat{n} ~\dot{f}_\nu|_{\rm scat}(\hat{n}) \approx 0.
\end{equation}

The true emission/absorption term in Eq.~\eqref{eq:rad-trans} is given by (see Ref.~\cite{Ali-Haimoud_2010b} for a derivation, which is easily generalized to arbitrary angular dependence):
\begin{eqnarray}
    \dot{f}_{\nu}|_{\rm em, ab} &=& p_{\rm ab} \frac{3 A_{2p, 1s}}{8 \pi \nu^2} \phi(\nu)\left( n_{1s} f_{(\rm em)} - n_{1s} f_\nu \right),\\
    f_{(\rm em)} &\equiv& p_{\rm ab}^{-1}\left(\frac{n_{2p}}{3 n_{1s}} - p_{\rm sc} \iint \frac{d^2 \hat{n}}{4 \pi} d \nu ~\phi(\nu) f_\nu(\hat{n})\right). \label{eq:fem}
\end{eqnarray}

\subsubsection{Net decay rate in the full-mixing limit}

We now consider the limit in which the characteristic distance traveled by Lyman-$\alpha$ photons between emission and (true) absorption is much longer than the wavelength of perturbations of interest. In that case, we may assume that the photon occupation number is approximately isotropic and homogeneous, $f_\nu(t, \vec{x}, \hat{n}) \approx \langle f_\nu \rangle$.


Inserting this approximation into the radiative transfer equation, and averaging over angles, we see that the gradient term $\hat{n} \cdot \nabla f_\nu$ and the scattering term (in the limit of negligible atomic motions) both drop out. Taking the spatial average of the equation (and recalling that we make the steady-state approximation and neglecting $\partial_t f_\nu$), we then obtain
\begin{equation}
   \partial_\nu \langle f_\nu \rangle \approx \langle \tau_{\rm ab} \rangle \phi(\nu) \left[ \langle f_\nu \rangle - f_{(\rm em)}^{\rm mix}\right], 
\end{equation}
where $\langle \tau_{\rm ab}\rangle $ is the average optical depth to true absorption, 
\begin{equation}
    \langle \tau_{\rm ab}\rangle  = \frac{3 A_{2p, 1s}}{8 \pi H \nu_\alpha^3} p_{\rm ab}\langle n_{1s} \rangle, \label{eq:<fnu>}
\end{equation}
where we approximated $\nu \approx \nu_\alpha$, the resonant frequency, and we have moreover defined
\barr
f_{(\rm em)}^{\rm mix} \equiv \frac{\av{n_{1s} f_{(\rm em)}}}{\av{n_{1s}}}
= p_{\rm ab}^{-1} \left(\frac{\av{n_{2p}}}{3 \av{n_{1s}}} - p_{\rm sc} \int d \nu~ \phi(\nu) \av{f_\nu} \right).\label{eq:fem-mix}
\earr
Equation \eqref{eq:<fnu>} is a simple linear 1st-order ODE, to be solved under the boundary condition of thermal equilibrium $\langle f_\nu \rangle \rightarrow e^{ - h \nu/T_\gamma} \approx e^{- E_\alpha/T_\gamma}$ at $\nu \rightarrow +\infty$ (neglecting feedback from higher-order lines, and again approximating $\nu \approx \nu_\alpha$ within the boundaries of the problem). It has the explicit solution
\begin{equation}
    \langle f_\nu \rangle = f_{(\rm em)}^{\rm mix}+ \left(e^{- E_\alpha/T_\gamma} - f_{(\rm em)}^{\rm mix} \right) \exp\left[- \langle \tau_{\rm ab} \rangle \int_{\nu}^{\infty} d \nu' \phi(\nu')\right].
\end{equation}
Integrating this equation over the line profile and using the definition of $f_{(\rm em)}^{\rm mix}$, Eq.~\eqref{eq:fem-mix}, we obtain, in the relevant limit that $\av{\tau_{\rm ab}} \gg 1$, 
\beq
\int d\nu ~\phi(\nu) ~\av{f_\nu} \approx \frac{\av{n_{2p}}}{3 \av{n_{1s}}} + \frac{p_{\rm ab}}{\av{\tau_{\rm ab}}} \left( e^{-E_\alpha/T_\gamma} - \frac{\av{n_{2p}}}{3 \av{n_{1s}}} \right).
\eeq
We are now in a position to compute the \emph{local} net Lyman-$\alpha$ decay rate: it is given by Eq.~\eqref{eq:dotn1s}, where $n_{2p}$ and $n_{1s}$ are \emph{local} abundances, but $f_\nu = \langle f_\nu \rangle$ is the homogeneous photon occupation number, that is,
\barr
   \dot{n}_{2p}|_{\rm Ly \alpha} &=& - A_{2p, 1s} \left[ n_{2p} - 3 n_{1s} \int d \nu ~\phi(\nu) \av{f_\nu} \right]\nonumber\\
  &=& - R_\alpha^{\rm mix} n_{1s} \left[ \frac{\langle n_{2s}\rangle}{\langle n_{1s} \rangle} - e^{- E_\alpha/T_\gamma} \right] \nonumber\\
  &&- 3 A_{2p, 1s} \left[ n_{2s} - n_{1s} \frac{\av{ n_{2s}}}{\av{n_{1s}}} \right], \label{eq:dotn1s-mix}
\earr
where we substituted $n_{2p} = 3 n_{2s}$ and we have defined
\begin{equation}
    R_\alpha^{\rm mix} \equiv \frac{8 \pi H}{\lambda_\alpha^3 \langle n_{1s} \rangle} = 1/\langle R_\alpha^{-1} \rangle.
\end{equation}
We see that, in the limit that the baryon density is homogeneous, so that $n_{1s} = \langle n_{1s} \rangle$ and $n_{2s} = \langle n_{2s} \rangle$, this net decay rate reduces to Eq.~\eqref{eq:Ly-alpha-std}, as it should. However, in the presence of inhomogeneities, the net decay rate can be \emph{significantly} different, especially due to the second term in the right-hand-side of Eq.~\eqref{eq:dotn1s-mix}, given that $A_{2p, 1s} \gg R_\alpha$.

\subsubsection{Effect of diffusion and bulk flows}

So far we have entirely neglected photon frequency diffusion due to resonant scattering off of thermally moving atoms, as well as the effect of bulk flows. The former effect is known to lead to $\mathcal{O}(4\%)$ corrections to the Lyman-$\alpha$ net decay rate at $z \approx 1100$, see e.g.~Ref.~\cite{Hirata_2009}, and we expect that correctly incorporating them would lead to similar corrections in the full-mixing limit. Since Lyman-$\alpha$ photons scatter many times between emission and true absorption, for perturbations much smaller than the Lyman-$\alpha$ re-absorption length scale, they scatter off many patches with incoherent peculiar velocities. Hence, on small scales bulk flows act effectively as an additional contribution to thermal motions. Therefore we expect that, in the full mixing limit, baryon peculiar velocities v can be effectively incorporated by substituting $T_b \rightarrow T_b +  m_p \av{{\rm v}^2}/3 \approx T_b \left( 1 + \langle {\rm v}^2 \rangle/3c_s^2\right)$ in the resonant scattering diffusion operator, where we took the limit $X_e \ll 1$ in $c_s^2$. Assuming the effect of frequency diffusion scales linearly with $T_b$ as long as it is perturbative, we expect bulk flows to start significantly affecting the net Lyman-$\alpha$ decay rate when $\av{{\rm v}^2}/ 3 c_s^2 \gtrsim 1/(4\%) \sim 25$, i.e.~when $\langle {\rm v}^2 \rangle^{1/2} \gtrsim 9 c_s$.

\subsection{Local recombination rate in the full mixing limit}

To find the local recombination rate, we must again solve for the abundance of excited hydrogen in the steady-state approximation. Instead of Eq.~\eqref{eq:n2s-ss}, we now have
\barr    
0 \approx \dot{n}_{2s} &=& \alpha_e n_e^2 - \beta_e n_{2s} - \Lambda_{2\gamma} \left(n_{2s} - n_{1s} e^{- E_\alpha/T_\gamma}\right) \nonumber\\
&-& R_\alpha^{\rm mix} n_{1s} \left[ \frac{\langle n_{2s}\rangle}{\langle n_{1s} \rangle} - e^{- E_\alpha/T_\gamma} \right] \nonumber\\
&-& A_{2p, 1s} \left[ n_{2s} - n_{1s} \frac{\av{n_{2s}}}{\av{n_{1s}}}\right]. \label{eq:n2s-ss-local}
\earr
We solve this equation in two steps. First, we take its spatial average. This eliminates the last term, and results in an equation for $\av{n_{2s}}$ identical to Eq.~\eqref{eq:n2s-ss}, with the substitutions $(n_e^2, n_{1s}, n_{2s}, R_\alpha) \rightarrow (\av{n_e^2}, \av{n_{1s}}, \av{n_{2s}}, R_\alpha^{\rm mix})$. 
with solution therefore identical to Eq.~\eqref{eq:n2s-sol} with the same substitutions. Next, we solve the local quasi-steady state equation \eqref{eq:n2s-ss-local} for $n_{2s}$, given $n_e^2, n_{1s}$ and their average values. In practice, given that $A_{2p, 1s} \gg \beta_e, R_\alpha, \Lambda_{2 \gamma}$, the solution is very well approximated by
\beq
n_{2s} \approx n_{1s} \frac{\av{n_{2s}}}{\av{n_{1s}}}.
\eeq
We can now finally compute the local net recombination rate, $\dot{n}_e|_{\rm rec} =
- (\alpha_e n_e^2 - \beta_e n_{2s})$, which, after substituting the solution for $n_{2s}$, becomes
\barr
\dot{n}_e|_{\rm rec} &=& - C^{\rm mix} \left(\alpha_e n_e^2 - \beta_e n_{1s} e^{- E_\alpha/T_\gamma}\right) \nonumber\\
&&- (1 - C^{\rm mix}) \alpha_e \left(n_e^2 - \frac{n_{1s}}{\av{n_{1s}}} \av{n_e^2} \right), \label{eq:dotne-mix}
\earr
where the full-mixing Peebles-C factor $C^{\rm mix}$ is given by Eq.~\eqref{Cfactor} with the substitution $R_\alpha \rightarrow R_\alpha^{\rm mix} = 1/ \av{R_\alpha^{-1}}$.

Equation \eqref{eq:dotne-mix} is one of our main new analytic results. The first term is a modification of the standard Peebles recombination rate with $C \rightarrow C^{\rm mix}$. The second term, proportional to $1 - C^{\rm mix}$, is a qualitatively new term, which accounts for the spatial mixing of Lyman-$\alpha$ photons. This term could in principle have a \emph{very} significant effect on the net recombination rate when $C^{\rm mix} \ll 1$. However, it averages to zero, hence should produce a relatively minor effect on the average free-electron density, as we will indeed confirm below.

\subsection{Illustration with simple 2-zone model}


To illustrate the difference between the no-mixing and full-mixing regimes, we consider a toy model in which the Universe consists of equi-probable regions with baryon overdensities $\Delta^{\pm} = \rho_b^{\pm}/\av{\rho_b} = n_{\rm H}^{\pm}/\av{n_{\rm H}} = 1 \pm \sqrt{b}$, where $b$ is the baryon clumping parameter, which we shall take to be constant $b = 0.5$.

In each case, we solve for the free-electron fraction $x_e \equiv n_e/n_{\rm H}$ in each zone, and then compute the ``global" free-electron fraction $X_e$, through
\beq
X_e \equiv \frac{\av{n_e}}{\av{n_{\rm H}}} = \frac12 \left[\Delta^+ x_e^+ + \Delta^- x_e^-\right].
\eeq
In the no-mixing limit, we simply solve two independent ODEs for the ionization fraction $x_e^{\pm}$ in each ``zone". In contrast, in the full-mixing limit, the 2 ODEs are coupled, since they each depend on average quantities such as $\av{n_{1s}} = \av{n_H} (1 - X_e)$.

\begin{figure*}[htbp]
\centering
    \includegraphics[width = 2\columnwidth]{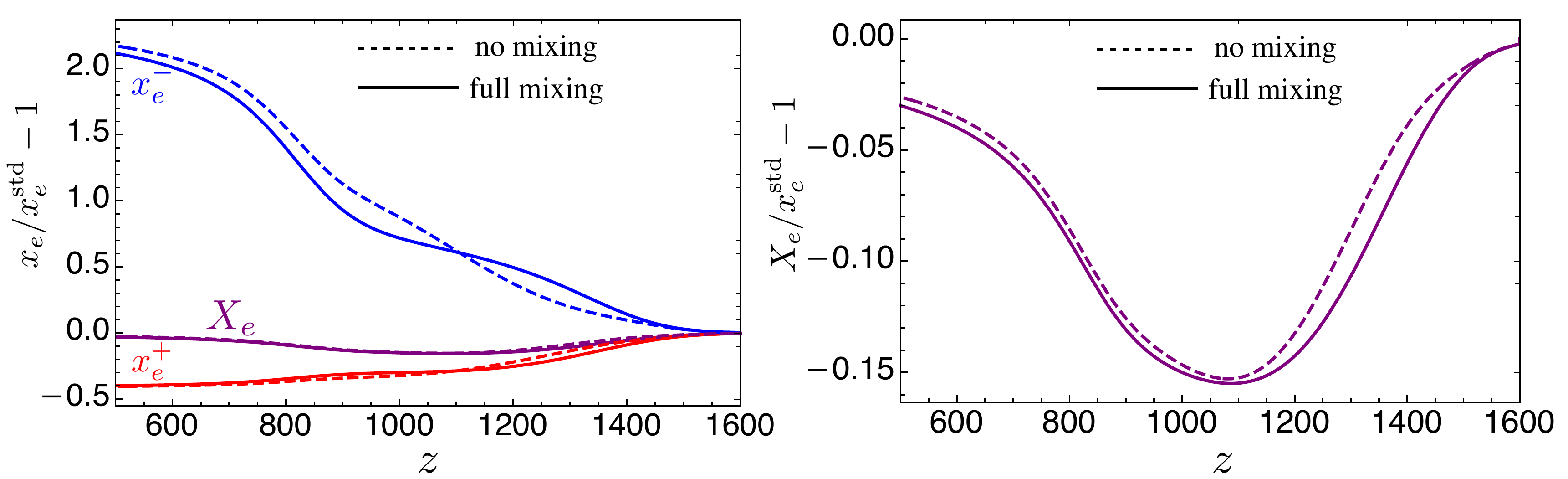}
    \caption{Change in the free-electron fraction relative to the standard history, in a simple toy model with two equi-probable zones with baryon densities $\rho_b^{\pm} = \av{\rho_b}(1 \pm \sqrt{b})$, with clumping parameter $b = 0.5$, in the no-mixing limit (dashed), and in the full-mixing limit (solid). \emph{Left}: separate evolution of $x_e^+$ (red), $x_e^-$ (blue) and $X_e \equiv \av{n_e}/\av{n_{\rm H}}$ (purple). \emph{Right}: zoom-in on the evolution of $X_e$ in the no-mixing and full-mixing limits. We see that the two limits give very similar results.}\label{fig:rec-2zone}
\end{figure*}

We show the solutions in the two limit cases in Fig.~\ref{fig:rec-2zone}. The left panel shows the separate evolutions of $x_e^\pm$ in the no-mixing and full-mixing limits, and the right panel shows the evolution of the global ionization fraction $X_e$ in both limits. We see that, first of all, the deviation of $X_e$ from the standard ionization fraction is typically much smaller than the deviations of the ionization fractions in each zones. Second, we see that the evolution of $X_e$ in the no-mixing and full-mixing limits are very similar. {We will however see that there are more significant deviations between these two limits in the realistic case of a Batchelor spectrum (Section V), 
particularly at lower redshifts.}

\section{A realistc calculation of Recombination in the Presence of
Primordial Magnetic Fields} \label{sec:realistic}

In this section we show results for the typical perturbations in the ionization
fraction during recombination in a magnetized Universe. We first present
some general results and then consider the best strategy to obtain precise results.
All calculations in this section assume non-helical PMFs with a Batchelor
spectrum and if not otherwise stated, are performed in the
limit of full mixing, i.e.~using Eq.~\eqref{eq:dotne-mix} for the net recombination rate. 

\subsection{General Trends}\label{sec:general-trends}

{PMF evolution in the early Universe is described by freely decaying MHD.
Individual magnetic modes excite fluid motions which then dissipate. Before
recombination those fluid motions are associated with baryon clumping which then decays again. 
As a first application of our code we determine when a particular 
magnetic mode of given proper scale $L$ and field strength $B$ is dissipated. We define this time approximately by the time of maximum clumping.}
In \cite{Banerjee:2004df} an analytic model based on numerical results was developed. It
found that magnetic modes dissipate when the eddy turn-over time
approximately equals the Hubble time, i.e.~when $L \approx \textrm{v}/H$,
where $\rm v$ are the peculiar motions excited by the
magnetic field. During recombination
the fluid undergoes a transition from viscous MHD, when the photon drag coefficient $\alpha$ defined in Eq.~\eqref{eq:alpha-def} is much greater than the expansion rate $H$, to turbulent MHD when $\alpha$ falls below $H$ due to the drop in electron density. 

\begin{figure}[htbp]
\centering
\includegraphics[width=0.48\textwidth]{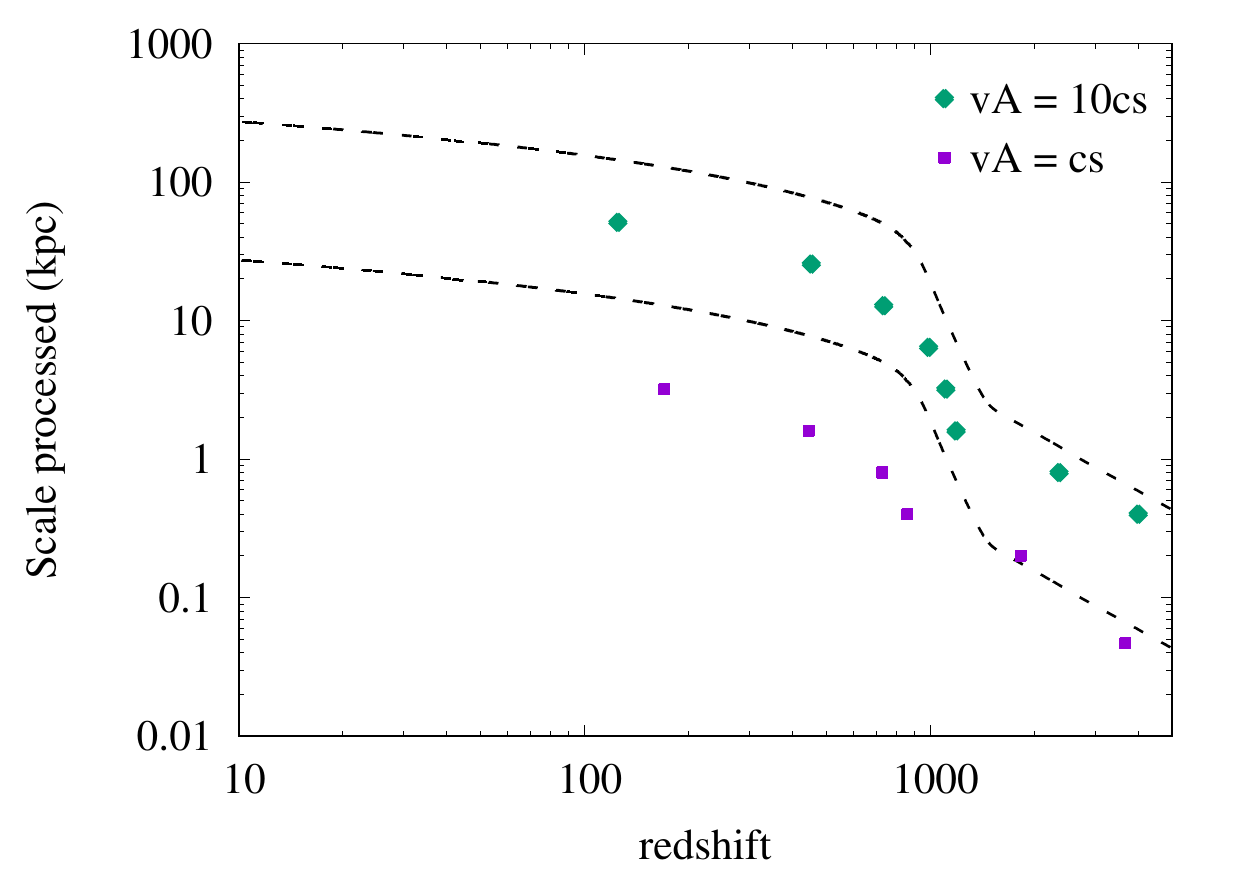}
\caption{Comparison of numerically obtained and theoretically inferred
redshifts $z_{diss}$ of the dissipation of magnetic modes as a function of their comoving scale, for two different field
strength, i.e. ${\rm v}_A =1$ and $10 c_s$ before 
recombination. The theoretical prediction (see text)
is given by the dotted lines, whereas the numerically determined redshifts are
shown by the green (${\rm v}_a = 10 c_s$) and purple (${\rm v}_a = c_s$) points. The numerical ``experiments" evolve one magnetic mode of a fixed comoving scale and the
redshift of dissipation is defined as the redshift of maximum clumping.
}
\label{fig:onemode}
\end{figure}

In the viscous regime $\alpha \gg H$ peculiar velocities
may be estimated by ${\rm v}\approx {\rm v_A^2}/(\alpha L)$, implying ${\rm v} \approx {\rm v_A}/\sqrt{\alpha/H}$, whereas in the turbulent regime, $\rm v \approx {\rm v_A}$. We can combine the two regimes by approximating ${\rm v} \approx {\rm v_A}/\sqrt{1 + \alpha/H}$, implying the following expression for $L$
\begin{equation}
L\approx \frac{\rm v_A}{\sqrt{H(\alpha + H)}}\, ,
\end{equation}
giving the length scale which is dissipated as a function of redshift and
magnetic field strength. In Fig.~\ref{fig:onemode} the dotted lines show
the prediction for two different field strengths. This can be compared to the results of the code when magnetic plane waves of wavelength $L$ are evolved from early times (the green and purple points). It is observed that the theoretical
approximation is good at early times $z > 1000$ but deviates from the
numerical results at lower redshift (or larger scales). 
Fig.~\ref{fig:onemode} is therefore useful to
find those magnetic modes which induce the maximum clumping shortly before recombination. The clumping induced by modes smaller/larger that this 
mode is already decaying/has not fully developed yet.

\begin{figure*}[htbp]
\centering
\includegraphics[width=0.48\textwidth]{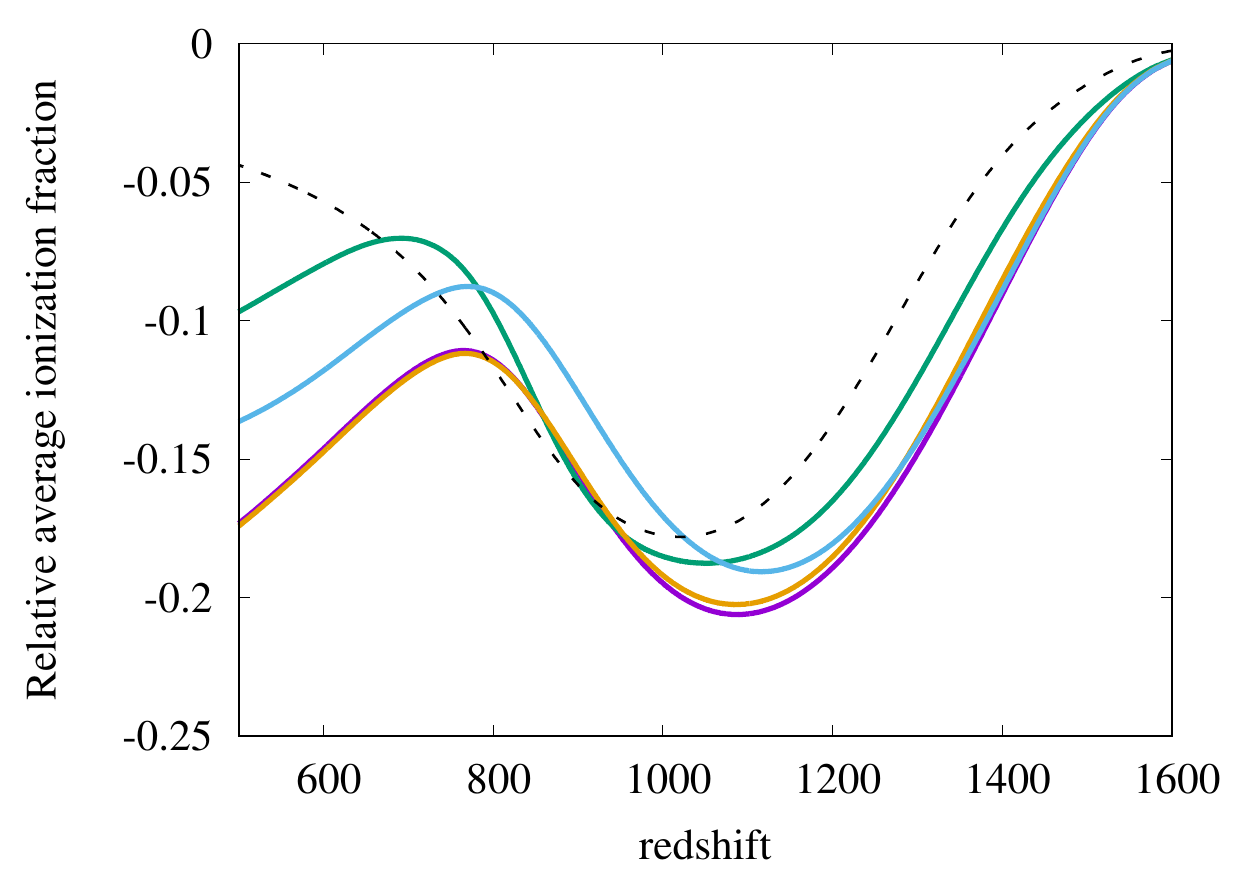}
\includegraphics[width=0.48\textwidth]{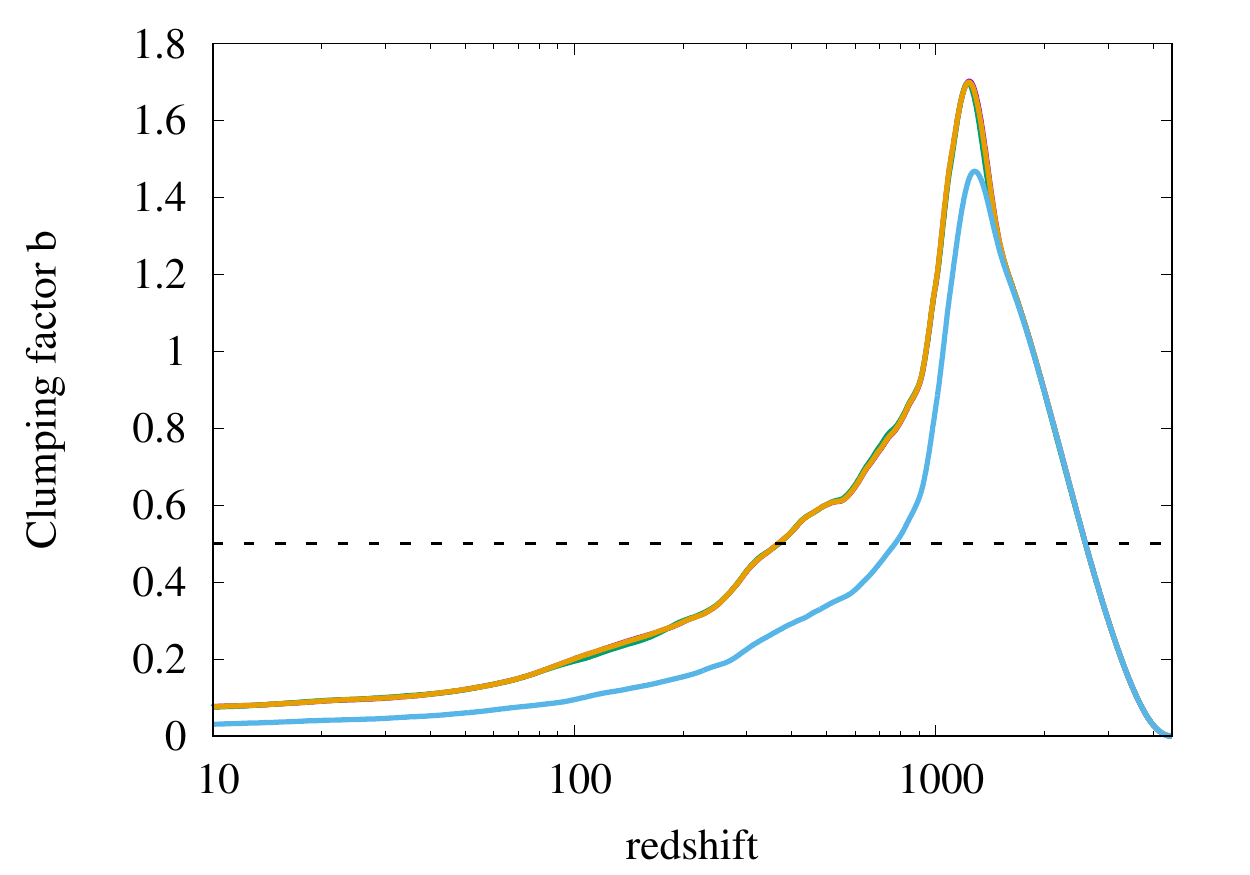}
\caption{Relative change of the global ionization fraction $X_e$ when compared to a homogeneous Universe without PMFs (\emph{left}) and evolution of the clumping factor (\emph{right}), 
for a number of $256^3$ numerical simulations of a non-helical PMF with Batchelor spectrum. Initial conditions
were chosen as ${\rm V_{A,rms}} = 12c_s$ and $L_{box}=24\,$kpc with all modes in the range 
$3\, {\rm kpc}\leq \lambda \leq 24\,$ kpc excited.  
In both panels, the different lines correspond to: full Lyman-$\alpha$ mixing, i.e. $f_{\alpha} = 1$ (purple), no
Lyman-$\alpha$ mixing, i.e. $f_{\alpha} = 0$ (green), Lyman-$\alpha$ mixing
as given by the MC results of the last section with mixing length scale $3$kpc
(orange), 
and full Lyman-$\alpha$ mixing but with the reduction of
speed of sound during recombination neglected (blue). For comparison, the dotted line
shows results of the M1 three-zone model (no Lyman-$\alpha$ mixing) for clumping factor $b=0.5$. It is seen that the reduction of speed of sound during recombination has a 
some impact on $b$, whereas details of the Lyman-$\alpha$ mixing do not. 
}
\label{fig:effects}
\end{figure*}

Next we consider the impact of individual effects on the ionization fraction
$X_e$ and the evolution of the clumping factor. We use the same PMF spectrum as
that with results shown in Figs.~\ref{fig:2D} and \ref{fig:Bdecay}, but omit particular effects. 
In Fig.~\ref{fig:effects} we
show results for the perturbation in the ionization fraction. The purple line shows results when
the reduction of speed of sound during recombination is taken into 
account and full Lyman-$\alpha$ photon mixing between different regions
is assumed. The orange line shows results when instead of full mixing $f_{\alpha}=1$, mixing according to the Monte-Carlo simulations in Section IV
(i.e. Figs. \ref{fig:MCtravel} with mixing "length" 3 kpc) is assumed. It is seen that the deviations between 
the two is very small. The blue line shows results for full mixing
but not taking into account of the reduction of speed of sound during recombination. Here a deviation of order 10 per cent is observed. Somewhat larger deviations are attained when Lyman-$\alpha$
photon mixing is neglected shown by the green line. Coincidentally, in this case $X_e$ is close to that of the three-zone model M1 with $b=0.5$. 

In the right panel of Fig.~\ref{fig:effects} we show the evolution of the clumping factor with the same different physical effects considered in the left panel. The reduction of speed of sound enhances clumping, whereas as expected, 
details of Lyman-$\alpha$ transport do not have an effect on clumping. It is noted that in all cases, the final field at redshift $z = 10$ is 
$4.38\times 10^{-2}$nG. Some residual clumping of $b \approx 0.1$ survives to low redshifts. 

The MHD simulations of the PMF evolution assume as initial condition a
particular random realization of the spectrum. It is important to know
if different random realizations result in deviating $X_e$ predictions, in particular, if there is realization variance. We have compared results for three different random realizations of the PMF field. We found that the global ionization fraction $X_e$ varies by at most $\sim 1 \%$ between different realizations. Hence the realization variance is well below the overall effect of PMFs on $X_e$, which is of order $20\%$. We also found that the clumping factor vary by at most $\sim 10\%$, well below peak clumping of order $b \sim 2$. We conclude that realization variance is of order $\sim 5\%$ of the effects we compute, and may be safely neglected. 

\subsection{Optimizing calculations for non-helical magnetic fields with a Batchelor spectrum} \label{sec:optimize}

It is our ultimate goal to calculate accurately the changes in ionization
fraction $\Delta X_e$ as a function of final surviving 
rms magnetic field $B_f$ strength. Here define $B_f$ as the total surviving
root-mean-square magnetic field at redshift $z=10$.  Though in
principle our code can accomplish this, in practice we have the
limitation of feasible numerical resolution. In particular, there are three
mutually possibly conflicting requirements on numerical resolution:
\vskip 0.1in
\begin{itemize}
\item Each individual Fourier mode has to be resolved sufficiently to obtain
convergence of results as a function of resolution. 
In Appendix \ref{app:convergence} we demonstrate that in order to have approximately $\sim 10\%$ accurate results in $\Delta X_e$ and $B_f$, individual modes have to be resolved at least with resolution $32^3$, preferably $64^3$.
\item All magnetic modes which contribute to $B_f$ and whose induced clumping
effects $\Delta X_e$ have to be included. Referring to 
Fig.~\ref{fig:onemode} we see that a larger range of scales is processed
before and during recombination, potentially requiring very large 
numerical resolutions.
\item There has to be a larger number of those magnetic modes which mostly influence $\Delta X_e$ (those which produce the peak in clumping shortly 
before recombination) as otherwise there could be large realization variance.
\end{itemize} 
To illustrate the point more clearly. Given a simulation box of physical
size $L_{box}$ we can populate randomly all modes $k_i = (2\pi/L_{box}) n_i$, where $n_i$ are integers with $i$ denoting spatial
direction. If we populate all modes $n = \sqrt{n_1^2+n_2^2+n_3^2}
\leq 8$, we need at least $256^3$ resolution in order to resolve all
modes with at least $32^3$. If we identify the $n_{rec}\approx 6 \pm 1$ modes with those which have the main impact on recombination there will be not much realization variance, as there are many modes 
$\sim n_{rec}^2\Delta n_{rec}\sim 72$. That was the case for all the simulations shown so far (c.f. discussion at the end of Sec.~\ref{sec:general-trends}). 
However, in that case we will not be able to populate ultraviolet modes
larger than $k_{max} = (4/3) k_{rec}$. Those ultraviolet modes will still have
an influence on recombination due the residual clumping they had produced and due to mode-mode coupling. Vice versa, if we take $k_{max} = 4 k_{rec}$ as to include enough UV modes, than there should be significant realization variance as there are only $\sim n_{rec}^2\Delta n_{rec}\sim 4$ of the most important modes for recombination. {We will investigate uncertainties due to missing ultra-violet modes and realization variance further below.}

We will now demonstrate that the easiest spectra to simulate are those of
phase-transition generated non-helical PMFs with Batchelor spectrum.
Considering inflationary fields with approximately scale-invariant spectra
we may directly refer to Fig.~\ref{fig:onemode}
to assess the range of scales required to simulate to obtain
an accurate result. Taking for example the case ${\rm v}_A = 10 c_s$, in order
to simulate all magnetic modes which dissipate between redshift $z \approx  5000$ and $z\approx  500$, we need to resolve all scales between 
$L\approx 0.3\,$kpc and $30\,$kpc, implying a dynamic range of $\sim 100$.
This would require simulations with resolution $3200^3$ which would require
1000 - 10000 times larger computer resources than used here\footnote{We note here that 2D simulations significantly overestimate the effects on recombination.}. On the other hand, as $B(L)\sim L^{-5/2}$ for
Batchelor spectra, the
dynamic range required is much compressed
compared to scale-invariant fields. 
Before recombination the mode dissipation redshift $z_{diss}$ scales as
$(B/L)^{2/3}$ such that modes a factor two smaller dissipate at a redshift
$z_{diss}$ a factor $2^{7/3}\approx 5$ larger. For scale-invariant fields the
factor is only $2^{2/3}\approx 1.6$. For fully helical fields with Batchelor
spectrum, the necessary dynamic range is also compressed, but not as much
as in the non-helical case, as the inverse cascade pushes fields to 
larger scales. 

\begin{table}[!h]
\label{T:1}
\begin{center}
\begin{ruledtabular}
\begin{tabular}{| c | c | c | c | c | c |c |}
Run & Size & $L_{min}$ (kpc) & $L_{max}$ (kpc) & ${\rm B_{rms}^{ini}}$ (pG) & ${\rm B_{rms}^{fin}}$ (pG) & $z_{diss}$\\
\hline

 1 & $256^3$ & 1.5 & 6. & $21.9$ & 8.54 & $< 1000$\\ \hline
 2 & $64^3$ & 1.03 & 1.03 & $51.0$ & 3.54 & $\approx 1000$\\ \hline
 3 & $64^3$ & 0.713 & 0.713 & $121$ & 2.77 & $\approx 1500$\\ \hline
 4 & $64^3$ & 0.492 & 0.492 & $327$ & 0.865 & $\approx 3300$\\ \hline
 5 & $64^3$ & 0.339 & 0.339 & $827$ & 0.627 & $\approx 7850$\\ \hline
 6 & $64^3$ & 0.224 & 0.224 & $2093$ & 0.738 & $\approx 18700$\\     

\end{tabular}
\end{ruledtabular}
\caption{Size, minimum scale resolved, maximum scale resolved, initial magnetic field strength, final magnetic field strength, and approximate dissipation redshift estimated from Fig.~\ref{fig:onemode} for
six smaller simulations Run 1 - Run 6.}
\end{center}
\end{table}

\begin{figure}[t]
\centering
\includegraphics[width=0.48\textwidth]{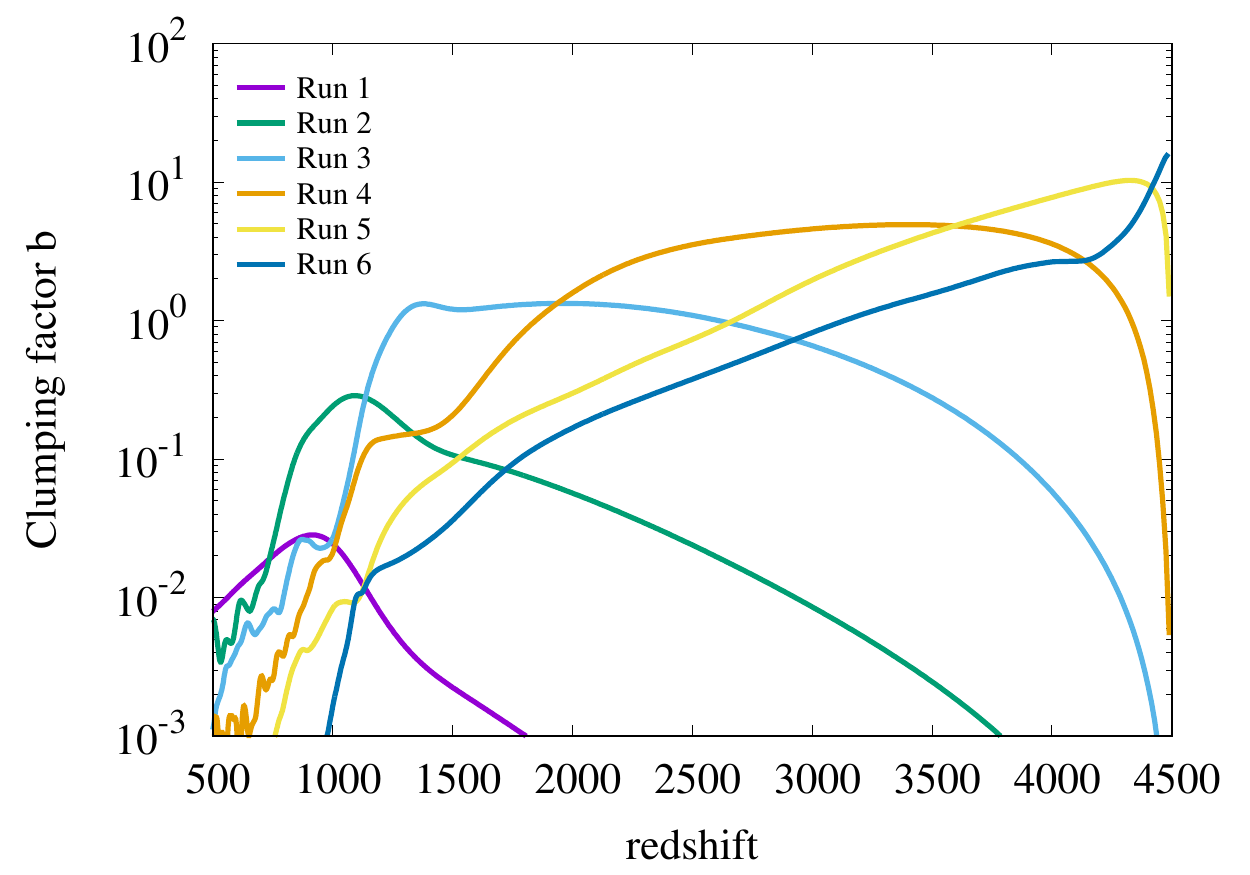}
\includegraphics[width=0.48\textwidth]{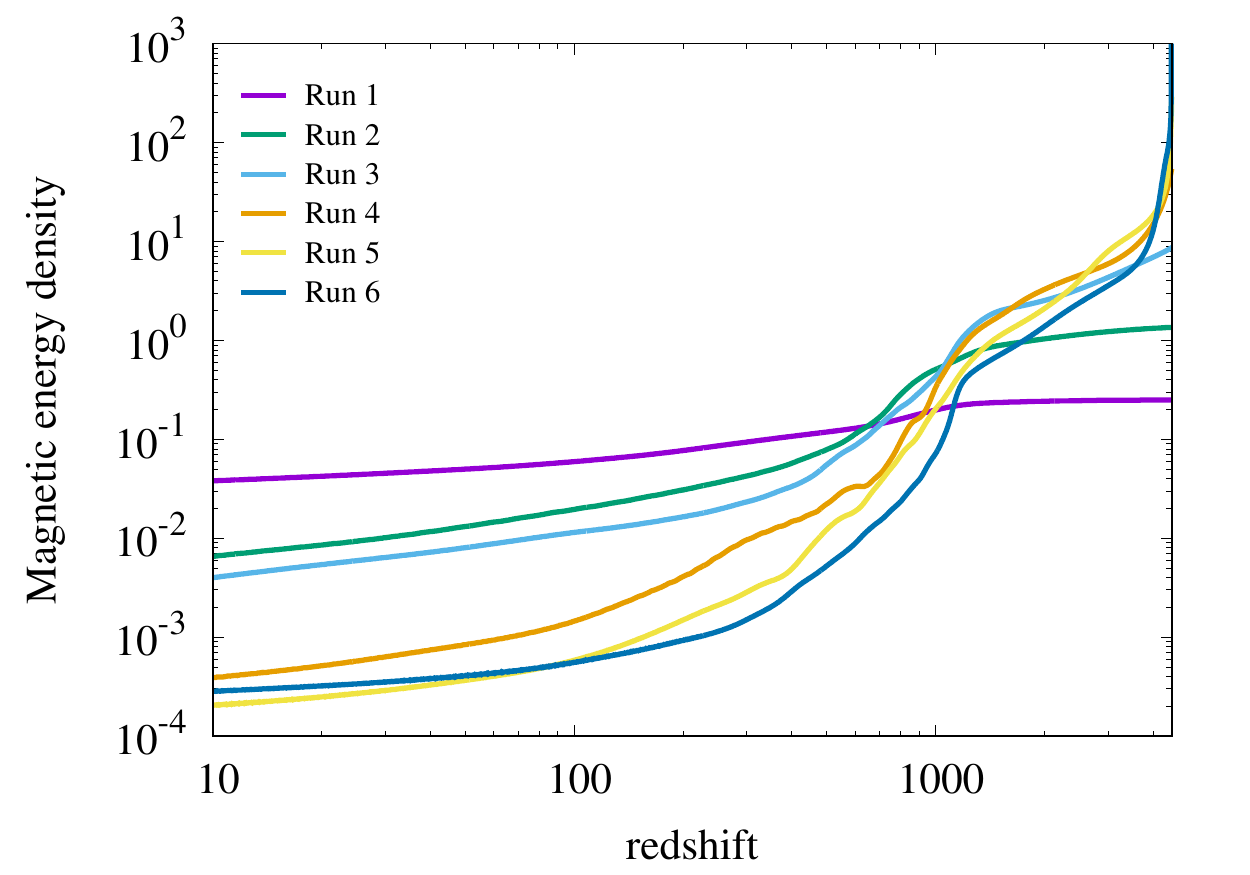}
\includegraphics[width=0.48\textwidth]{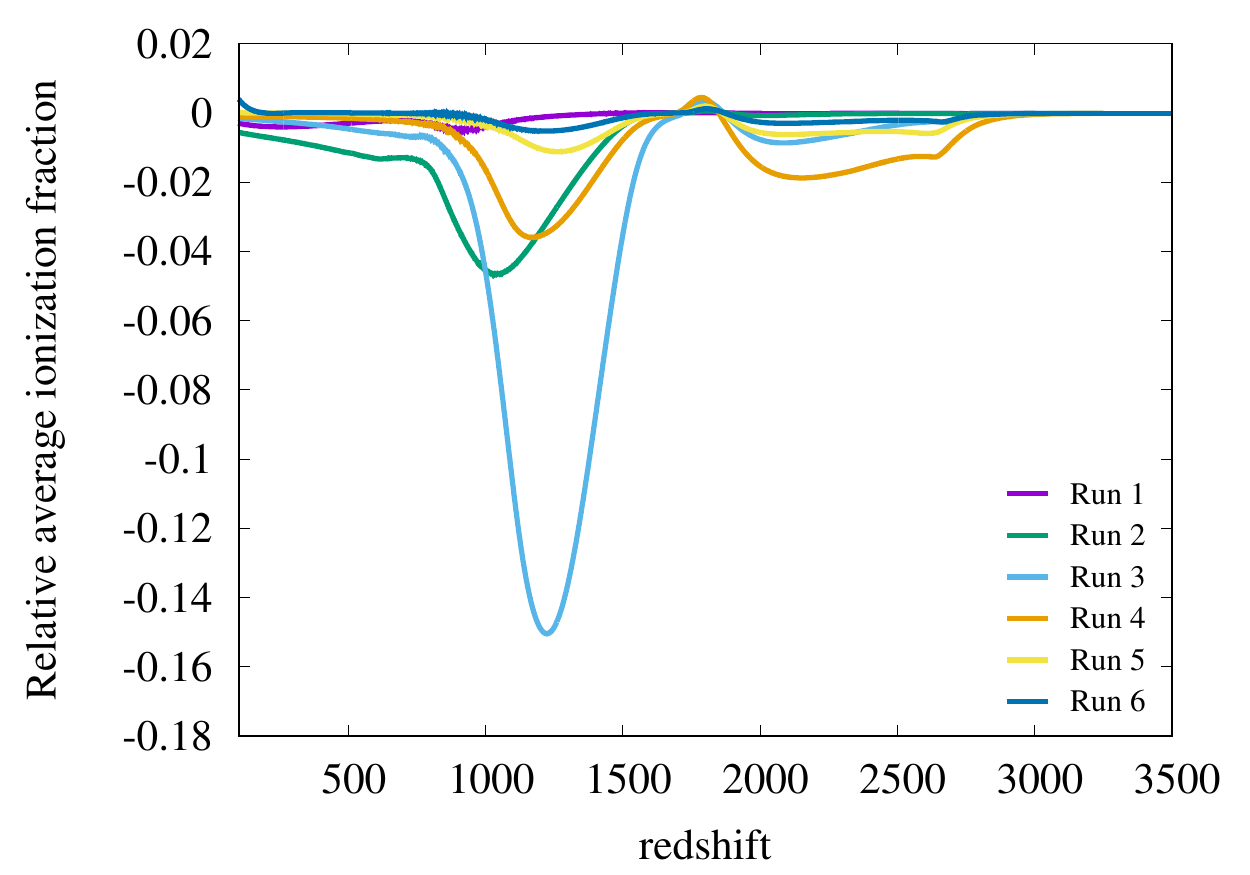}
\caption{Redshift evolution of the clumping factor (\emph{top}), magnetic energy density (\emph{middle}) and fractional change to the global ionization fraction (\emph{bottom}) for the simulations Run 1 - Run 6.}
\label{fig:Runs1-6}
\end{figure}

We will from now on focus on the non-helical Batchelor case. In order to
establish which scales need to be resolved in the simulation, we perform a series of smaller simulations. The scales resolved and initial magnetic field
strengths $B_{rms}^{ini}$ of these simulations are shown in Table I.
Field strength and length scales in all these simulations
were chosen as to mimic the Batchelor spectrum,
i.e. $B(L)\sim L^{-5/2}$, i.e. Run 1 presents the largest scales with 
smallest $B_{rms}^{ini}$ and Run 6 presents the smallest scales with
largest $B_{rms}^{ini}$. Obtaining results for these simulations allows us,
to assess, in the absence of mode-mode coupling, which modes contribute most to the final field strength $B_{rms}^{fin}$ and which modes most to the perturbation in $\Delta X_e$. Results for the evolution of clumping factor,
magnetic energy density, and ionization fraction are shown 
in Fig.~\ref{fig:Runs1-6}. As expected, the peak in the clumping occurs at
successively higher redshifts as one goes from Run 1 to Run 6
\footnote{Strictly speaking should Run 5 and Run 6 be started
at higher initial redshift. Due to the strength of the photon drag
at early times this is however computationally very expansive. We have verified that changing the initial redshift to somewhat higher values does not significantly change the evolution at lower redshift.}. Run 3 produces
the largest clumping right before hydrogen recombination, the "neighboring" runs
Run 2 and Run 4 also lead to some significant clumping shortly before
hydrogen recombination. This compares well with the analytic estimate of
"dissipation" redshift also given in Table I, defined as the redshift of peak clumping for a mode. These are
between $z_{diss} \approx 1000$ and $3300$ for runs Run 2 - Run 4. In the bottom panel of Fig.~\ref{fig:Runs1-6} we see that indeed $X_e$ is mostly
affected by Run 3, and to a lesser degree by Run 2 and Run 4. In the absence
of mode-mode coupling effects, one could imagine that only resolving the scales resolved in Run 2 to Run 4, would already lead to a fairly accurate estimate
for $\Delta X_e$. This corresponds to a required dynamic range of only two.
Inspecting the middle panel of Fig.~\ref{fig:Runs1-6} one infers that the final magnetic
field is dominated by the larger modes resolved in Run 1 with a subdominant contribution from Run 2 and Run 3.

\begin{figure}[htbp]
\centering
\includegraphics[width=0.48\textwidth]{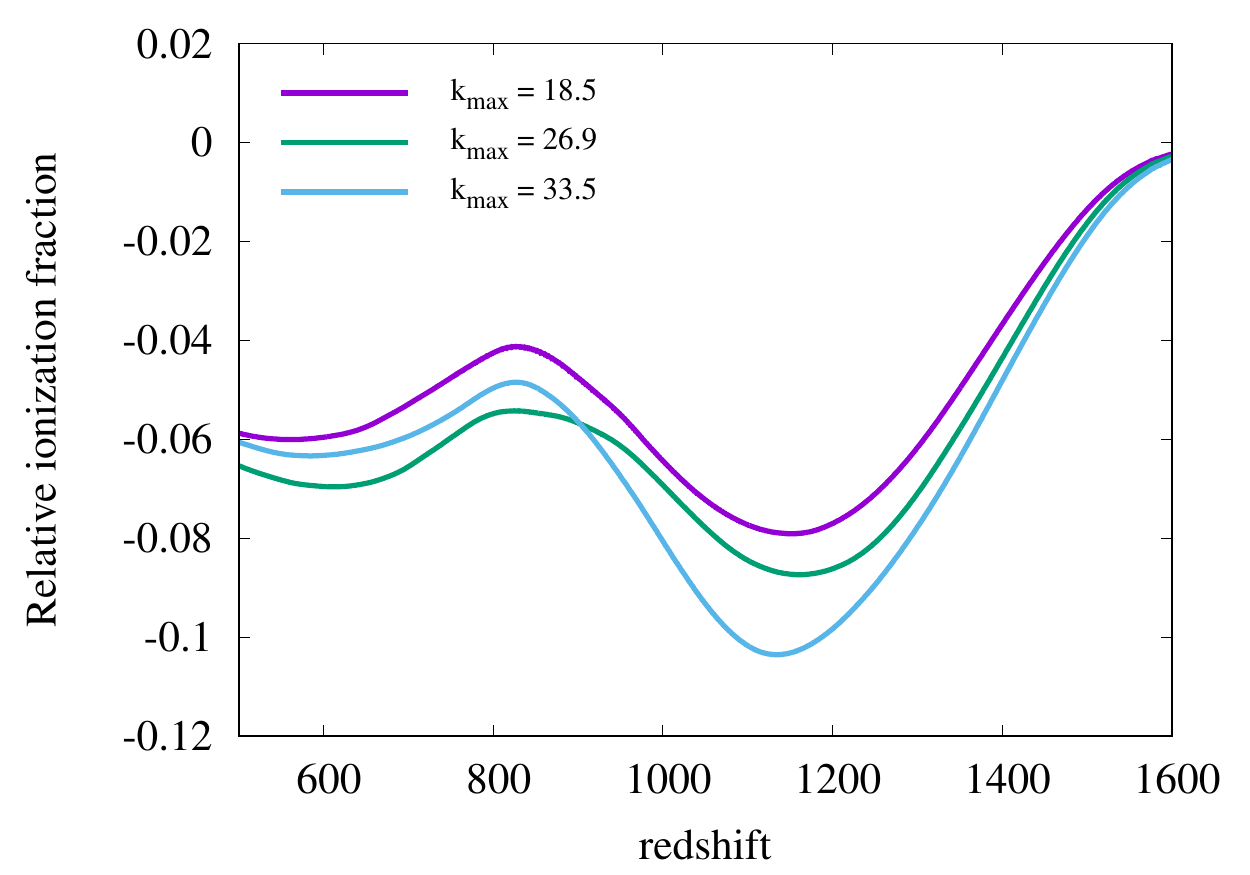}
\caption{Dependency of $\Delta X_e$ on the ultraviolet cut-off of the 
magnetic field spectrum. Each curve represents the average $\Delta X_e$
of four different random realizations of a PMF with Batchelor spectrum.
It is seen that including higher-$k$ modes, which dissipate well before
recombination (i.e. at $z_{diss}\approx 2\times 10^4$ for Run 9 and 
$z_{diss}\approx 3.3\times 10^4$ for Run 10)
do have impact on $\Delta X_e$ due to mode-mode coupling.}
\label{fig:highk}
\end{figure}

\begin{table}[!h]
\label{T:2}
\begin{center}
\begin{ruledtabular}
\begin{tabular}{| c | c | c | c | c |}
Run & Size & $L_{min}$ (kpc) & $L_{max}$ (kpc) & ${\rm B_{rms}^{ini}}$ (pG) \\
\hline

 7 & $256^3$ & 1. & 8. & $60.4$ \\ \hline
 8 & $256^3$ & 0.339 & 1.5 & $900$ \\ \hline
 9 & $256^3$ & 0.234 & 1.5 & $2278$ \\ \hline
 10 & $256^3$ & 0.1875 & 1.5 & $3966$ \\

\end{tabular}
\end{ruledtabular}
\caption{Size, minimum scale resolved, maximum scale resolved, and initial magnetic field strength 
for Run 7 - Run 10.}
\end{center}
\end{table}

This suggests the following strategy to obtain relatively precise results.
The final magnetic field strength may be obtained by a simulation of the
relevant larger scales, whereas $\Delta X_e$ may be obtained by an independent simulation of the relevant smaller scales. In Table II we show the size, resolved scales, and initial magnetic field for simulations to obtain the
final field (Run 7) and the impact on $X_e$ (Run 8 - Run 10).
Here Run 8 resolves all scales of Run 2 - Run 5. 
In Fig.~\ref{fig:highk} we show averaged results of
three realizations of each Run 8 - Run 10. It is seen that including even smaller modes than initially deemed necessary from the results of 
Run 2 - Run 6 have impact on $\Delta X_e$. There is a significant difference
in $\Delta X_e$ between Run 8 and Run 9, which becomes even larger between
Run 8 and Run 10. The smallest modes in Run 10 have 
$z_{diss} \approx 3.3\times 10^4$ and due to mode-mode coupling effects
still have an important effect on $X_e$. Such small modes 
naively thought to dissipate at high redshift should therefore still be included. {We are currently not able to assess if the inclusion of even smaller modes further enhances the $X_e$ reduction.}

\begin{figure}[htbp]
\centering
\includegraphics[width=0.48\textwidth]{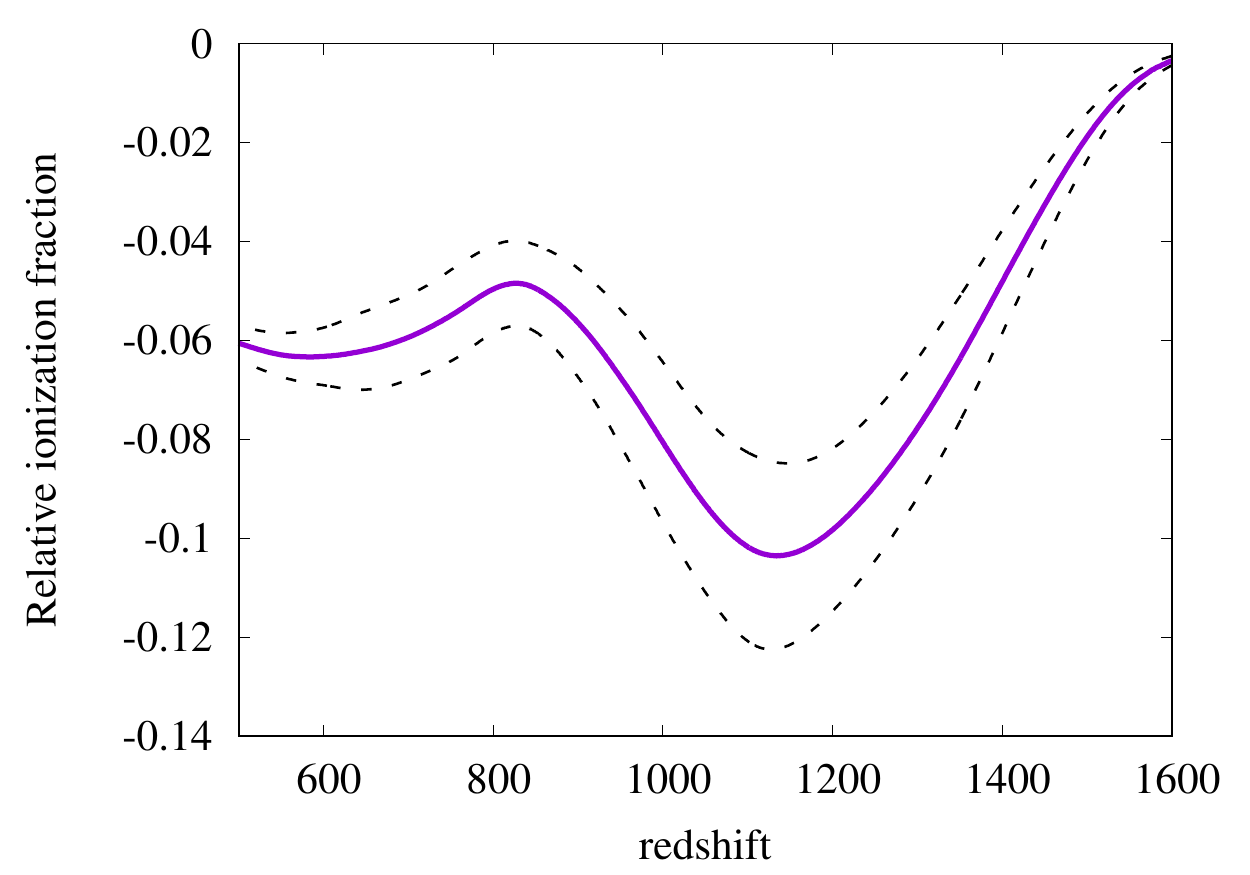}
\caption{The average $\Delta X_e$  and one-sigma ranges
of three different random realizations of a PMF with Batchelor spectrum Run 10. It is seen that random realization uncertainties are substantial, around $\sim 20\%$. {This is in contrast do the small realization variance found for 
the results displayed in Fig. \ref{fig:effects}, since those simulations included a large number of modes dissipating during recombination, but did hardly include important ultra-violet modes.}}
\label{fig:standard_deviation}
\end{figure}

As it is necessary to include more UV modes, the $n_{rec}$ for Run 10 is
only two. We therefore expect significant realization variance. 
In Fig.~\ref{fig:standard_deviation} the average $\Delta X_e$ is shown for
Run 10, albeit based only on four different realizations. The dotted lines show the average plus, minus one standard deviation. It is observed that realization variance is significant, $\sim 20\%$, such that it is preferable to reduce the error due to realization variance by averaging the result of a larger number of
simulations with different realizations.

    \section{The combined effects of plasma heating and baryon clumping} \label{sec:heating}

It is well known \cite{Sethi:2004pe} that the decay of magnetic turbulence induces a heat
source for the baryons which may influence the ionization fraction $X_e$ 
substantially at lower redhifts. Similar holds for ambipolar diffusion at even
lower redshifts. An increase of the baryon temperature $T_b$ over that of the CMB $T_{\rm CMB}$
lowers the temperature-dependant recombination rate which leads to higher residual ionization fractions. The evolution of the baryon temperature with time $t$ is given by \eqref{eq:Tb-evolution}. At redshifts $z\approx 1000$ the ionization fraction is still fairly high
$X_e\approx 0.1$ such that baryon cooling by electron Thomson scattering on CMB photons
keeps $T_b$ very close to $T_{\rm CMB}$. However, at slightly lower redshifts
$z\approx 900$ when $X_e\approx 10^{-2}$ the heating of baryons due to the dissipation of magnetic fields can significantly increase the baryon 
temperature. This in turn reduced the temperature dependent recombination 
rate which leads to higher residual $X_e$. 

\begin{figure}[htbp]
\centering
\includegraphics[width=0.48\textwidth]{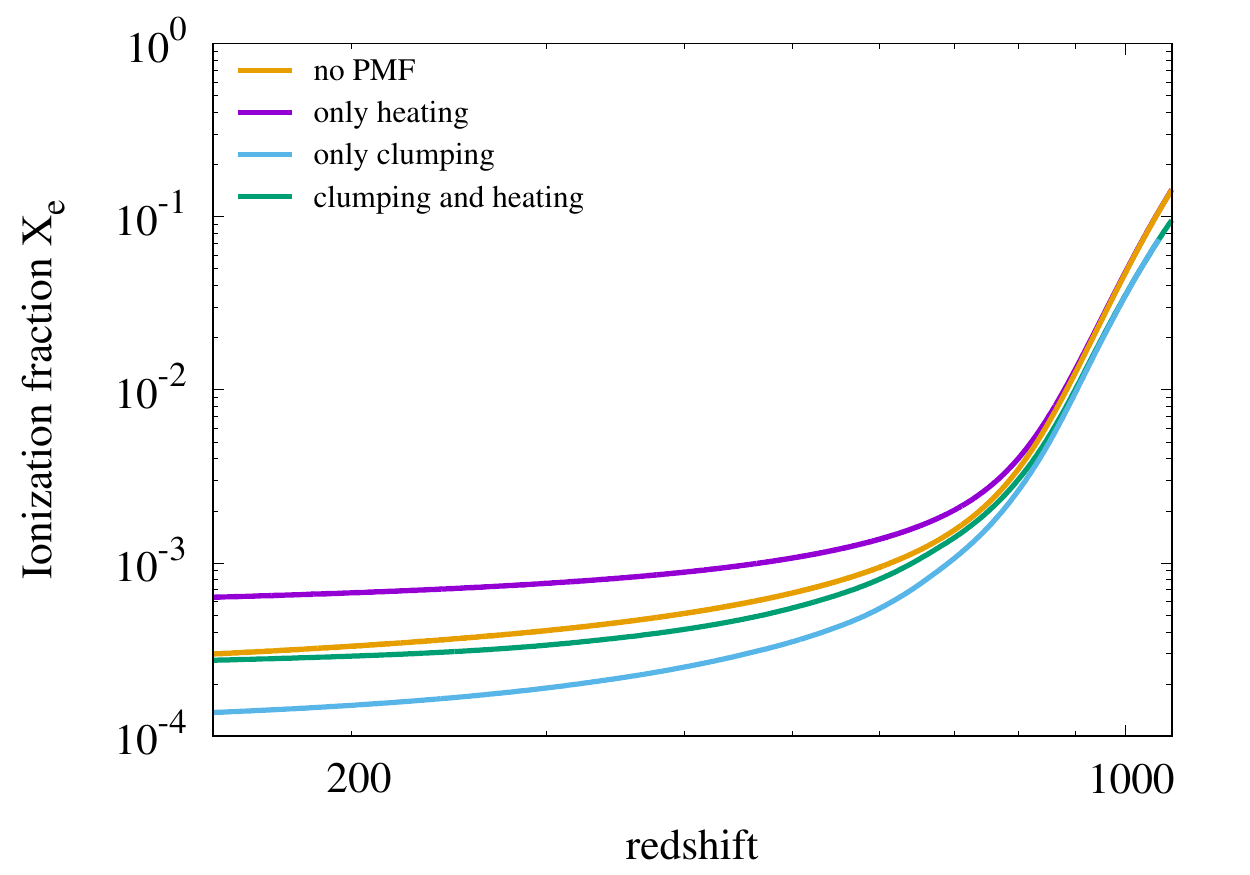}
\caption{The ionization fraction $X_e$ as a function of redshift for four different cases: the homogeneous case without clumping and dissipation (orange), the inhomogeneous case due to baryon clumping, but without dissipation (blue),
the homogeneous case without clumping but with dissipation (purple), and the
realistic inhomogeneous case with clumping and dissipation (green).}
\label{fig:diss}
\end{figure}

A higher residual $X_e$ in the redshift range $\sim 400-900$ induces a suppression of the CMB anisotropies at higher multipoles due to residual scatterings of photons out of the beam. It also introduces a feature at small multipoles $l\sim 100-200$. These effects may be constrained by high-precision CMB data, as has been done in \cite{Kunze:2014eka,Chluba:2015lpa,Planck:2015zrl,Paoletti:2018uic,Paoletti:2022gsn}. These authors used an analytic expression
from \cite{Sethi:2004pe} to approximate the heating rate due to magnetic field dissipation.
The only study which attemps to compute the heating rate numerically \cite{Trivedi:2018ejz}
could not simulate at higher magnetic field strength, as they were facing the
complication of baryon clumping, which many MHD solvers have problems with.
We introduced heating of baryons due to magnetic field dissipation in the
modified ENZO code. To this end we use a uniform heating rate
~\footnote{Magnetic- and kinetic- energy dissipation here occurs over numerical
viscosity, which is a common practice in MHD simulations in 
astrophysics \cite{Kritsuk:2011ki}. In principle we could have computed the local dissipation rate
when using the standard energy evolution equation in {\sl ENZO} amended by
CMB cooling/heating, as {\sl ENZO} conserves energy. 
However, we do not think that the approximation of applying a global average homogeneous heating rate could change our conclusions below drastically.
}
\begin{equation}
\Gamma = \frac{1}{\rm V}\frac{d}{d t}\int d{\rm V}\biggl( \frac{1}{2} \rho_b {\rm {\bf v}}^2 + \frac{1}{8\pi} {\bf B}^2\biggr)\, ,
\end{equation}
which allows us to evolve the local baryon temperature. We did not consider the less important heating due to ambipolar diffusion.
We are thus able, for the first time, to treat both effects, clumping and 
heating at the same time. 

Results are shown in Fig.~\ref{fig:diss}. Here a particularly strong magnetic field with Batchelor spectrum was chosen, with remaining magnetic field strength of $1.2\,$nG at $z = 1088$ leading to a final field of $0.13\,$nG at 
redshift $z = 10$. According to \cite{Paoletti:2022gsn}, 
which give the limit $B < 0.18\,$nG at
$z = 1088$ such a field should be clearly ruled out due to heating. It should
also be ruled out due to clumping, if the three-zone model M1 with older data
is used \cite{Jedamzik:2018itu}, due to modifications of $X_e$ at 
$z\approx 1100$ rather than at $z\approx 400 - 900$.
In Fig.~\ref{fig:diss} one sees the ionization fraction for four different simulations.
The orange line shows $X_e$ for a homogeneous Universe with no magnetic fields.
The purple line shows $X_e$ when the heating rate is extracted from the MHD
simulation and used in a homogeneous Universe. This corresponds to the 
evolution of $X_e$ computed in \cite{Paoletti:2022gsn} with the difference that the heating rate
is not taken from an analytic approximation but computed directly in the code. 
This case seems clearly ruled out by current CMB data. The blue line shows 
$X_e$ when heating is not taken into account, but the baryon inhomogeneities,
i.e. clumping, 
are. In this case the average $X_e$ is lower than in the no magnetic field 
case, and taking only the low $z$ evolution into account the model should not 
be ruled out. The same holds for the realistic case, when clumping and 
heating is taken into account, which is shown by the green line. The realistic
case in fact is very close to the no magnetic field case at low redshifts. Baryon heating due to magnetic field dissipation increases $X_e$ at lower
redshift, whereas the clumping of baryons decreases $X_e$ with the net effect
being only small change from the no magnetic field case. In fact we have performed many more simulations with varying magnetic field strengths including the effects of clumping and heating and in
none of these simulations have we ever found an increased $X_e$.
We conclude therefore,
that current limits on PMFs from baryon heating do not apply, as they have not
taken into account clumping. We stress, that constraints on this model come
from clumping, and the associated changes of $X_e$ around the peak of the visibility function at $z\approx 1090$, inducing a change in the sound horizon, modified Silk damping, and a change in the width of the visibility function,
but not from hydrodynamic heating. 

\section{Conclusions} \label{sec:conclusion}

In this paper we presented an in depth analysis of the effects
of primordial magnetic fields on the process of cosmic recombination. It had
been shown priorly that PMFs of sufficient magnitude lead to non-linear 
inhomogeneities in the baryons (clumping) on small
scales before recombination, and that this clumping could result in a partial solution to the cosmic Hubble
tension. However, prior analysis was in large parts based on approximative 
three zone models and omitted a number of physical effects. 
For our analysis
we employed full 3D MHD simulations including the effects of photon drag and cosmic
expansion. These MHD simulations were coupled to a new recombination code, sufficiently accurate to obtain $\sim 10-20\%$ accurate results in the 
relative perturbation of the ionization fraction $\Delta X_e/X_e$ due to the
existence of PMFs. We employed detailed Monte-Carlo (MC) simulations of Lyman-$\alpha$ photon propagation in space and frequency. These simulations
establish that for the typical peculiar flows found in the MHD simulations
for observationally allowed PMF strengths,  
the Lyman-$\alpha$ photon escape fraction due to redshifting is not much changed
from that in an unmagnetized Universe. However, the MC simulations also established
that Lyman-$\alpha$ photons are almost fully mixed between underdense- and overdense- regions such that the recombination process is no further local. 
We derived an analytic result for the recombination process in an inhomogeneous Universe with fully mixed Lyman-alpha photons, which leads to a reduction of the average ionization fraction compared to the non-mixed case. We investigated the
dependence of results for PMFs with Batchelor spectrum on ultra-violet magnetic
modes and found some unexpected dependency. 


Our study may serve as the theoretical foundation of a
precise comparison between the theory of recombination with PMFs and CMB observations.

\acknowledgments
We acknowledge Levon Pogosian for a multitude of useful conversations and much emotional support to conclude this lengthy project. We also acknowledge many useful exchanges with Jens Chluba, as well as Andrey Saveliev.
This research was enabled in part by support provided by WestGrid (www.westgrid.ca) and the Digital Research Alliance of Canada (alliancecan.ca). This work was partially supported by the U.S. Departement of
Energy SLAC Contract No. DE-AC02-76SF00515. YAH is a CIFAR-Azrieli Global Scholar and acknowledges support from the Canadian Institute for Advanced Research (CIFAR), and is grateful to be hosted by the USC departement of physics and astronomy while on sabbatical.

\appendix

\section{Monte Carlo simulations of the Lyman-$\alpha$ escape fraction}
\label{sec:MC}

The thermally averaged cross section~\footnote{Note that in this section all units are natural, i.e. $\hbar = c = k = 1$. In case numerical values are given they refer to values at redshift $z=1100$, close to the standard recombination redshift.} 
for a Lyman-$\alpha$ photon
to scatter of a hydrogen atom in the $1s$ ground state is given by

\begin{equation}
\langle\sigma\rangle_{\rm th} = \frac{1}{(2\pi)^3}\int {\rm d}\Omega\, {\rm d}p\,
p^2\, \biggl(\frac{2\pi}{m T}\biggr)^{3/2} {\rm exp}\biggl(\frac{-p^2}{2 m T}\biggr) 
\sigma ({\bf {p}})\, ,
\label{eq1}
\end{equation}
where $m$ is hydrogen mass.
Here the cross section is given by
\begin{equation}
\sigma = \frac{3\lambda^2_{\alpha}}{8\pi}\, \frac{\Gamma_{\alpha}^2}
{(\omega_{r}-\omega_{\alpha})^2+\Gamma_{\alpha}^2/4}\, ,
\end{equation}
where $\omega_{r}$ and $\omega_{\alpha}$ are photon circular frequency in the atomic rest frame and the Lyman-$\alpha$ circular frequency, respectively,
$\lambda_{\alpha}$ is the Lyman-$\alpha$ 
wavelength, and $\Gamma_{\alpha}$ is the
decay rate of the excited $2p$ state back to the ground state.  
Here $\omega_r$ relates to the photon frequency $\omega$
in the gas rest frame 
\begin{equation}
\omega_r = \omega \biggl( 1 - \frac{p_\|}{m} - U_\|\biggr)\, ,
\end{equation}
where $p_\| = {\bf\hat{k}}\cdot{\bf p}$ and 
$U_\| = {\bf\hat{k}}\cdot{\bf U}$ are the atomic momentum and fluid velocity
parallel to the photon direction given by the unit vector ${\bf\hat{k}}$.
Introducing the new variable
\begin{equation}
x = \frac{(\omega-\omega_{\alpha})}{\omega_{\alpha}{\rm v}_{\rm th}}
\end{equation}
where ${\rm v}_{\rm th} = \sqrt{2T/m}$ as well as $\tilde{p} = p/\sqrt{2 m T}$
one can rewrite the thermally averaged cross section as follows
\begin{eqnarray}
\langle\sigma\rangle = \biggl(\frac{1}{\sqrt{\pi}}\int {\rm d}\tilde{p}_{\perp 1}\,
{\rm exp}(-\tilde{p}^2_{\perp 1})\biggr)
\biggl(\frac{1}{\sqrt{\pi}}\int {\rm d}\tilde{p}_{\perp 2}\,
{\rm exp}(-\tilde{p}^2_{\perp 2})\biggr) \nonumber \\
\biggl(\frac{1}{\sqrt{\pi}}\int {\rm d}\tilde{p}_{\|}\,
\frac{{\rm exp}(-\tilde{p}^2_{\|})\sigma_0\Gamma_{\alpha}^2}
{\omega_{\alpha}^2v_{\rm th}^2(x-\tilde{p}_\| -U_\|/v_{\rm th})^2
+\Gamma_{\alpha}^2/4}\biggr)
\label{eq:MC}
\end{eqnarray}
where $\sigma_0 = 3\lambda_{\alpha}^2/8\pi$.
Eq.~\ref{eq:MC} has been written as an equation immediately usable for
Monte Carlo methods. First note that the first two terms in brackets
are unity. The third term in brackets is thus simply 
$\langle\sigma\rangle$. In the Monte Carlo we try to sample 
one-dimensional cumulative
distributions between zero and unity by randomly generated numbers between 
zero and unity. Eq.~\ref{eq:MC} can be used for that to obtain the
likely hydrogen $(\tilde{p}_{\perp 1},\tilde{p}_{\perp 2},\tilde{p}_{\| })$ on
which the photon scatters. Given those, one may compute the new
frequency of the photon after scattering. After performing three
successive Lorentz transformations, first from the gas rest frame at
emission to the gas rest frame at absorption, than from the gas rest
frame at absorption to the atomic rest frame, and than vice versa, 
and assuming that ${\rm v}_{\rm th},\, U\, \ll 1$ and that atomic recoil can be 
neglected, we
find 
\begin{equation}
x_{\rm out} \approx x_{\rm in} - \frac{(U_\|^{\rm abs}-U_\|^{\rm em})}{v_{\rm th}} - \tilde{p}_\|
(1-\mu) + \tilde{p}_{\perp}\sqrt{1-\mu^2}{\rm cos}(\beta)\, ,
\label{eq:inout}
\end{equation}
Here the rescaled photon frequencies $x_{\rm in}$ and $x_{\rm out}$ refer to 
the frequency in the gas rest frame at the location of emission and the frequency
after re-scattering in the gas rest frame at the location of absorption, whereas
$U_\|^{\rm abs}$ and $U_\|^{\rm em}$ are the fluid velocities parallel to
the photon momentum at the location of absorption and emission, respectively.
Furthermore, 
$\tilde{p}_{\perp} = \sqrt{\tilde{p}_{\perp 1}^2+\tilde{p}_{\perp 2}^2}$ is
the total perpendicular rescaled momentum, $\mu = \hat{k}_{\rm in}\cdot\hat{k}_{\rm out}$ is the cosine of the scattering angle between the incoming and outgoing photons, and $\beta$ is another scattering
angle between zero and $2\pi$. We note that beta always has a flat 
distribution and $\mu$ is flat in the interval $[-1,1]$ for
isotropic scattering, but not for dipole scattering.

After absorption of a line photon the hydrogen atom is in a metastable
$2p$ state. In most cases
it will spontaneously de-excite into the $1s$
state by re-emission of a Lyman-$\alpha$ photon with frequency 
$x_{\rm out}$. However, on rare occasions $\sim 10^{-7}$ 
the metastable $2p$ state will
be photo-ionised by CMBR photons or decay via the two-photon transition
\footnote{On less rare occasions $\sim 10^{-4}$ the hydrogen atom will be further excited by CMBR photons into $n>2$ levels. These higher excitations will with near unity probability eventually de-xcite into the ground state with the generation a new Lyman-$\alpha$ photon. We do not treat this process in the current Monte-Carlo.}.
It is evident that the small probability for reionsation makes
a Monte-Carlo very challenging, as a larger number of interactions have 
to be followed before the fate of the Lyman-$\alpha$ photon is known.

Our Monte-Carlo simulation proceeds as follows
\begin{itemize}
\item (1) Inject a photon with frequency $x=0$ at a random location
(we checked explicitly that injecting it from a Voigt profile does not
change results)
\item (2) Compute the thermally averaged cross section $\sigma$
and the photon mean free path $l_{\gamma} = 1/(\sigma n_H)$ where $n_H$ is the
neutral hydrogen density
\item (3) Advance the photon over a path length $Dl = \epsilon l_{\gamma}$. Here values of $\epsilon =10^{-1}-10^{-2}$ are 
used and results are independent of it. 
\item (4) Redshift the photon by an amount $\Delta x = - H Dl/{\rm v}_{th}$
\item (5) Determine probalistically if the photon scatters after having 
traveled distance $Dl$. This is the case when a random number between zero
and unity is larger that ${\rm exp}(-\epsilon)$.
\item (6) If not return to (2), if yes proceed to (7)
\item (7) Determine via random number if (a) a Lyman-$\alpha$ photon is re-emitted (i.e re-scattering of the Lyman-$\alpha$ photon or the much less
likely alternatives (b) the intermediate $n=2$ state is reionised or (c)
the state decays via a two-photon transition. In case (a) go to (8), in case
(b) and (c) proceed to the next injected photon (1) where in (b) 
no net recombination has occurred, and in (c) it has. 
\item (8) Using three random numbers determine via Eq.~\ref{eq:MC} the probable $\tilde{p}_\|$, $\tilde{p}_{\perp 1}$, and $\tilde{p}_{\perp 2}$ of the scattering atom.  Using two more random numbers determine $\mu$ and $\beta$  
and compute the new gas rest frame frequency $x_{\rm out}$ from 
Eq.~\ref{eq:inout}. Return to (2).

\end{itemize}

Lyman-$\alpha$ photons which are lost since they redshifted out on the red 
wing out of resonance may be identified as the loop does not leave anymore
the range (2)-(6). These photons never reach (7), thus never scatter anymore,
and their frequency becomes ever smaller. Accounting for the numbers of
different scenarios it is straightforward to compute the well-known Peebles factor $C$ Eq.~\ref{Cfactor}. 

\begin{figure}[htbp]
\centering
\includegraphics[width=0.48\textwidth]{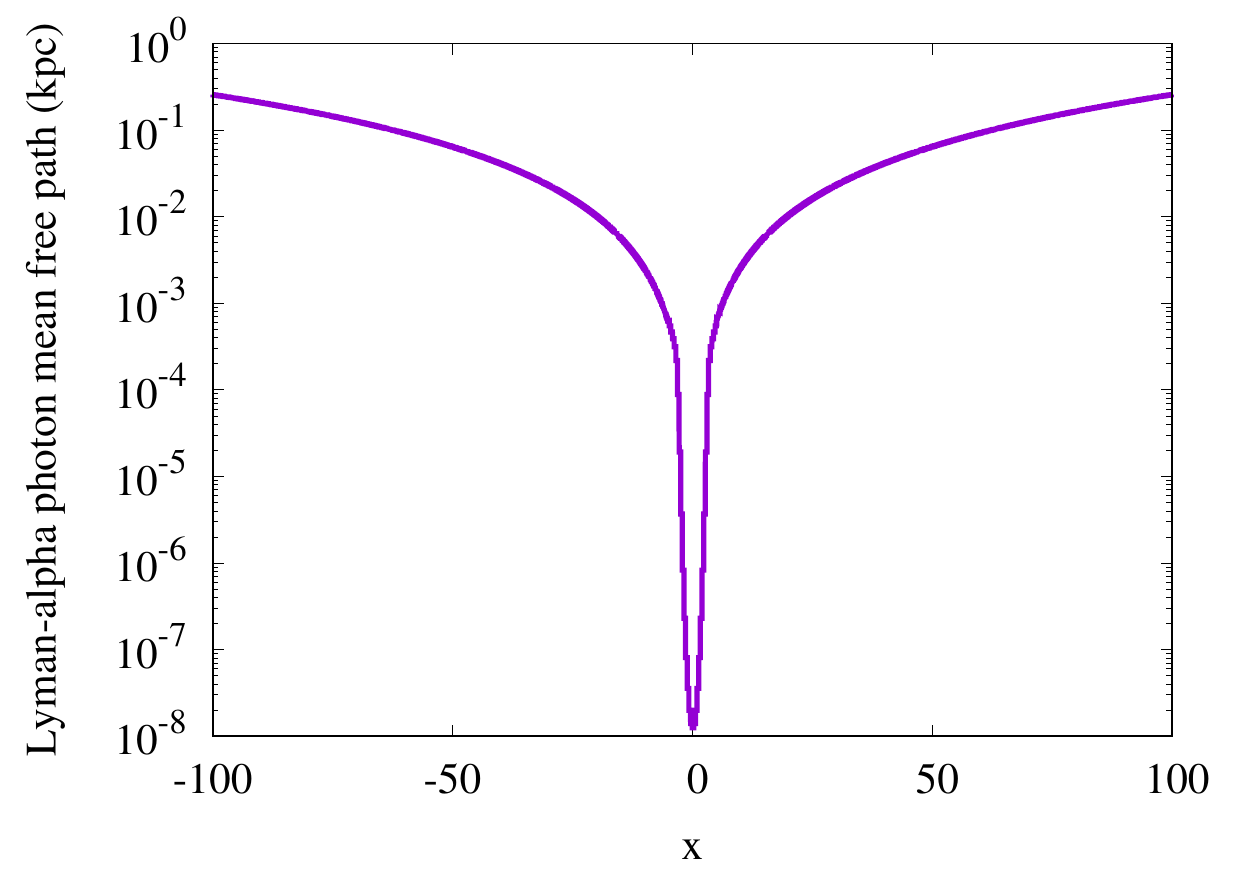}
\caption{Comoving Lyman-$\alpha$ photon mean free path in as a function
of photon frequency $x$ at redshift $z\approx 1100$.}
\label{fig:lalpha}
\end{figure}

The mean free path as a function of frequency is shown in 
Fig.~\ref{fig:lalpha}. It is
seen that it is extremely small (e.g. compare for example to the horizon
$l_H\sim 100\,$Mpc. Taking into account that 
$\sigma\sim 1/x^2$ on the wing, even photons on the extreme thermal red wing,
i.e. $x\sim -10^5$ may still scatter. However, at such small $x$ 
between each scattering the frequency redshifts further, reducing
the cross section even more, such that the photons with exceedingly high
probability become inert (except if reionisation happens during those
few last scatterings).
One may estimate the $x$ at which the probability for the Lyman-$\alpha$ photon
becoming inert for the $1s$-$2s$ transition is approximately fifty percent.
One can show \cite{Dijkstra:2017lio} that on the wing the probability distribution function
$P(x_{\rm out})$ one obtains from Eq.~\ref{eq:inout} is approximately given by a Gaussian with width unity and centered on $x_{\rm out} = x_{\rm in} - 1/x_{\rm in}$. That is $x$ performs a random walk during the many scatterings with,
however, a drift term which pulls it back to the center of the line $x=0$.
This drift term competes with the drift away from the center of the line due to
redshifting. Demanding approximate equality between these two drift terms
\begin{equation}
-\frac{H}{\sigma(x_{50})n_H {\rm v}_{th}} \approx \frac{-1}{x_{50}}
\end{equation}
one obtains at which $x_{50}$ the probability is approximately $50\%$ to
loose the photon. We obtain $x_{50}\approx 60$ which is confirmed by the
simulations. At this frequency the mean free path is approximately 
comoving $0.1\,$kpc
and as the frequency hovers often for many scatterings around $x_{50}$ the
distance traveled is easily in the larger than kpc regime, the scale on which magnetic fields excite inhomogeneities in density and flow.

\begin{figure}[htbp]
\centering
\includegraphics[width=0.48\textwidth]{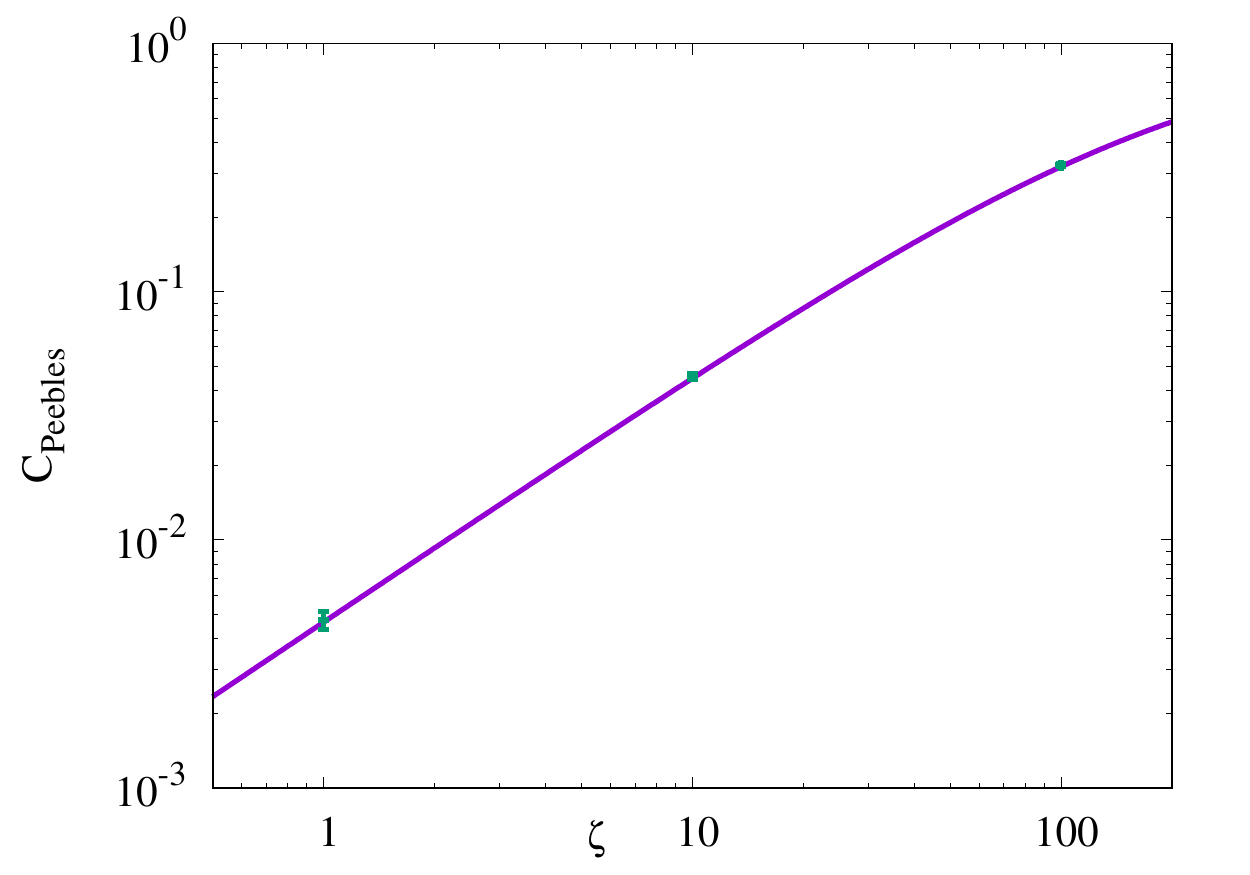}
\caption{Comparison of the theoretical and Monte-Carlo derived
value of the $C$ factor 
Eq.~\ref{Cfactor} in a homogeneous Universe, employing $\Lambda_{2\gamma} = 0$, as a function of $\zeta$,
where the cosmological expansion is given by $\zeta H$. Here $H$ is the correct
Hubble rate at recombination. The solid line shows the theoretical value whereas the points with error bars show results of the MC simulation.}
\label{fig:Cfact}
\end{figure}

We present next how the results of our Monte-Carlo simulation in
a homogeneous Universe compares to the theoretical $C$-factor. We consider
this as our test problem for the routine of these somewhat challenging
simulations (the evolution of one photon takes on average $\sim 20\,$seconds).
Fig.~\ref{fig:Cfact} shows the MC results compared to Eq.~\ref{Cfactor} for varying
Hubble constants, setting 
$\Lambda_{2\gamma}$ to zero~\footnote{Our simulations reproduce well
photon loss due to the $2s$-$1s$ two $\gamma$ transition}. 
It is seen that the comparison is close to perfect. This and the test displayed
in Fig.~\ref{fig:MCescape} for Lyman-$\alpha$ escape in the presence of
bulk flows gives us confidence that our Monte-Carlo simulation is reliable.

\section{Helium recombination} \label{app:helium}

The details of helium recombination are not crucial for the purposes of this work: the first helium recombination takes place around $z \sim 6000$, before the beginning of our simulations, and during the second helium recombination the free-electron fraction is limited to the range $1 \leq x_e \leq 1.08$, regardless of the details of the recombination process. Changes to the ionization history during helium recombination are also less critical to CMB anisotropies. We therefore defer a detailed study to future work, and in the meantime have implemented a simple 3-level model for helium, accounting only for transitions from the singlet $2P$ state (the equivalent of the hydrogen Lyman-$\alpha$ line), and for hydrogen continuum opacity, but neglecting transitions from the triplet $2P$ state. We adopt the simple model of Ref.~\cite{Kholupenko:2007qs}. We also ran an MC simulation for the Lyman-$\alpha$ transition in Helium and found an even larger mixing fraction (i.e.~smaller propagation distances) than for hydrogen, so that Helium recombination can also be assumed to proceed in the ``full-mixing" regime. We implemented the full-mixing regime in a similar fashion as for hydrogen.

\section{Convergence Study}
\label{app:convergence}

\begin{figure}[h]
\centering
\includegraphics[width=0.48\textwidth]{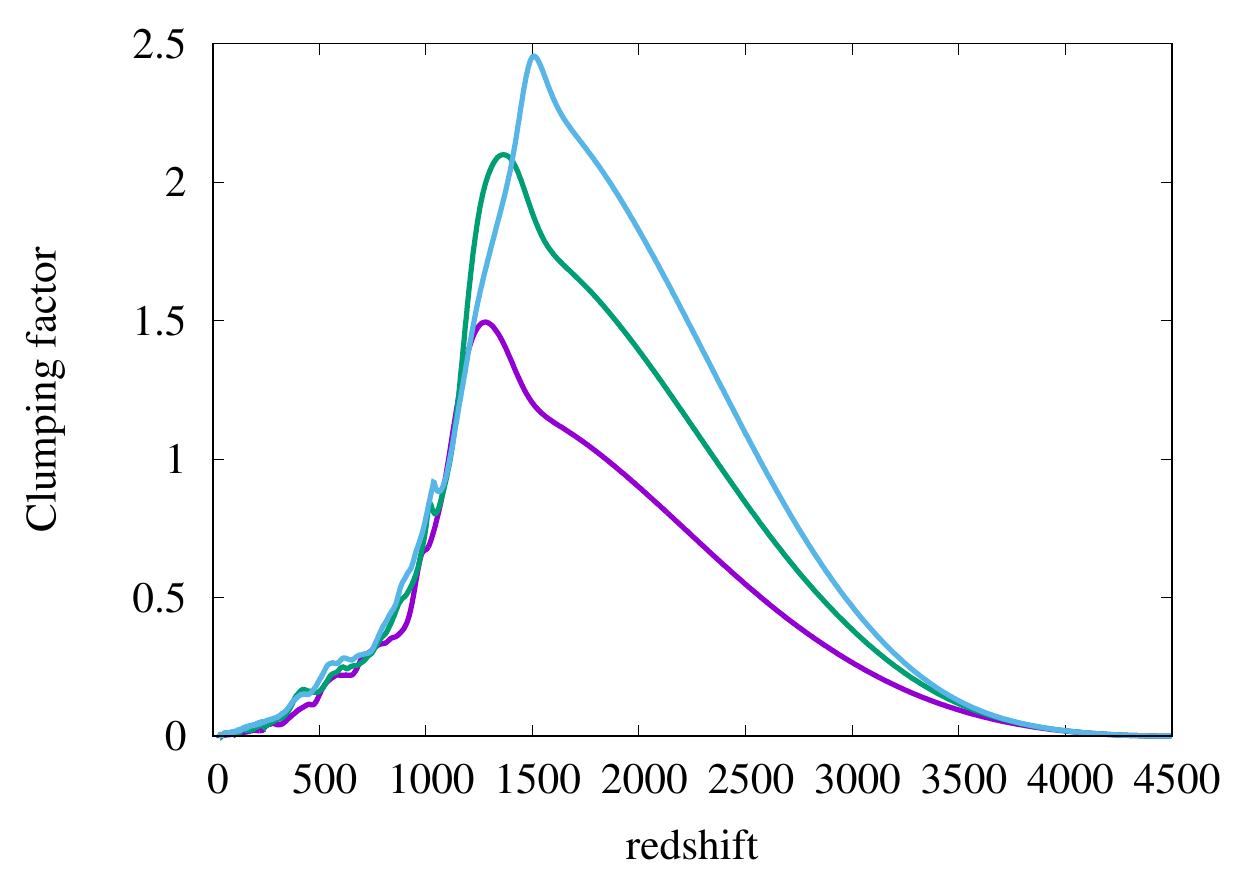}
\includegraphics[width=0.48\textwidth]{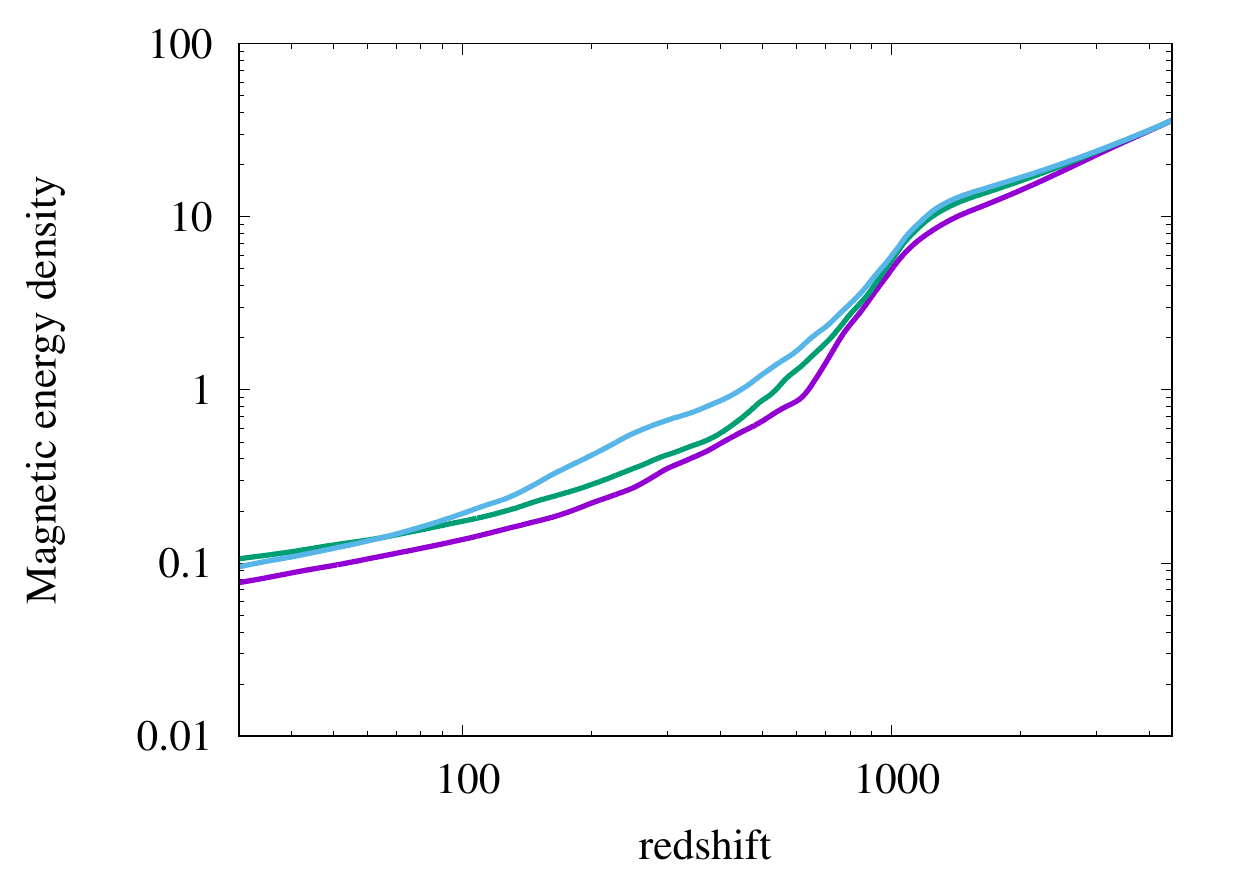}
\includegraphics[width=0.48\textwidth]{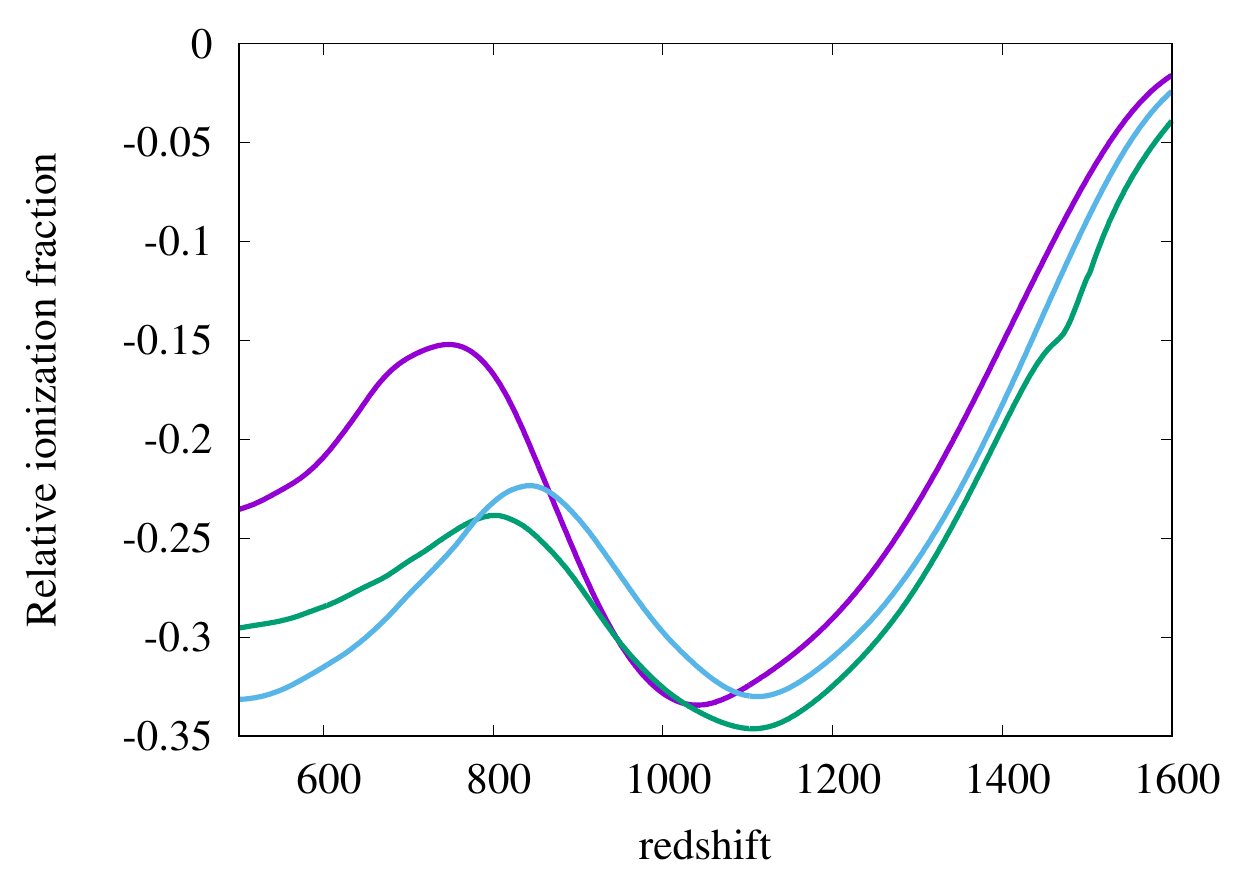}
\caption{Evolution of the clumping factor $b$ (\emph{top}), magnetic energy density (\emph{middle}) and change to the global free-electron fraction (\emph{bottom}) for different resolutions of the simulation. The simulation box is 6kpc and ${\rm v}_A^{rms} = 30 c_S$. All modes with $k\leq 2$ have been randomly excited. Here a mode is given by ${\rm v}_A(x) = {\rm v}_A^k {\rm sin} (2\pi k x + \phi_k)$ in a simulation box $0\leq x \leq 1$ and
${\rm v}_A^k$ and $\phi_k$ are random amplitude and phase for the mode $k$. A non-helical 
magnetic field with a Batchelor spectrum has been assumed. 
The pink, green, and blue lines are for resolutions $64^3$, $128^3$, and $256^3$, respectively.}
\label{fig:conv1}
\end{figure}



In this appendix we briefly study the convergence of results with the number
of employed zones. Fig. \ref{fig:conv1} shows the evolution of clumping factor, magnetic energy density, and ionization
fraction $X_e$ for a particular PMF spectrum and amplitude for three different
resolutions, $64^3$, $128^3$, and $256^3$. As modes only up to $k=2$ are 
excited, these modes are effectively resolved by $32$, $64$, and $128$ zones,
respectively. From Fig. \ref{fig:conv1} it is seen that convergence in the
clumping factor is not attained. In particular, the higher the resolution, the
higher the clumping factor. This is because the clumping factor is dominated
by the very few highest density regions, which can only be resolved with a large number of zones. On the
other hand, Fig.~\ref{fig:conv1} shows that approximate convergence in magnetic energy density
and $X_e$ is attained even at lower resolution. This is possible because those very high-density regions, which are not properly resolved, recombine very 
early independently of the exact overdensity. As they are rare as well, no 
significant impact on $X_e$ is observed. We infer that $32$ or $64$ zones per
Fourier mode should give somewhat accurate results.







\clearpage
\bibliography{pmf}

\end{document}